\documentclass[fleqn,usenatbib]{mnras}
\usepackage{graphicx}	
\usepackage{amsmath}	
\usepackage[dvipsnames]{xcolor}  
\usepackage{amssymb}	
\usepackage{newtxtext,newtxmath}    
\usepackage{ulem}
\usepackage{soul}
\usepackage{macros_vdb} 
\usepackage{fontawesome}
\usepackage{arydshln}

\usepackage{orcidlink}

\begin{document}
%

\title[BASILISK II. Satellite Kinematics in SDSS]
      {BASILISK II. Improved Constraints on the Galaxy-Halo Connection from Satellite Kinematics in SDSS}

\author[Mitra et al.]{%
   Kaustav Mitra$^1$\thanks{E-mail: kaustav.mitra@yale.edu}\orcidlink{0000-0001-8073-4554},
   Frank~C.~van den Bosch$^1$\orcidlink{0000-0003-3236-2068},
   Johannes U. Lange$^{2,3,4}$\orcidlink{0000-0002-2450-1366}
\vspace*{8pt}
\\
   $^1$Department of Astronomy, Yale University, PO. Box 208101, New Haven, CT 06520-8101\\
   $^2$Department of Physics, American University, 4400 Massachusetts Avenue NW, Washington, DC 20016, USA\\
   $^3$Department of Physics, University of Michigan, Ann Arbor, MI 48109, USA\\ 
   $^4$Leinweber Center for Theoretical Physics, University of Michigan, Ann Arbor, MI 48109, USA\\
}


\date{}

\pagerange{\pageref{firstpage}--\pageref{lastpage}}
\pubyear{2024}

\maketitle

\label{firstpage}


\begin{abstract}
\Basilisk is a novel Bayesian hierarchical method for inferring the galaxy-halo connection, including its scatter, using the kinematics of satellite galaxies extracted from a redshift survey. In this paper, we introduce crucial improvements, such as updated central and satellite selection, advanced modelling of impurities and interlopers, extending the kinematic modelling to fourth order by including the kurtosis of the line-of-sight velocity distribution, and utilizing satellite abundance as additional constraint. This drastically enhances \Basilisc's performance, resulting in an unbiased recovery of the full conditional luminosity function (central and satellite) and with unprecedented precision. After validating \Basilisc’s performance using realistic mock data, we apply it to the SDSS-DR7 data. The resulting inferences on the galaxy-halo connection are consistent with, but significantly tighter than, previous constraints from galaxy group catalogues, galaxy clustering and galaxy-galaxy lensing. Using full projected phase-space information, \Basilisk breaks the mass-anisotropy degeneracy, thus providing precise global constraint on the average orbital velocity anisotropy of satellite galaxies across a wide range of halo masses. Satellite orbits are found to be mildly radially anisotropic, in good agreement with the mean anisotropy for subhaloes in dark matter-only simulations. Thus, we establish \Basilisk as a powerful tool that is not only more constraining than other methods on similar volumes of data, but crucially, is also insensitive to halo assembly bias which plagues the commonly used techniques like galaxy clustering and galaxy-galaxy lensing. 
\end{abstract} 




\section{Introduction}
\label{sec:intro}

According to the current cosmological paradigm, the vast majority of all galaxies form and reside in extended dark matter haloes \citep[][]{White.Rees.78, MBW10}. Halo occupation modelling tries to use observational constraints on the population of galaxies in order to infer the statistical link between the galaxy properties (mainly their luminosity or stellar mass) and the properties of the dark matter haloes (mainly some measure of halo mass) in which they reside \citep[see][for a review]{Wechsler.Tinker.18}.  The resulting `galaxy-halo connection' provides valuable insight regarding the formation and evolution of galaxies, and benchmarks to calibrate, compare and validate semi-analytic models \citep[eg.][]{Somerville.etal.2008b, Stevens.etal.2016} and simulations \citep[eg.][]{Crain.etal.2015, Munshi.etal.2013, Stinson.etal.2013}. In addition, since it describes the link between the light we see and the mass that governs the dynamical evolution of the Universe, it is a powerful tool that allows astronomers to constrain cosmological parameters using the observed distribution of galaxies \citep[e.g.,][]{Yang.etal.04, Seljak.etal.05, Tinker.etal.05, Yoo.etal.06, Cacciato.etal.09}.

Arguably, the most straightforward method to infer the galaxy halo connection, and one that has become extremely popular, is subhalo abundance matching \citep[hereafter SHAM,][]{Kravtsov.etal.04, Vale.Ostriker.06, Conroy.etal.06, Reddick.etal.13}. It matches the ordered list of galaxies (typically ranked by stellar mass or luminosity) to that of subhaloes (typically ranked by their peak or infall mass)\footnote{Rather than abundance matching individual galaxies to subhaloes, one can also match the abundance of galaxy groups (identified using some group finder) to dark matter host haloes \citep[e.g.,][]{Yang.etal.05, Yang.etal.07}.}. In this mostly non-parametric method, one usually allows for some amount of scatter (a free parameter) in the rank-order matching, to have realistic spread in the stellar mass - halo mass relation \citep[e.g.,][]{Behroozi.etal.10}. A key advantage of SHAM over other methods, discussed below, is that it only requires stellar mass (or luminosity) measurements of the galaxies. However, an important downside is that it relies crucially on the assumption that both the galaxy sample and the (sub)halo sample (typically taken from a $N$-body simulation) are complete. Hence, SHAM cannot be applied to subsamples of galaxies (i.e., samples of emission line galaxies, or galaxies selected by colour). In addition, even if the galaxy sample is complete, the (sub)halo catalogues used, which are typically extracted from numerical simulations, suffer from incompleteness due to artificial disruption \citep[][]{vdBosch.etal.18a, vdBosch.etal.18b} and failures of subhalo finders \citep{Han.etal.12, Diemer.etal.24}. This can significantly impact the galaxy-halo connection inferred via SHAM \citep[][]{Campbell.etal.18}. 

These problems can be overcome using data that more directly constrains halo mass. The two most commonly used methods are galaxy clustering \citep[e.g.,][]{Berlind.Weinberg.02, Yang.etal.03, vdBosch.etal.07, Zheng.etal.07, Zehavi.etal.11} and galaxy-galaxy lensing \citep[e.g.,][]{Guzik.Seljak.02, Mandelbaum.etal.06, Mandelbaum.etal.16, Leauthaud.etal.17}. The former relies on the fact that more massive haloes are more strongly clustered \citep[][]{Mo.White.96}; hence, the clustering strength of a given population of galaxies informs the characteristic mass of the haloes in which they reside. Unfortunately, its reliability is hampered by the finding that halo clustering strength depends not only on mass but also on secondary halo properties \citep[e.g.,][]{Gao.etal.05, Wechsler.etal.06, Dalal.etal.08, Lacerna.Padilla.11, Salcedo.etal.18}, something that is collectively referred to as halo assembly bias. Galaxy-galaxy lensing, which is a manifestation of weak gravitational lensing, uses the tangential shear distortions of distant background galaxies around foreground ones in order to constrain the halo masses of the latter \citep[e.g.,][]{Brainerd.etal.96, Hoekstra.etal.01, Sheldon.etal.04, Mandelbaum.etal.06}. Although, in principle, a fairly direct probe of halo mass, this method requires tedious shape measurements of faint background sources, which can be prone to effects like blending and intrinsic alignment. Typically the background sources lack spectroscopic redshifts, which can also cause systematic errors in the modelling of their measured shear distortions. In addition, on large scales the 2-halo term of the lensing shear is also impacted by the same assembly bias issues that plague clustering.

Another method that can be used to constrain the galaxy-halo connection, but which has hitherto been severely under-utilized, is satellite kinematics. It uses measurements of the line-of-sight velocities of satellite galaxies with respect to their corresponding central galaxy in order to constrain the gravitational potential, and hence the mass, of the host halo\footnote{Famously, the same principle was used by \cite{Zwicky.33} in order to infer the presence of dark matter in the Coma cluster}. With the exception of large galaxy groups and clusters, individual central galaxies typically only have a few spectroscopically detected satellites. Consequently, it is common to combine the satellite velocity measurements from a large stack of central galaxies in order to estimate an average satellite velocity dispersion, which in turn is used to infer an average host halo mass using either a virial mass estimator or a simple Jeans model \citep[e.g.,][]{Zaritsky.etal.93, Brainerd.Specian.03, Prada.etal.03, Norberg.etal.08, Wojtak.Mamon.13}.

It has often been argued that satellite kinematics is not a reliable mass estimator for any combination of the following reasons: (a) satellite galaxies are not necessarily virialized tracers \citep[][]{Wang.etal.17, Wang.etal.18, Adhikari.etal.19}, (b) their orbits may well be anisotropic \citep[][]{Diemand.etal.04, Cuesta.etal.08, Wojtak.Mamon.13}, resulting in a well-known mass-anisotropy degeneracy \citep[][]{Binney.Mamon.82}, and (c) the stacking that is used implies `mass-mixing' (i.e., combining the kinematics of haloes of different masses), which muddles the interpretation of the data. In addition, the selection of centrals and satellites from a redshift survey is unavoidably impacted by impurities, incompleteness and interlopers, further complicating the analysis. Despite these concerns, a number of studies have progressively improved satellite kinematics and have shown that it can yield reliable, as well as precise, constraints on the galaxy-halo connections\footnote{and on the masses of individual clusters \citep[e.g.,][]{Biviano.etal.06, Munari.etal.13, Saro.etal.13, Old.etal.15, AguirreTagliaferro.etal.21}}. In particular, \cite{vdBosch.etal.04} demonstrated that by selecting centrals and satellites using iterative, adaptive selection criteria the impact of impurities and interlopers can be minimized.  \cite{More.etal.09a} has shown that by combining different weighting schemes one can accurately account for mass mixing, and even constrain the scatter in the stellar mass-halo mass relation \citep[also see][]{More.etal.09b, More.etal.11}. This was significantly improved upon by \cite{Lange.etal.19a, Lange.etal.19b} who demonstrated that kinematics of satellite galaxies from a large redshift survey such as the Sloan Digital Sky Survey \citep[SDSS][]{York.etal.00} can yield constraints on the galaxy-halo connection that are complementary to, and competitive with, constraints from galaxy clustering and/or galaxy-galaxy lensing. 
\citet{Wojtak.Mamon.13} were the first to analyse satellite kinematics while accounting for orbital anisotropy. Using a method first developed by \cite{Wojtak.etal.08, Wojtak.etal.09} they were able to simultaneously constrain halo mass, halo concentration and orbital anisotropy, albeit without accounting for mass mixing.

\citet[][hereafter \papI]{vdBosch.etal.19} developed \Basilisc, a Bayesian hierarchical inference formalism that further improves on the ability of satellite kinematics to constrain the galaxy halo connection. Unlike previous methods, \Basilisk does not resort to stacking the kinematics of satellite galaxies in bins of central luminosity, and does not make use of summary statistics, such as satellite velocity dispersion. Rather, it leaves the data in the raw form and computes the corresponding likelihood. Consequently, it can simultaneously solve for halo mass and orbital anisotropy of the satellite galaxies, while properly accounting for scatter in the galaxy-halo connection. In addition, \Basilisk can be applied to flux-limited, rather than volume-limited samples, greatly enhancing the amount and dynamic range of the data. 

Paper~I also tested and validated \Basilisk against mock data sets of varying complexity, and demonstrated that it yields unbiased constraints on the parameters specifying the galaxy-halo connections. However, in order to speed up the analyses, all those tests where performed using mock data samples that were only about $1/8$ the size of the full SDSS sample analysed here.  When we ran \Basilisk on full-sized mocks instead, the model parameter uncertainties shrank considerably, as expected, revealing several significant discrepancies that turned out to be systematic. This necessitated a number of modifications to \Basilisc, which we present in the first half of this paper. Most notably, we introduce significant improvements to the treatment of interlopers (i.e., galaxies that are selected as satellites but that do not reside in the same dark matter halo as the central), allowing for both a population of splash-back galaxies \citep[][]{Diemer.Kravtsov.14, Adhikari.etal.14, More.etal.15a, ONeil.etal.22} and a large scale infall population akin to the well-known \cite{Kaiser.87} effect. In addition, we slightly modify the cylindrical selection criteria in order to improve the purity of our sample (i.e., reduce misclassification of satellites as centrals), we assure that the selection of secondaries around each individual primary is volume-limited, and we forward-model the contribution of impurities that arise from haloes in which the brightest galaxy is a satellite rather than the central. We also let go of the oversimplified assumption that the satellite velocity profile along any given line-of-sight is Gaussian, as was done in \papI. Rather, we now use the fourth-order Jeans equations to model  the kurtosis of the line-of-sight velocity distribution (LOSVD). This enables more accurate modelling of the full phase-space distribution of satellite galaxies, and allows \Basilisk to break the mass-anisotropy degeneracy. Finally, we also replace fitting binned statistics of centrals with zero (detected) satellites, as done in \papI, with a more general, Bayesian hierarchical modelling of the number of satellites around each central. Although  this data on satellite abundances does not yield {\it direct} kinematic constraints on halo mass, it greatly helps to constrain the overall galaxy-halo connection. 

The goal of this paper is threefold: (i) showcase the advancements in satellite kinematics methodology that we have introduced in \Basilisc, and highlight its improved performance when tested against realistic SDSS-like mock data; (ii) apply \Basilisk to SDSS DR7 data to simultaneously constrain the conditional luminosity functions of central and satellite galaxies, the satellite velocity anisotropy and satellite radial distribution, all with unprecedented precision, and compare those with previous constraints on halo occupation statistics; and (iii) establish \Basilisk as a powerful method to infer the galaxy-halo connection which is free of halo assembly bias effects, and that is even more constraining than commonly used techniques like galaxy clustering and galaxy-galaxy lensing, when applied on data of similar volumes.

Throughout this paper we adopt the flat Planck18 $\Lambda$CDM cosmology \citep[][]{Planck.18}, which has matter density parameter $\Omega_\rmm = 0.315$, power spectrum normalization $\sigma_8=0.811$, spectral index $n_\rms = 0.9649$, Hubble parameter $h = (H_0/100\kmsmpc) = 0.6736$ and baryon density $\Omega_\rmb h^2 = 0.02237$.\footnote{These are the TT,TE,EE$+$lowE$+$lensing best-fit values assuming a base-$\Lambda$CDM cosmology.}

\section{Sample Selection}
\label{sec:sample_selection}

\subsection{Selecting central-satellite pairs}
\label{sec:selection}

The first step in analysing satellite kinematics is to select a sample of centrals and their associated satellites from a redshift survey. Unfortunately, this selection is never perfect; one undoubtedly ends up selecting some bright satellites as centrals (we refer to these as `impurities') and not every galaxy selected as a satellite actually resides in the same host dark matter halo as the corresponding central (those that don't are referred to as `interlopers'). In what follows, we therefore use `primaries' and `secondaries' to refer to galaxies that are selected as centrals and satellites, respectively.

\begin{figure}
\centering
\includegraphics[width=0.48\textwidth]{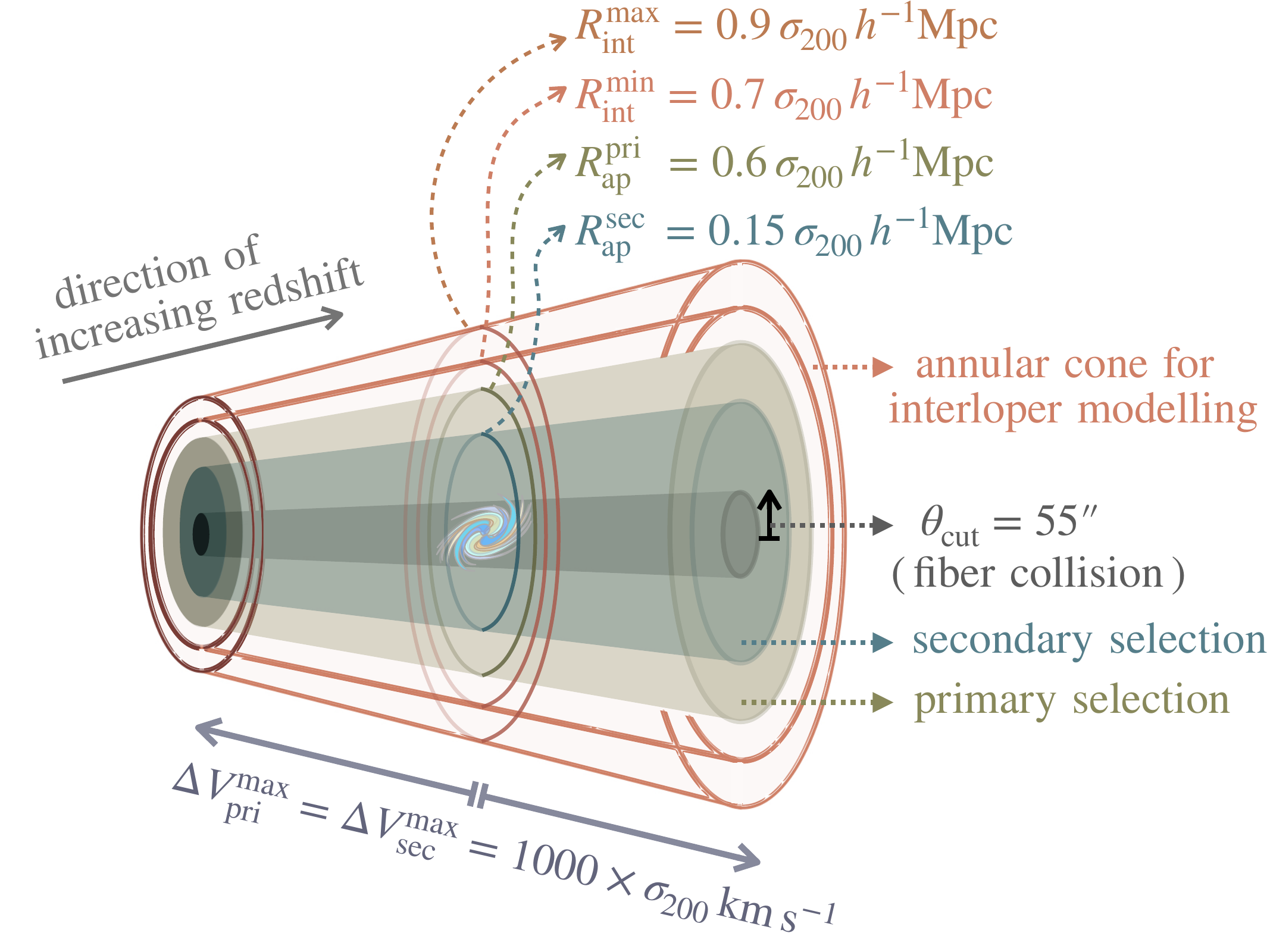}
\caption{Schematic showing the conical cylinders used for the selection of primaries and secondaries, and for modelling the interloper velocity profile. Redshift increases from left to right, and the depth of these cones are expressed as a velocity difference with respect to the primary galaxy on which the cones are centered. Note that all radii listed here are computed at the redshift of the primary. see the text for details.}
\label{fig:conical_cylinders}
\end{figure}

A galaxy at redshift $z$ is considered a potential primary if it is the brightest galaxy in a conical volume of opening angle $\Theta_{\rm ap}^{\rm pri} \equiv \Rh/d_\rmA(z)$  centered on the galaxy in question, and extending along the line-of-sight from $z-(\Delta z)^{\rm pri}$ to $z+(\Delta z)^{\rm pri}$. Here $d_\rmA(z)$ is the angular diameter distance at redshift $z$, and $(\Delta z)^{\rm pri} = (\dVh/c) \, (1+z)$. The parameters $\Rh$ and $\dVh$ specify the primary selection cone. Following \cite{Lange.etal.19a}, we select the primaries in a rank-ordered fashion, starting with the most luminous galaxy in the survey. Any galaxy located inside the selection cone of a brighter galaxy is removed from the list of potential primaries. All galaxies fainter than the primary and located inside a similar cone, but defined by $\Rs$ and $\dVs$, centred on the primary are identified as its secondaries.  Note that, although it is common to refer to these selection volumes as `cylinders', a convention we also adopted in \papI, in actuality the selection volumes are frustums of cones. In order to rectify this confusing nomenclature, in this paper we refer to them as `selection cones' (see Fig.~\ref{fig:conical_cylinders}).

The four parameters $\Rh$, $\dVh$, $\Rs$, and $\dVs$ control the completeness and purity of the sample of primaries and secondaries. Increasing $\Rh$ and/or $\dVh$, boosts the purity among primaries (i.e., it reduces the number of satellites erroneously identified as centrals), but reduces the overall completeness. Similarly, decreasing $\Rs$ and/or $\dVs$ reduces the number of interlopers, but at the costs of a reduced number of satellites, which are the dynamical tracers of interest. Since brighter primaries typically reside in larger, more massive haloes, it is advantageous to scale the sizes of the selection cones with the luminosity of the primary \citep[][]{vdBosch.etal.04}. In particular, we adopt $\Rh = 0.6 \,\sigma_{200} \mpch$, $\Rs = 0.15 \, \sigma_{200} \mpch$, and $\dVh = \dVs = 1000\,\sigma_{200} \kms$. Here $\sigma_{200}$ is a rough measure of the satellite velocity dispersion in units of $200\kms$, which, following \citet{vdBosch.etal.04} and \citet{More.etal.09b}, we take to scale with the luminosity of the primary as
\begin{equation}\label{sig200}
\log \sigma_{200} = 0.04 + 0.48 \log L_{10} + 0.05 (\log L_{10})^2\,,
\end{equation}
where $L_{10} = L / (10^{10} \Lsunh)$, and $\sigma_{200}$ is allowed to take a maximum value of 4. The values of $\Rh$ and $\Rs$ correspond to roughly $1.65$ and $0.4$ times the halo virial radius, respectively, while the value for $\dVs$ is large enough to include the vast majority of all satellites around primaries of the corresponding luminosity. Note that the numerical values in these selection criteria are tuned in order to optimize the selection of primaries and secondaries against impurities and interlopers. In particular, they are slightly different from the values we adopted in \papI. As detailed in Appendix~\ref{App:selcrit}, this is done in order to reduce the fraction of impurities that are neither true centrals, nor the brightest satellites in their corresponding host haloes. These impurities are particularly difficult to account for in our forward-modelling approach and can cause a small but systematic overestimate of the scatter in the relation between halo mass and central luminosity.

The SDSS redshift catalogue, to which we apply \Basilisk in this study, is a flux-limited survey. As emphasized in \papI, an important advantage of \Basilisk over earlier studies of satellite kinematics, is that it is not limited to volume-limited subsamples, thereby greatly boosting the number of primaries and secondaries to be used in the analysis. However, in order to facilitate proper modelling of the number of secondaries (true satellites and interlopers) we need to assure that the selection of secondaries around each individual primary {\it is} volume limited. This is something that was not implemented in \papI, but which turned out to be important in order to avoid a systematic bias in the inferred faint-end slope of the satellite luminosity function. This effect was not significant in the smaller mock data samples used to test \Basilisk in \papI, but could no longer be overlooked using data sets comparable in size to the SDSS data used here.

\begin{figure}
\centering
\includegraphics[width=0.47\textwidth]{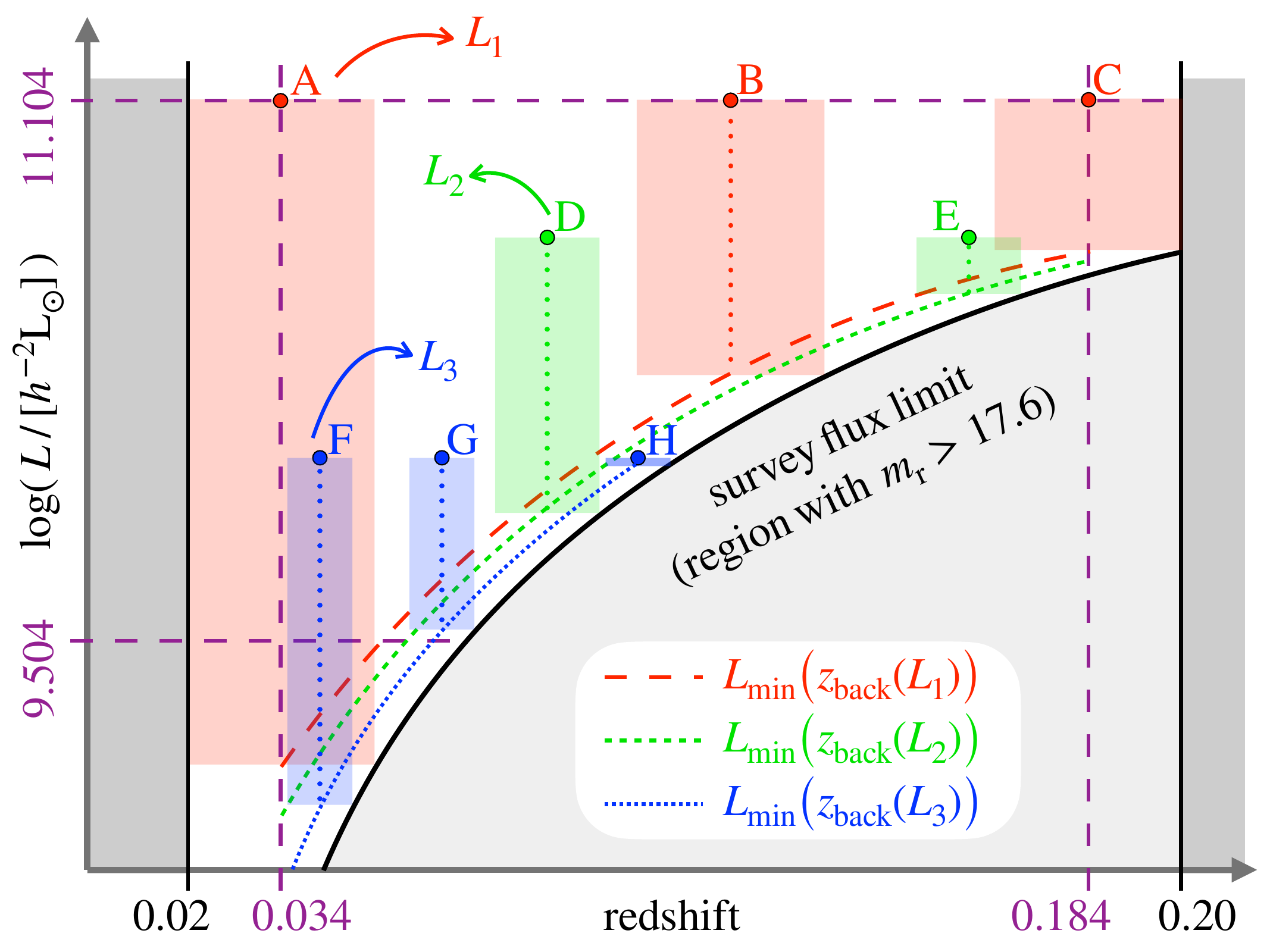}
\caption{Illustration showing the luminosity and redshift cuts that play a role in our selection
of primaries and secondaries. Coloured dots indicate primaries while the corresponding shaded rectangles indicate the volume-limited luminosity and redshift ranges of their secondaries. see the text for a detailed explanation.}
\label{fig:logL_z_selection}
\end{figure}

In this paper, we limit our analysis to primaries in the luminosity range $9.504 \leq \log(\Lc/[h^{-2}\Lsun]) \leq 11.104$, corresponding to $-19 \leq M_r^{0.1} - 5\log h \leq -23$. Here $M_r^{0.1}$ is the absolute magnitude in the SDSS $r$-band $K$+$E$ corrected to $z=0.1$. In addition, we only use data in the redshift range $0.02\leq z \leq 0.20$.  Note that the selection cone used to identify secondaries around a primary of luminosity $L_{\rm pri}$ at redshift $z_{\rm pri}$ extends from $\zf$ to $\zb$, given by
\begin{align}\label{zback}
&\zf = z_{\rm pri} -  \frac{\Delta V_{\rm max}^{\rm sec}(L_{\rm pri}) }{ c} (1 + z_{\rm pri}) \, \, \, \, {\rm and} \nonumber \\
&\zb = z_{\rm pri} +  \frac{\Delta V_{\rm max}^{\rm sec}(L_{\rm pri}) }{ c} (1 + z_{\rm pri}) \,.
\end{align}
Hence, as depicted in Fig.~\ref{fig:logL_z_selection} we are only complete in the selection of secondaries with luminosities $\Ls \geq L_{\rm min}(\zb)$. Here $L_{\rm min}(z)$ is the minimum luminosity of galaxies at redshift $z$ that make the apparent magnitude limit of our survey data ($m_r = 17.6$; see \S\ref{sec:SDSS}). In order to assure a complete, volume-limited selection of secondaries around each primary, secondaries fainter than $L_{\rm min}(\zb)$ are discarded. In addition, in order to assure that the entire secondary selection cone around a given primary fits within the flux-limits of the SDSS data, we require that  $\Lpri > L_{\rm min}(\zb)$. Finally, the redshifts of the primaries are restricted to $0.034 \leq z_{\rm pri} \leq 0.184$, such that the $\zf$ and $\zb$ of the most luminous primaries fit within the $0.02 \leq z \leq 0.20$ limits of the entire sample. 

To elucidate this further, Fig.~\ref{fig:logL_z_selection} illustrates the bounds on the selection of primaries and secondaries.  The solid, vertical lines at $z=0.02$ and $z=0.20$ mark the minimum and maximum redshifts of the entire sample, while the dashed, vertical lines at $z=0.034$ and $z=0.184$ mark the redshift limits allowed for primaries. Dashed, horizontal lines at $\log L = 11.104$ and $\log L = 9.504$ mark the luminosity cuts for primaries. The solid circles, labelled A to H, represent hypothetical primaries of three different luminosities $L_1$, $L_2$ and $L_3$ (indicated by three different colours). The shaded rectangle associated with each primary indicates the allowed luminosity-redshift ranges of its corresponding secondaries. Since the redshift extent of the secondary selection cone scales with the luminosity of the primary, the shaded regions of fainter primaries have a smaller $\Delta z$-extent, as is evident from the figure. Note that these shaded regions extend down to where the apparent magnitude of the secondaries at the back-end of the selection cone is equal to the magnitude limit of the survey, which can be  significantly lower than $\log L = 9.504$, specifically for the primaries that are relatively nearby (e.g., primaries A and F).

Primary A is at the minimum allowed redshift for primaries, $z_{\rm min}^{\rm pri} = 0.034$, such that the `front' end of its secondary selection cone is equal to the minimum redshift of our survey data (i.e., $\zf=0.02$). Similarly, primary C is located at the maximum allowed redshift for primaries, $z_{\rm max}^{\rm pri} = 0.184$, and has $\zb = 0.20$.  Primary H is also special in that it has the highest redshift possible given its luminosity. Had it been any farther away, the far end of its secondary selection cone would stick outside of the SDSS flux limit, resulting in incompleteness. The three dashed and dotted curved lines, labelled as $L_{\rm min}(\zb(L_{1/2/3}))$, show the lower luminosity limits for secondaries as a function of $\zpri$, corresponding to each of the three different primary luminosities represented in the figure. For example, primaries of luminosity $L_2$ (like primary D and E) can not have secondaries fainter than the middle dashed curve in green. This ensures that their secondaries are individually volume limited around each of those primaries.

\subsection{Survey incompleteness}
\label{sec:incompleteness}

As any spectroscopic redshift survey, the SDSS data, from which our sample of primaries and secondaries derives, suffers from spectroscopic incompleteness due to fibre collisions and other failure modes \citep[see][for details]{Blanton.etal.05}. Each galaxy in the survey is assigned a spectroscopic completeness, $\mathcal{C}_{\rm spec}$, which indicates the fraction of spectroscopic targets in the angular region of the galaxy in question with a useful spectrum. In order to avoid primaries in regions with poor spectroscopic completeness, we remove all primaries with $\mathcal{C}_{\rm spec} < 0.8$.

If a primary is close to the edge of the survey, such that its secondary selection cone sticks partially outside of the survey footprint, or if the secondary selection cone overlaps with a masked region, the number of secondaries may be underestimated. In order to account for this, we proceed as follows. For each primary we uniformly distribute $\sim 5\times 10^4$ particles in its secondary selection cone. We then compute the fraction, $w_{\rm app}$, of those particles that are located inside the angular footprint of the SDSS, accounting for both survey edges and masked areas. In what follows we use $w_{{\rm app},i}$ to denote the aperture completeness of galaxy $i$. In order to avoid primaries with a poor aperture completeness, we remove all primaries with $w_{\rm app} < 0.8$.

As demonstrated in \citet{Lange.etal.19a}, it is important to correct satellite kinematics data for  fibre-collision induced incompleteness. In the SDSS, spectroscopic fibres cannot be placed simultaneously on a single plate for objects separated by less than $\vartheta_{\rm fc} \equiv 55''$ \citep{Blanton.etal.03a}. Although some galaxies are observed with multiple plates, yielding spectroscopic redshifts even for close pairs, roughly 65\% of galaxies with a neighbour within $55''$ lack redshifts due to this fibre collision effect. In order to correct the data for the presence of fibre collisions, we follow \citet{Lange.etal.19a} and start by assigning each fibre-collided galaxy the redshift of its nearest neighbour \citep[see][]{Blanton.etal.05, Zehavi.etal.05}. Note that we only use these during the {\it identification} of primaries. Once the selection is complete, all fiber-collided primaries and secondaries are removed from the sample\footnote{As shown in \cite{Lange.etal.19a}, including fibre-collided galaxies during the selection of primaries significantly reduces sample impurity.}. In addition, each galaxy is assigned a spectroscopic weight, $w_{\rm spec}$, that is computed as follows. For each galaxy we first count the number of neighboring galaxies, $n$, brighter than $m_r = 17.6$ within a projected separation less than $55''$. Next, for all galaxies in the survey with $n$ neighbours, we compute the fraction, $f_{\rm spec}$, of those neighbours that have been successfully assigned a redshift. Finally, all galaxies with $n$ neighbours are then assigned a spectroscopic weight equal to $w_{\rm spec} = 1 / f_{\rm spec}$. 

In order to correct for aperture incompleteness and fibre collisions, \Basilisk down-weights the expectation value for the number of secondaries around primary $i$ (see equation~[\ref{lamtot}] below), using the following correction factor:
\begin{equation}\label{fcorr}
f_{{\rm corr},i} = \frac{\Nsi}{\sum_{j=1}^{\Nsi} w_{{\rm spec},ij}} \, w_{{\rm app},i}\,.
\end{equation}
Here $w_{{\rm spec},ij}$ is the spectroscopic weight for secondary $j$ associated with primary $i$. Since correcting for fibre collisions is extremely difficult on scales below the fibre-collision scale, we remove all secondaries with $R_\rmp < R_{\rm cut}(\zpri) \equiv d_\rmA(\zpri) \, \vartheta_{\rm fc}$. Hence, the secondary selection volumes used in the end are conical frustums with a central hole with an opening angle of $55''$ (see Fig.~\ref{fig:conical_cylinders}). As shown in \paperI and \citet{Lange.etal.19a}, this combined approach of down-weighting the model predictions for the number of secondaries and ignoring secondaries below the fibre-collision scale accurately accounts for incompleteness arising from fibre-collisions in the SDSS.

\section{Observables}
\label{sec:observables}

Here we describe the various observables used by \Basilisk in order to constrain the galaxy-halo connection. These include (i) accessible 2D phase-space parameters of primary-secondary pairs (line-of-sight velocity and projected separation), which contains the information regarding the kinematics of satellite galaxies, (ii) statistics regarding the number of secondaries per primary (including primaries with zero secondaries), which helps to constrain the halo occupation statistics, and (iii) the galaxy luminosity function. The following subsections discuss each of these observables in detail.

\subsection{Satellite kinematics}
\label{sec:satkin}

For each primary-secondary pair in the sample we compute their projected
separation 
\begin{equation}\label{Rpdef}
\Rp = d_\rmA(\zpri) \, \vartheta\,,
\end{equation}
and their line-of-sight velocity difference
\begin{equation}\label{dvdef}
\dV = c \, \frac{(\zsec - \zpri)}{1 + \zpri}\,.
\end{equation}
Here $\zpri$ and $\zsec$ are the observed redshifts of the primary and secondary, respectively, $c$ is the speed of light, and $\vartheta$ is the angular separation between the primary and secondary on the sky.

As detailed in \papI, the main data vector used in \Basilisk is given by 
\begin{equation}\label{dataprim}
\bD_{\rm SK} = \bigcup\limits_{i=1}^{N_{+}} \, \left( \{\dVij, \Rij | j=1,...,\Nsi \} | \Lci, \zci, \Nsi\right)\,.
\end{equation}
where the union is over all $N_{+}$ primaries with at least one secondary. Here $\Nsi$ is the number of secondaries associated with primary $i$, and it is made  explicit that $\Lci$, $\zci$, and $\Nsi$ are only treated as {\it conditionals} for the data $\{\dVij, \Rij | j=1,...,\Nsi \}$. In other words, we consider $\Lci$, $\zci$ and $\Nsi$ as `given' and shall not use the distributions of these quantities as constraints on our satellite kinematics likelihood. Rather, \Basilisk uses the number densities of {\it all} galaxies as additional constraints (see \S\ref{sec:galnumdens}). The main reason for doing so is to make the method less sensitive to the detailed selection of primaries, which is difficult to model in detail. In particular, as discussed in \papI, this approach makes \Basilisk fairly insensitive to details regarding the $\sigma_{200}(L)$ relation (equation~[\ref{sig200}]) used to define the selection cones.

\subsection{Number of secondaries}
\label{sec:Pnull}

The data vector $\bD_{\rm SK}$ described above only contains primaries with at least one secondary. The complementary data vector $\bD_0 = (\{\Lci,\zci\} \, | \,i=1,2,...,N_0)$ lists all $N_0$ primaries with zero spectroscopically detected secondaries. Even though $\bD_0$ contains no kinematic data, it still provides additional constraints on the galaxy-halo connection, in particular regarding the occupation statistics of satellite galaxies. In \paperI we utilized this information by computing the fraction, $P_0 = N_0 /(N_0 + N_{+})$, of primaries, in a given bin of $\log \Lpri$ and $\zpri$, that have zero secondaries. Here $N_0$ is the number of `lonely primaries' with zero detected secondaries, and $N_{+}$ is the number of primaries that have at least one secondary. As discussed in \papI, this $P_0$ statistic provides valuable constraints on the galaxy-halo connection. However, upon closer examination we found that the binning used in this method causes small, but systematic errors in the inference. Using smaller bins was not able to solve this problem, which is why we ultimately opted for the following alternative, unbinned approach. 

In line with \Basilisc's philosophy to leave the data as much as possible in its raw form, rather than computing $P_0$ on a $(\log \Lpri, \zpri)$-grid, we use the following raw data vector as constraint on the model:
\begin{equation}\label{dataNs}
\bDNs = \bigcup\limits_{i=1}^{N_{\rm NS}} \left( \Nsi | \Lci, \zci \right)\,.
\end{equation}
Here the union is over (a random subset of) {\it all} $\Npri = N_0 + N_{+}$ primaries, independent of how many secondaries they have (i.e., including the primaries with zero secondaries). Since $N_0 \gg N_{+}$, computing the likelihood $\bDNs$ for all $\Npri$ primaries is much more time-consuming than computing the likelihood for the satellite kinematics data vector (eq.~[\ref{dataprim}]). Therefore, we only use a downsampled, random subset of $N_{\rm NS} = \calO(N_{+})$ primaries, where each primary has a probability equal to $N_{+}/\Npri$ to be included. In the case of the SDSS data set described in \S\ref{sec:SDSS} this probability is $0.094$. This downsampling assures that the computation of the likelihood for $\bDNs$ has a CPU requirement that is comparable to that for $\bD_{\rm SK}$. We emphasize that our constraints are primarily driven by the satellite kinematics data. Hence, this down-sampling of the satellite abundance data has no significant impact on our constraining power of the central galaxy-halo connection or the orbital anisotropy of the satellite galaxies. It only slightly broadens the posterior constraints for some of the parameters characterizing  
galaxy-halo connection of satellite galaxies.

\subsection{Galaxy Number Densities}
\label{sec:galnumdens}

The final observable that we use to constrain the galaxy-halo connection is the galaxy luminosity function, which provides important additional constraints on the CLF \citep[e.g.,][]{Yang.etal.03, vdBosch.etal.03, Cooray.Milosavljevic.05, Cooray.06}, and therefore helps to tighten the posterior in our inference problem. We use the number density of galaxies in 10 bins of 0.15~dex in luminosity, ranging from $10^{9.5}$ to $10^{11} \Lsunh$. These are computed using the corresponding, volume-limited subsamples, carefully accounting for the SDSS DR-7 footprint. In what follows, we refer to the data vector representing these 10 number densities as $\bD_{\rm LF}$. The covariance matrix of this data is computed using a jackknife estimator. In particular, we apply a recursive routine\footnote{{\faGithub} https://github.com/rongpu/pixel\_partition} developed by \citet{Zhou.etal.21}, that takes into account the survey mask and window, and iteratively constructs $\mathcal{N}$ maximally compact, equal-area partitions of the survey footprint \citep[see also][]{Wang.etal.22}. We adopt $\mathcal{N}=100$ which is large enough to capture the covariance in the survey while also being small enough to assure that each subregion still hosts an adequate number of galaxies.\footnote{We apply a Hartlap correction factor \citep[][]{Hartlap.etal.07} to the inverse of the covariance matrix to account for the relatively small number of jackknife samples, but note that this has a negligible impact.}

\section{Methodology}
\label{sec:method}

We analyse the data described above using the Bayesian, hierarchical satellite kinematics code \Basilisc, which is described in detail in \papI. Here we briefly summarize its salient features and introduce a few modifications that improve \Basilisc's performance.

\Basilisk uses an affine invariant ensemble sampler \citep[][]{Goodman.Weare.10} to constrain the posterior distribution, 
\begin{equation}
P(\btheta | \bD) \propto \calL(\bD | \btheta) \, P(\btheta)\,.
\end{equation}
Here $\bD = \bD_{\rm SK} + \bDNs + \bD_{\rm LF}$ is the total data vector, $\btheta$ is the vector that describes our model parameters, $P(\btheta)$ is the prior probability distribution on the model parameters, and $\calL(\bD|\btheta)$ is the likelihood of the data given the model. The latter consists of three parts: the likelihood $\calL_{\rm SK}$ for the satellite kinematics data $\bD_{\rm SK}$, the likelihood $\calL_{\rm NS}$ for the numbers of secondaries as described by the data vector $\bDNs$, and the likelihood $\calL_{\rm LF}$ for the luminosity function data $\bD_{\rm LF}$. In what follows we briefly describe the computation of each of these three different likelihood terms in turn. However, we first describe the model that we use to characterize the galaxy-halo connection.

\subsection{Galaxy-halo connection model}
\label{sec:hod}

\subsubsection{Conditional luminosity function}
\label{sec:CLF}

The galaxy occupation statistics of dark matter haloes are modelled using the conditional luminosity function (CLF), $\Phi(L|M,z) \, \rmd L$, which specifies the average number of galaxies with luminosities in the range $[L-\rmd L/2, \, L + \rmd L/2]$ that reside in a halo of mass $M$ at redshift $z$ \citep[][]{Yang.etal.03, vdBosch.etal.03}. In particular, we write that
\begin{equation}
    \Phi(L|M,z) = \Phi_\rmc(L|M) + \Phi_\rms(L|M)\,.
\end{equation}
Here and throughout the rest of the paper, subscripts `c' and `s' refer to central and satellite, respectively, and we assume that the CLF is redshift independent, at least over the redshift range considered here ($0.02 \leq z \leq 0.20$).

\begin{figure}
\centering
\includegraphics[width=0.47\textwidth]{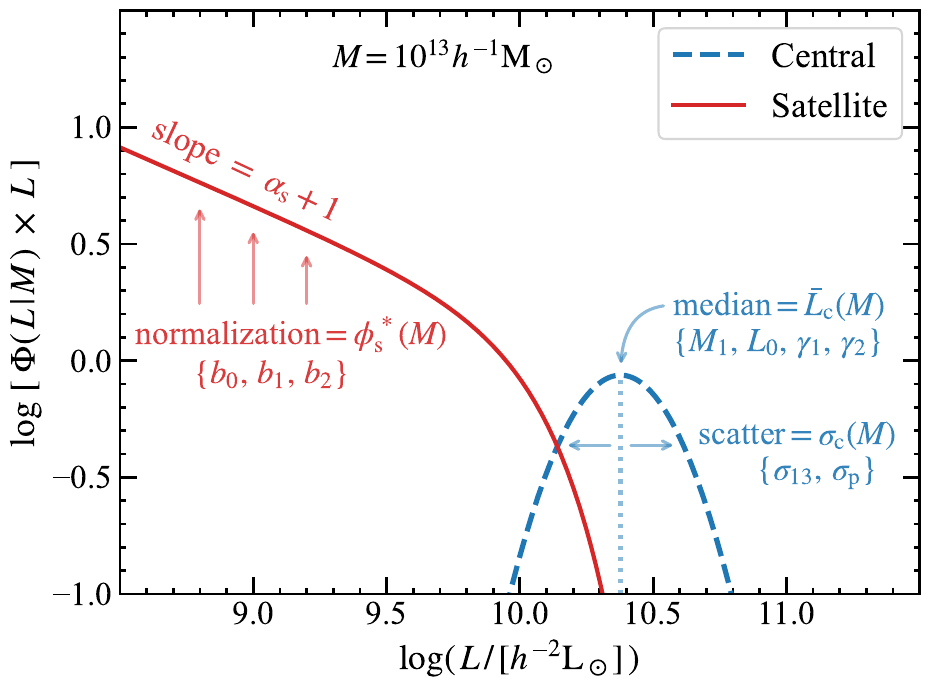} 
\caption{Illustration of the CLF used to characterize the galaxy-halo connection. Blue dashed and brown solid curves show the CLFs for central and satellite galaxies, respectively, in haloes of mass $M = 10^{13} h^{-1} {\rm M}_\odot$ used to construct the Tier-3 mock data discussed in \S\ref{sec:validation}. The different parameters that characterize the exact shape of the CLF are listed inside curly brackets. }
\label{fig:CLF}
\end{figure}

The CLF of centrals is parametrized using a log-normal distribution (see blue, dashed curve in Fig.~\ref{fig:CLF}),
\begin{equation}\label{CLFcen}
\Phi_\rmc (L | M) \rmd L = \frac{\log e}{\sqrt{2\pi \sigma_\rmc^2}} \exp \left[ -\left(\frac{\log L - \log\bar{L}_\rmc}{\sqrt{2} \sigma_\rmc} \right)^2\right] \frac{\rmd L}{L}.
\end{equation}
The mass dependence of the median luminosity, $\bar{L}_\rmc$, is parametrized by a broken power-law: 
\begin{equation}\label{averLc}
\bar{L}_\rmc (M) = L_0 \frac{(M / M_1)^{\gamma_1}}{(1 + M / M_1)^{\gamma_1 - \gamma_2}}.
\end{equation}
which is characterized by three free parameters; a normalization, $L_0$, a characteristic halo mass, $M_1$, and two power-law slopes, $\gamma_1$ and $\gamma_2$. 

Motivated by the fact that several studies suggest that the scatter, $\sigma_\rmc$, increases with decreasing halo mass \citep[e.g.,][]{Sawala.etal.17, Matthee.etal.17, Pillepich.etal.18, Wechsler.Tinker.18, Lange.etal.19b}, we allow for a mass-dependent scatter using
\begin{equation}\label{scatter}
\sigma_\rmc(M) = \sigma_{13} + \sigma_\rmP (\log\Mh-13)
\end{equation}
Hence, the scatter is characterized by two free parameters, a normalization, $\sigma_{13}$, that specifies the intrinsic scatter in $\log\Lc$ in haloes of mass $\Mh = 10^{13}\Msunh$, and a power-law slope $\sigma_\rmP$. Note that this is slightly different from the parametrization adopted in \papI.

For the satellite CLF we adopt a modified Schechter function (see red curve in  Fig.~\ref{fig:CLF}): 
\begin{equation}\label{satCLF}
\Phi_\rms (L | M) = \frac{\phi_\rms^*}{L_\rms^*} \left( \frac{L}{L_\rms^*} \right)^{\alpha_\rms} \exp \left[ - \left( \frac{L}{L_\rms^*} \right)^2 \right].
\end{equation}
Thus, the luminosity function of satellites in haloes of a given mass follows a power-law with slope $\alpha_\rms$ and with an exponential cut-off above a critical luminosity, $L_\rms^*(M)$. Throughout we adopt
\begin{equation}
L_\rms^*(M) = 0.562 \, \bar{L}_{\rmc} (M)\,.
\end{equation}
which is motivated by the results from galaxy group catalogues \citep[see][and \papI]{Yang.etal.09}. As in \citet{Lange.etal.19b}, we assume a universal value for the faint-end slope of the satellite CLF, $\alpha_{\rm s}$, independent of halo mass. Finally, the normalization $\phi_\rms^*(M)$ is parametrized by 
\begin{equation}\label{satCLFnorm}
\log \left[ \phi_\rms^*(M) \right] = b_0 + b_1 \log M_{12} + b_2 (\log M_{12})^2.
\end{equation}
where $M_{12} = M/(10^{12}\Msunh)$. Note that this characterization of the CLF is very similar to that adopted in a number of previous studies \citep[][]{Cacciato.etal.09, Cacciato.etal.13, More.etal.09b, vdBosch.etal.13, Lange.etal.19a, Lange.etal.19b}. All CLF parameters, along with parameters that characterize the satellite velocity anisotropy, and nuisance parameters used for interloper modelling, are listed in Table~\ref{table:SDSS_params}. It also includes the best-fitting values and $1\sigma$ confidence intervals for all the parameters, obtained by fitting the SDSS-DR7 data.

\subsubsection{Spatial distribution of satellites}
\label{sec:method_satellite_profile}

Throughout we assume that the radial distribution of satellite galaxies is given by a spherically symmetric, generalized NFW (gNFW) profile
\begin{equation}\label{nsatprof}
n_{\rm sat}(r|M,z) \propto \left( \frac{r}{\calR \, r_\rms} \right)^{-\gamma} \left( 1 + \frac{r}{\calR \, r_\rms} \right)^{\gamma - 3}\,.
\end{equation}
Here $\calR$ and $\gamma$ are free parameters and $r_\rms$ is the scale radius of the dark matter halo, which is related to the halo virial radius via the concentration parameter $\cvir = \rvir / r_\rms$. This gNFW profile has sufficient flexibility to adequately describe a wide range of radial profiles, from satellites being unbiased tracers of their dark matter halo ($\gamma = \calR = 1$), to cored profiles that resemble the radial profile of surviving subhaloes in numerical simulations ($\gamma=0$, $\calR \sim 2$). This also brackets the range of observational constraints on the radial distribution of satellite galaxies in groups and clusters \citep[e.g.,][]{Carlberg.etal.97, vdMarel.etal.00, Lin.etal.04, Yang.etal.05a, Chen.08, More.etal.09b, Guo.etal.12a, Cacciato.etal.13, Watson.etal.10, Watson.etal.12}.

\subsection{Satellite Kinematics}
\label{sec:modelSK}

The data vector for the satellite kinematics is given by eq.~(\ref{dataprim}) and contains the projected phase-space coordinates $\dV$ and $\Rp$ of all secondaries (satellite galaxies plus interlopers) associated with the $N_+$ primaries (centrals plus impurities). We make the reasonable assumption that the data for different primaries is independent. Additionally, for a primary with more than one secondary, we assume that the phase-space distribution of the secondaries are not correlated with each other. The latter may not be entirely justified, given that satellites are often accreted in groups, which can bias halo mass estimates \citep[][]{Old.etal.18}. We emphasize, though, that the majority ($71\%$ in the case of the SDSS data discussed in \S\ref{sec:SDSS}) of primaries that contribute to the satellite kinematics data only have a single secondary. In addition, tests based on realistic simulation-based mocks (see \S\ref{sec:validation}) indicate that any potential correlations between satellites (subhaloes) that occupy the same host halo can safely be ignored (i.e., do not cause a significant systematic error in our inference). Hence, we have that
\begin{equation}\label{LikelihoodSK}
\begin{split}
\calL_{\rm SK} & \equiv \calL(\bD_{\rm SK}|{\btheta}) \\
& = \prod\limits_{i=1}^{N_{+}} \, \prod\limits_{j=1}^{\Nsi} P(\dVij, \Rij | \Lci, \zci, \Nsi, \btheta)\,. 
\end{split}
\end{equation}
Here, $P(\dV, \Rp | \Lpri, \zpri, \Ns)$ is the probability that a secondary galaxy in a halo at redshift $\zpri$, with a primary of luminosity, $\Lpri$, and with a total of $\Ns$ detected secondaries has projected phase-space parameters $(\dV, \Rp)$. For true satellites, the probability is computed assuming that satellite galaxies are a virialized, steady-state tracer of the gravitational potential well in which they orbit (see \S\ref{sec:pdvrp}). Throughout, we assume dark matter haloes to be spherical and to have NFW \citep{Navarro.etal.97} density profiles characterized by the concentration-mass relation of \cite{Diemer.Kravtsov.15} with zero scatter. Hence, host haloes are completely specified by their virial mass, $\Mh$, alone\footnote{Throughout this paper, we define virial quantities according to the virial overdensities given by the fitting formula of \cite{Bryan.Norman.98}.}, which implies that we can factor the likelihood as
\begin{equation}\label{Mmarg}
\begin{split}
\calL_{\rm SK} & = \prod\limits_{i=1}^{N_{+}} \, \int \rmd \Mh \, P(\Mh | \Lci, \zci, \Nsi) \, \times \\
 & \;\;\;\;\;\;\;\;\;\;\;\;\;\;\;\;\;\;\;\;\;\;\;\prod\limits_{j=1}^{\Nsi} P(\dVij, \Rij|\Mh, \Lci, \zci)\,.
\end{split}
\end{equation}
This equation describes a marginalization over halo mass, which serves as a latent variable for each individual primary, accentuating the hierarchical nature of our inference procedure. Note that the `prior' for halo mass is informed by $\Lpri$, $\zpri$, and $\Ns$ according to the model $\btheta$.  Using Bayes theorem, we have 
\begin{equation}\label{ProbMass}
P(M|L,z,\Ns) = \frac{P(\Ns|M,L,z) \, P(M,L,z)}{\int \rmd M \, P(\Ns|M,L,z) \, P(M,L,z)}\,.
\end{equation}
In what follows we discuss each of the conditional probability functions required to compute $\calL_{\rm SK}$ in turn.

\subsubsection{The probability $P(\Ns|M,\Lpri,\zpri)$}
\label{sec:PNsMLz}

The number of secondaries, $N_\rms$, associated with a particular primary consists of both satellites (galaxies that belong to the same dark matter host halo as the primary), and interlopers (those that do not). Throughout we assume that the number of interlopers and the number of satellite galaxies are independent, and that both obey Poisson statistics. As shown in \papI, this implies that
\begin{equation}\label{PNsec}
P(N_\rms|M,L,z) = \frac{\lambda^{N_{\rms}}_{\rm tot} \, \rme^{-\lambda_{\rm tot}}}{N_\rms!}\,,
\end{equation}
where
\begin{equation}\label{lamtot}
\lambda_{\rm tot} = f_{\rm corr}  \times [\lambda_{\rm sat} + \lambda_{\rm int}]\,,
\end{equation}
is the expectation value for the number of secondaries, corrected for fibre collision and aperture incompleteness using the correction factor of equation~(\ref{fcorr}), and with $\lambda_{\rm int}(L,z)$ and $\lambda_{\rm sat}(M,L,z)$ as the expectation values for the numbers of interlopers and satellites, respectively.

The expectation value for the number of satellites brighter than the magnitude limit $L_{\rm min}(\zb)$, in a halo of mass $M$ at redshift $\zpri$, that fall within the aperture used to select secondaries around a primary of luminosity $\Lpri$, is given by
\begin{equation}\label{lambdasat}
 \lambda_{\rm sat}(M,\Lpri,\zpri) = f_{\rm ap}(M,\Lpri,\zpri) \, \int\limits_{L_{\rm min}}^{\infty} \Phi_\rms(L|M) \, \rmd L
\end{equation}
Note that $L_{\rm min}$ is a function of $\zb$ which in turn is a function of $\Lpri$ and $\zpri$ (see \S\ref{sec:selection}). Here $\Phi_\rms(L|M)$ is the satellite component of the CLF given by equation~(\ref{satCLF}) and $f_{\rm ap}$ is the aperture fraction, defined as the probability for true satellites to fall within the secondary selection cylinder specified by $\Rs$ and $\dVs$. Given that $\dVs$ is much larger than the extent of the halo in redshift space, we have that
\begin{equation}\label{fap}
\begin{split}
f_{\rm ap}(M&, \Lpri, \zpri) = \\
& 4 \pi \int\limits_0^{\rvir} \bar{n}_{\rm sat}(r|M,\zpri) 
\, \Big[\zeta(r,\Rmax) - \zeta(r,\Rmin)\Big] \, r^2 \, \rmd r\,.
\end{split}
\end{equation}
Here $\rvir = \rvir(M,\zpri)$ is the virial radius of the halo in question, and $\Rmax = \Rs(\Lpri)$ and $\Rmin = R_{\rm cut}(\zpri)$ are the outer and inner radii of the conical volume used to select the secondaries. The function $\bar{n}_{\rm sat}(r|M,z)$ is the average, radial profile of satellites around haloes of mass $M$ at redshift $z$, normalized such that
\begin{equation}
4 \pi \int\limits_0^{\rvir} \bar{n}_{\rm sat}(r|M,z) \, r^2 \, \rmd r = 1\,,
\end{equation}
and
\begin{equation}\label{zetafunc}
\zeta(r, R) = \begin{cases}
1 &\quad\text{if } r \leq R \\
1 - \sqrt{1 - R^2 / r^2} &\quad\text{otherwise.} \\ 
\end{cases}
\end{equation}
More specific expressions for $\lambda_{\rm sat}(M,L,z)$ and $f_{\rm ap}(M,L,z)$ are provided in \papI.

For the interlopers, one naively expects the abundance to be proportional to the number density of galaxies within the relevant range of luminosities and the volume of the secondary selection cone. However, being biased tracers of the mass distribution, galaxies are highly clustered which typically will boost the number density of galaxies in the vicinity of a bright primary. Moreover, this clustering strength is known to depend on halo mass, galaxy luminosity and redshift \citep[e.g.,][]{Mo.White.96, Zehavi.etal.11}, and to be affected by peculiar velocities, in particular due to large-scale infall \citep[][]{Kaiser.87}. We bypass the intricate complexities involved with modeling this clustering on small scales by modeling the expectation value for the number of interlopers as the product of an effective `bias', $b_{\rm eff}$, and the expectation value for the number of galaxies with $L_{\rm min}(\zb) < L < \Lpri$ in a randomly located conical selection volume, $V_{\rm cone}(\Lpri, \zpri)$:
\begin{equation}\label{interlopermodel}
 \lambda_{\rm int}(\Lpri,\zpri) = b_{\rm eff} \times V_{\rm cone} \times \bar{n}_{\rm gal}\,.
\end{equation}
where each term on the right-hand side is a function of $\{ \Lpri,\zpri \}$. Here
\begin{equation}\label{phitot}
\bar{n}_{\rm gal}(\Lpri, \zpri) = \int_{L_{\rm min}}^{\Lpri} \rmd L \int_0^{\infty} \Phi(L|M) \, n(M,\zpri) \, \rmd M
\end{equation}
is the average number density of galaxies at redshift $\zpri$ with luminosity in the range $[L_{\rm min},\Lpri]$,
with  $n(M,z)$ the halo mass function at redshift $z$, computed using the fitting function of \citet{Tinker.etal.08}, and 
\begin{equation}\label{Vcyl}
V_{\rm cone}(\Lpri,\zpri) = \pi \, \left[\Rmax^2 - \Rmin^2\right] \, \frac{2 \dVs}{H(\zpri)} \, (1+\zpri)^3
\end{equation}
with $H(z)$ the Hubble parameter\footnote{Note, there was a typo in Eq.~(22) in \papI, where the power-law index of $(1+\zc)$ was 2, rather than the correct 3}. The effective bias is modelled as
\begin{equation}\label{beff}
b_{\rm eff}(\Lpri,\zpri) = \eta_0 \, \left(\frac{\Lpri}{10^{10.5}\Lsunh}\right)^{\eta_1} \, \left( 1+\zpri \right)^{\eta_2}
\end{equation}
where $\eta_0$, $\eta_1$, and $\eta_2$ are three free nuisance parameters that fully specify our interloper bias model, and that are constrained simultaneously with all other physical parameters. This model has proved to be sufficiently flexible to accurately model the full complexity of interloper abundance in realistic simulation-based mock data (see Section~\ref{sec:inf3}).

\subsubsection{The probability $P(M,\Lpri,\zpri)$}
\label{sec:PMLz}

The function $P(M,\Lpri,\zpri)$ describes the probability distribution function of primaries as a function of host halo mass, luminosity and redshift, and can be written as
\begin{equation}\label{PMLz}
P(M,\Lpri,\zpri) = P(\Lpri|M,\zpri)  \, n(M,\zpri) \, \calC(M,\Lpri,\zpri) \,.
\end{equation}
where $n(M,z)$ is the halo mass function \citep{Tinker.etal.08} and $\calC(M,L,z)$ is a `completeness', to be defined below. As in \papI, if we assume that all primaries are true centrals, then we have that $P(\Lpri|M,\zpri) = \Phi_\rmc(\Lpri|M)$. However, in reality some primaries are misidentified satellites, and such impurities need to be accounted for. In \paperI we argued that the impact of these impurities is sufficiently small that it can be ignored. Although this was indeed the case for the small mock data sets used there, the impact of impurities can no longer be ignored when using data sets similar in size to the SDSS data analysed here. In fact, detailed tests showed that they can systematically bias the inferred scatter in the relation between halo mass and central luminosity, and we therefore devised the following scheme in order to account for impurities.

The vast majority of all impurities in realistic SDSS-like mocks (such as the Tier-3 mock described in \S\ref{sec:validation}) are those satellite galaxies which happen to be the brightest galaxy in their halo (even brighter than their central). In what follows, we refer to these as Type-I impurities. Since primaries are by definition the brightest galaxies in their selection cones, such brightest-halo-galaxy (hereafter BHG) satellites typically end up being selected as primaries, rather than their corresponding central. In rare cases, a primary is neither a true central, nor a BHG satellite. We refer to these as Type-II impurities, which arise, for example, if the true central is the BHG but is absent from the SDSS survey data, either because of fiber collisions or because it falls  outside the window of the SDSS footprint. As detailed below, Type I impurities can be accounted for in our new theoretical modelling. However, Type-II impurities are virtually impossible to model accurately. Using detailed mock data sets, we therefore tuned our selection criteria in order to minimize the contribution of Type-II impurities. In particular, we found that we were able to significantly reduce their frequency by slightly enlarging the volume of the primary selection cone as described in \S\ref{sec:selection}; In particular, the new criteria reduce the fraction of Type-II impurities from $\gtrsim 1\%$ when using the old selection criteria used in \papI, to $\sim 0.5\%$ with our new selection criteria. More importantly, in mock data, the new selection criteria predominantly eliminate the presence of Type-II impurities that are extreme outliers of the average relation between halo mass and primary luminosity, and which are the main culprits for causing mild systematic errors in the inferred galaxy-halo connection (specifically in the scatter, $\sigma_\rmc$). Detailed tests with mock data, presented in \S\ref{sec:validation} below, show that our new primary selection criteria sufficiently suppress the impact of Type-II impurities that it allows for unbiased estimates of the galaxy-halo connection (at least for a survey the size of SDSS).

Therefore, in what follows, we ignore Type-II impurities and assume that primaries are either true centrals or BHG satellites (i.e., Type-I impurities). Hence, we have that
 \begin{equation}\label{P_Lpri_given_M}
 \begin{split}
 P(\Lpri|M,z) & =  P(\Lc=\Lpri|M,z) P(L_{\rm bs}<\Lpri|M,z)
 \\
 &+ P(\Lc<\Lpri|M,z) P(L_{\rm bs}=\Lpri|M,z).
 \end{split}
 \end{equation}
Here $P(L_{\rm bs}<L|M,z)$ is the probability that the brightest satellite in a halo of mass $M$ at redshift $z$ has a luminosity less than $L$, which is given by
\begin{equation}\label{P_Lbs_lt_L}
P(L_{\rm bs}<L|M,z) = \exp\left[ - \Lambda(L|M,z) \right]\,.
\end{equation}
Here $\Lambda(L|M,z)$ is the expectation value for the number of satellites brighter than $L$ in a halo of that mass and redshift, which in turn is given by
\begin{equation}\label{Lambda_for_L_M}
\Lambda(L|M,z) = \int_{L}^{\infty} \rmd L' \, \Phi_\rms(L'|M)
\end{equation}
Differentiating  $P(L_{\rm bs}<L|M,z)$ with respect to luminosity yields:
\begin{equation}
P(L_{\rm bs}=L|M,z) = \Phi_{\rm s}(L|M,z) \, e^{-\Lambda(L|M,z)}
\end{equation}
The two other terms that appear in equation~(\ref{P_Lpri_given_M}) are  $P(\Lc=\Lpri|M,z)$, which is simply equal to the central CLF, $\Phi_\rmc(\Lpri|M)$, and its cumulative distribution, which is given by
\begin{equation}
P(\Lc<L|M,z) = 
\int_{0}^{L} \rmd \Lc' \, \Phi_\rmc(\Lc'|M)
\end{equation}

The expression for $P(\Lpri|M,\zpri)$ given by equation~(\ref{P_Lpri_given_M}), when substituted in equation~(\ref{PMLz}), accurately forward models the impact of the vast majority of impurities. 

The final ingredient we need is an expression for the completeness $\calC(M,L,z)$, which is defined as the fraction of haloes of mass $M$ at redshift $z=\zpri$ with a central {\it or brightest satellite} of luminosity $\Lpri$ that falls within the survey volume of the SDSS, and that is selected as a primary by our selection criteria.  In general we have that $\calC(M,L,z) = \calC(M|L,z) \calC_0(L,z)$. As is evident from equation~(\ref{ProbMass}), the modelling in \Basilisk is independent of $\calC_0$, which drops out (see also \papI). In other words, we only need to account for the halo mass dependence of the completeness. As shown in Appendix~\ref{App:compl}, this mass-dependence is already accounted for by our forward-modelling of the Type-I impurities. Hence, we set $\calC(M,L,z) = 1$ throughout.

\subsubsection{The probability $P(\dV,\Rp|M,\Lpri,\zpri)$}
\label{sec:pdvrp}

In order to model the line-of-sight kinematics of the secondaries we proceed as follows. Since secondaries consist of both true satellites and interlopers, which have distinct phase-space distribution, we write
\begin{equation}\label{PdVall}
\begin{split}
P(\dV, \Rp|M,L,z) = & f_{\rm int}\, P_{\rm int}(\dV, \Rp|L,z) \, + \\
& [1 - f_{\rm int}] \, P_{\rm sat}(\dV, \Rp|M,L,z) \,
\end{split}
\end{equation}
with the interloper fraction defined as
\begin{equation}\label{fint}
f_{\rm int} = f_{\rm int}(M,L,z) = \frac{\lambda_{\rm int}(L,z)}{\lambda_{\rm tot}(M,L,z)}\,,
\end{equation}
where $\lambda_{\rm int}$ and $\lambda_{\rm tot}$ have been individually defined in \S\ref{sec:PNsMLz}. We first describe how we compute $P_{\rm sat}(\dV, \Rp|M,L,z)$ (in \S\ref{sec:Psat}) before detailing our treatment of interlopers (\S\ref{sec:Pint}).

\subsubsection{The phase-space distribution of satellites:}
\label{sec:Psat} 

In computing the joint 2D probability $P_{\rm sat}(\dV, \Rp|M,L,z)$, we assume that the baryonic matter of the central galaxy has a negligible impact on the kinematics of its satellite galaxies\footnote{We address the accuracy of this assumption, which is common to virtually every study of satellite kinematics, in a forthcoming paper (Baggen et al., in prep).}, and we model the satellites as tracers in a pure dark matter halo which is fully characterized by its halo mass and concentration. Throughout, we use the median concentration-halo mass relation of \citet{Diemer.Kravtsov.15}, and we emphasize that our modelling is fairly insensitive to the exact choice of the concentration-mass relation within reasonable bounds of its theoretical uncertainty. We also assume the central galaxy to be located at rest at the centre of the halo. As shown in \papI, relaxing this assumption by allowing for non-zero velocity bias for centrals has negligible impact on \Basilisc's inference. 

Under these assumptions we have that
\begin{equation}\label{split}
P_{\rm sat}(\dV, \Rp|M,L,z) = P(\Rp|M,L,z) \, P(\dV|\Rp,M,z)\,.
\end{equation}
with
\begin{equation}\label{PrP}
P(\Rp|M,L,z) = \frac{2\, \pi\, \Rp \, \bar{\Sigma}(\Rp|M,z)}{f_{\rm ap}(M,L,z)}\,.
\end{equation}
Here $f_{\rm ap}$ is defined in equation~(\ref{fap}), and
\begin{equation}\label{ProjNsat}
\bar{\Sigma}(\Rp|M,z) = 2 \int_{\Rp}^{r_{\rm sp}(M,z)} \bar{n}_{\rm sat}(r|M,z) \frac{r\,\rmd r}{\sqrt{r^2 - R^2}}\,,
\end{equation}
is the projected, normalized number density distribution of satellite galaxies.

In \papI, we made the simplified assumption that the line-of-sight velocity distribution, $P(\dV|\Rp,M,z)$, is a Gaussian, which is completely characterized by the line-of-sight velocity dispersion $\sigma_{\rm los}(\Rp|M,L,z)$. However, there is no a priori reason why the LOSVD should be Gaussian. In fact, the detailed shape of the LOSVD contains valuable information regarding the velocity anisotropy \citep[e.g.,][]{DeJonghe.87, Gerhard.91, Wojtak.Mamon.13}, which we aim to constrain using \Basilisc. In this work we therefore improve upon \paperI by extending our modelling of the kinematics to fourth-order and by describing $P(\dV|\Rp,M,z)$ as a generalised Gaussian with a projected velocity dispersion, $\sigma_{\rm los}(\Rp|M,L,z)$, and a line-of-sight kurtosis, $\kappa_{\rm los}(\Rp|M,L,z)$. The projected, line-of-sight velocity dispersion is related to the intrinsic, radial velocity dispersion, $\sigma_r^2(r|M,z)$, according to 
\begin{equation}\label{sigmalos}
\begin{split}
\sigma^2_{\rm los}(\Rp |M,z) = \frac{2}{\bar{\Sigma}(\Rp)} & \int_{\Rp}^{r_{\rm sp}(M,z)} \left[ 1 - \beta(r|M) \frac{\Rp^2}{r^2} \right] \times \\
&  \bar{n}_{\rm sat}(r|M,z) \, \sigma_r^2(r|M,z) \, \frac{r \, \rmd r}{\sqrt{r^2 - \Rp^2}\,,}
\end{split}
\end{equation}
where $\sigma_r^2(r|M,z)$ follows from the second order Jeans equation for a spherically symmetric NFW halo \citep[see equation~(50) in \papI, and][for more details]{Binney.Mamon.82}. Here, the local anisotropy parameter
\begin{equation}\label{eqn:beta}
\beta(r|M) \equiv 1 - \dfrac{\sigma^2_\rmt(r|M)}{2 \, \sigma^2_\rmr(r|M)}
\end{equation}
relates the tangential ($\sigma_\rmt$) and radial ($\sigma_\rmr$) velocity dispersions. For our fiducial model we assume that $\beta$ is independent of both radius and halo mass, and we constrain this `average' velocity anisotropy using the satellite kinematics data. In \S~\ref{sec:anisotropy} we discuss the implications of adopting more flexible models in which the anisotropy parameter is allowed to depend on halo mass. Note that the upper-integration limit of equations~(\ref{ProjNsat}) and~(\ref{sigmalos}) is set to $r_{\rm sp}(M,z) = f_{\rm sp} r_{\rm vir}(M,z)$, instead of $r_{\rm vir}(M,z)$, to account for a population of splash-back galaxies (see \S\ref{sec:Pint}). 

The projected, fourth moment of the LOSVD at projected separation $\Rp$ is given by
\begin{equation}\label{eqn:v4los}
\begin{split}
    \overline{v_{\rm los}^4}(R_\rmp) = & \dfrac{2}{\overline{\Sigma}(R_\rmp)}  \int_{R_\rmp}^{r_{\rm sp}}   \left[1-2\beta R_\rmp^2/r^2 + \tfrac{1}{2} \beta (1+\beta) R_\rmp^4 / r^4 \right] \times \\& \overline{v_r^4}(r|M,z) \, n_{\rm sat}(r|M,z) \dfrac{r \, \rmd r}{\sqrt{r^2-R_\rmp^2}}\,,
\end{split}
\end{equation}
where $\beta$ is $\beta(r|M)$ in general, and $\overline{v_r^4}(r|M,z)$ follows from the fourth-order spherical Jeans equation \citep{Lokas.02, Lokas.Mamon.03}, which for radius-independent anisotropy is given by:
\begin{equation} \label{v4local}
\begin{split}
    \overline{v_r^4}(r&|M,z) = \dfrac{3 \, G}{r^{2\beta} \, \bar{n}_{\rm sat}(r|M,z)} \times \\
    & \int_r^{r_{\rm sp}} {\rm d}r' \, (r')^{2\beta} \, \bar{n}_{\rm sat}(r'|M,z) \,  \sigma_r^2(r'|M,z) \, \dfrac{M(r')}{r'^2}
\end{split}
\end{equation}
Here $M(r)$ is the enclosed mass of the spherical NFW halo inside radius $r$. Given the fourth-order line-of-sight velocity moment, we can compute the projected kurtosis as
\begin{equation}
\kappa_{\rm los}(R_\rmp|M,z) = \overline{ v_{\rm los}^4 } (R_\rmp|M,z) \, / \,  \sigma_{\rm los}^4  (R_\rmp|M,z).
\end{equation}

Finally, in order to account for non-zero redshift errors in the data, the line-of-sight velocity dispersion is modified according to $\sigma_{\rm los} \to \sqrt{\sigma^2_{\rm los} + 2 \sigma^2_{\rm err}}$, with $\sigma_{\rm err} = 15 \kms$ the typical SDSS redshift error \citep[][]{Guo.etal.15b}. Having computed both the velocity dispersion and kurtosis, we model the detailed shape of the LOSVD, $P(\dV|\Rp,M,z)$, using a symmetric (all odd moments are equal to zero), generalized form of the normal distribution, known as the \citet{Langdon.1980} distribution\footnote{The Langdon distribution is often used to characterize the non-Maxwellian velocity distribution of electrons heated due to the inverse-Bremsstrahlung process \citep[e.g.,][]{Matte.etal.88,Mora.Yahi.82}.}:
\begin{equation}
 P_\rmL(\dV) = \dfrac{1}{2 \Gamma(1/m)} \, \dfrac{m}{a_m} \, \exp\left(-|\dV/a_m|^{m} \right)\,.
\end{equation}
Here the parameters $a_m$ and $m$ are related to the variance, $\sigma^2$, and the kurtosis, $\kappa$, according to 
\begin{equation}\label{Langdonpar}
\sigma^2 =  a_m^2 \dfrac{\Gamma(3/m)}{\Gamma(1/m)} \,\,\,\, {\rm and} \,\,\,\, \kappa = \dfrac{\Gamma(5/m) \Gamma(1/m) }{\Gamma^2(3/m)}
\end{equation}
The reason for using this particular distribution function is purely one of convenience; $P_\rmL(\dV)$ has a nice analytic closed form, is simple to compute, has all the features required of a probability distribution (normalized and positive-definite), and includes the Gaussian as a special case ($m=2$).

In \Basilisc, we use equation~(\ref{Langdonpar}) to compute $a_m$ and $m$ from $\sigma^2_{\rm los}(\Rp |M,z)$ and $\kappa_{\rm los}(\Rp |M,z)$, after which we compute 
\begin{equation}\label{LOSVD}
P(\dV|\Rp,M,z) = P_\rmL(\dV) \,
\dfrac{\Gamma(1/m)}{\Gamma(1/m) - \Gamma(1/m, (\dVs/a_m)^m)}
\end{equation}
which is properly normalized such that its integral from $-\dVs$ to $\dVs$ is unity.

\subsubsection{The phase-space distribution of interlopers} 
\label{sec:Pint}

In \paperI we assumed that interlopers have a constant projected number density and a uniform distribution in line-of-sight velocities, so that $P_{\rm int}(\dV, \Rp |L,z) =   \Rp /  [\dVs \, (\Rmax^2 - \Rmin^2)]$. Here $\Rmax = \Rs(L)$ and $\Rmin = R_{\rm cut}(z)$ are the outer and inner radii of the conical volume used to select secondaries around primaries of luminosity $L$ at redshift $z$ and $\dVs$ is the corresponding line-of-sight depth (see \S\ref{sec:selection}).

However, as discussed in detail in \papI, a subset of the interlopers are either infalling or splash-back galaxies and have kinematics that are very similar to the true satellites. Assuming that the velocity distribution of interlopers is uniform ignores this `kinematically coupled interloper population', which causes \Basilisk to overestimate the number of satellite galaxies. Although the resulting offsets were modest for the smaller mock samples studied in \papI, they cause a significant, systematic bias (predominantly in the satellite CLF parameters $b_0$, $b_1$ and $b_2$) when using larger samples. This motivated us to develop a more sophisticated treatment for the phase-space distribution of interlopers. 

Based on a detailed assessment of interlopers in our mock data sets (see \paperI and \S\ref{sec:validation} for details), we  now model the interlopers as consisting of three fairly distinct populations: (i) a population of `splash-back galaxies' associated with the host halo of the primary, and extending out to a distance $r_{\rm sp}$ from the primary, (ii) a roughly uniform background population of `true' interlopers that are uncorrelated with the primary, and (iii) an `infalling' population of interlopers, located outside of the splash-back radius. This infall motion, on large linear scales, is responsible for redshift space distortions in clustering data known as the Kaiser effect \citep{Kaiser.87}.

We assume that the phase-space distribution of splash-back galaxies can be modelled similar to that of the satellites; i.e., they follow the same $n_{\rm sat}(r|M,z)$, extrapolated to beyond the halo's virial radius, and their kinematics obey the same Jeans equations. The only difference is that they are located between the host halo's virial radius, $r_{\rm vir}$, and a splash-back radius $r_{\rm sp} \equiv f_{\rm sp} r_{\rm vir}$. In order to account for a population of splash-back galaxies we simply change the upper-integration limit of equations~(\ref{fap}), (\ref{ProjNsat}), (\ref{sigmalos}), (\ref{eqn:v4los}) and~(\ref{v4local}) from $r_{\rm vir}(M,z)$ to $r_{\rm sp}(M,z)$. Throughout we adopt $f_{\rm sp} = 2$, which is motivated by estimates of the splash-back radius in simulations \citep[][]{Diemer.17, Mansfield.etal.17}. In addition, detailed tests with mock data sets (see \S\ref{sec:validation}) show that \Basilisk yields unbiased estimates of the velocity anisotropy parameter $\beta$ for $f_{\rm sp}\gtrsim 1.5$. We find that not accounting for splash-back galaxies (i.e., setting $f_{\rm sp}=1$) results in a weak bias of $\beta$, without significantly affecting any of the other parameters. On the other hand, setting a much larger value for splash-back radius, like $f_{\rm sp}=3$, yields posteriors that are indistinguishable from those for $f_{\rm sp}=2$. Hence, our choice of $f_{\rm sp}=2$ is reasonable and our results are robust against modest changes in the adopted value of $f_{\rm sp}$.

We assume that both the uncorrelated `background' interlopers (bg) as well as the `infalling' interlopers (inf) have a uniform angular distribution on the sky, such that their phase-space distribution can be written as
\begin{equation}\label{PdVint}
P_{\rm int}(\dV, \Rp) = \frac{2 \Rp}{(\Rmax^2 - \Rmin^2)} \, \left[P_{\rm bg}(\dV) + P_{\rm inf}(\dV)\right]\,,
\end{equation}
where the term in square brackets is normalized such that its integral from $- \dVs$ to $+ \dVs$ is unity. Since the secondary selection volume is conical in shape, the backside has a larger volume than the front (see Fig.~\ref{fig:conical_cylinders}). Note that all the secondaries of any given primary have luminosities above a fixed threshold. Therefore, due to the conical selection volume, we expect the velocity distribution of the uncorrelated background interlopers, $P_{\rm bg}(\dV)$, to increase with $\dV$. In particular, $P_{\rm bg}(\dV)$ is proportional to the comoving volume of the corresponding velocity slice of the secondary selection cone, and we therefore adopt
\begin{equation}\label{Pbg}
P_{\rm bg}(\dV) = \frac{3}{H(z')} \, \frac{d^2(z')}{d^3(\zb) - d^3(\zf)} \, (1+z_{\rm pri})
\end{equation}
Here  $z' = z_{\rm pri} + (1+z_{\rm pri}) \dV/c$ and $d(z)$ is the comoving distance out to redshift $z$.
\begin{figure}
\centering
\includegraphics[width=0.48\textwidth,angle=0]{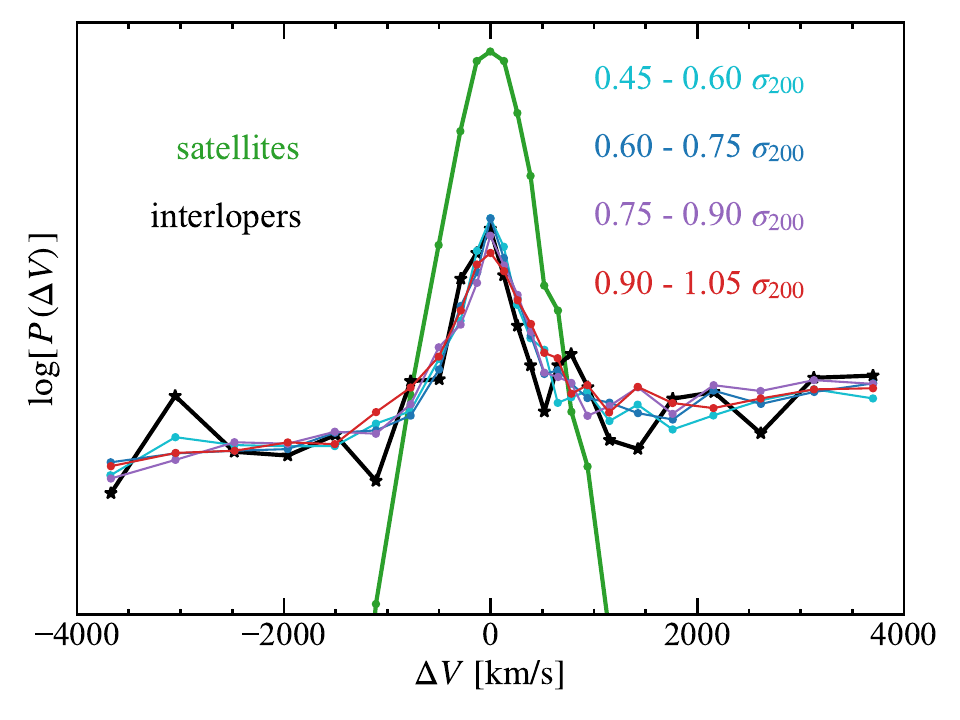}
\caption{Normalized line-of-sight velocity distribution (LOSVD) of galaxies around Tier-3 primaries in a narrow redshift and luminosity bin ($\zpri = 0.065 \pm 0.01$ and $\log \Lpri = 10.35 \pm 0.1$). The thick green and black lines are the LOSVDs of true satellites and of interlopers, respectively. Note that the vertical scale is logarithmic to highlight the high-velocity wings of the interloper distribution. The thin coloured lines are the LOSVDs of tertiary mock galaxies in annular rings around primaries with radial bins ranging from $\{R_{\rm p,min}, R_{\rm p,max}\} = \{0.45, 0.60\}$ to $\{0.90, 1.05\} \times \sigma_{200} h^{-1}{\rm Mpc}$, as indicated. All histograms are individually normalized to highlight their relative shapes. As is evident, the LOSVDs of these tertiaries closely resemble those of the interlopers. This empirical fact indicates that one can model the LOSVD of interlopers as that of `tertiary' galaxies in any of these annular rings, and it is the data-driven method we introduce in \Basilisk (see the text for details).}
\label{fig:interloper_modelling}
\end{figure}

We have experimented with modelling the line-of-sight velocity distribution of the infalling population of interlopers, $P_{\rm inf}(\dV)$, using the linear \citep[][]{Kaiser.87} model, but this did not yield sufficiently accurate results. We therefore opted for a semi-empirical approach. Tests with mock data (see \S\ref{sec:validation}) show that $P_{\rm inf}(\dV)$ is accurately fit by a Gaussian, $P_{\rm inf}(\Delta V) =  A_{\rm inf} \, \rme^{-\frac{1}{2}(\Delta V^2 / \sigma_{\rm inf}^2)}$, with $A_{\rm inf}$ and $\sigma_{\rm inf}$ free parameters that vary from primary to primary. Rather than trying to devise an analytical model for these parameters, we use the following data-driven approach. Around each primary we select a set of `tertiary' galaxies in a conical volume similar to that used for the secondaries, but at larger projected distances from the primary. More specifically, the tertiary selection cone is specified by an inner projected radius $R_{\rm min}^{\rm int} = 0.6 \, \sigma_{200} \mpch$ and an outer projected radius $R_{\rm max}^{\rm int} = 0.9 \, \sigma_{200} \mpch$, and by the same redshift depth as the secondary selection cone (shown by the outermost hollow annular cone in Fig.~\ref{fig:conical_cylinders}). Tests with mock data (see Fig.~\ref{fig:interloper_modelling}) indicate that (i) the line-of-sight velocity distribution of these tertiaries is virtually indistinguishable from that of the infalling interlopers among the secondaries, (ii) the results are insensitive to the exact radii of the tertiary selection cone\footnote{A weak dependence of $\sigma_{\rm inf}$ on projected radius is apparent in Fig.~\ref{fig:interloper_modelling}, and has also been noted by \citet{Mamon.etal.10}. We have experimented with implementing such a $R_\rmp$-dependence and extrapolating this
to the radial interval of the secondary selection cone, but found that this had a negligible impact on the inference.}, and (iii) less than one percent of the tertiaries are actual satellites of the primary. The latter indicates that for most primaries $R_{\rm min}^{\rm int}$ lies well outside the virial radius of the host halo of the primary, as required. We assume that $A_{\rm inf}$ and $\sigma_{\rm inf}$ are each quadratic functions of $\log(L_{\rm pri})$ and $\zpri$ (including the cross-term), and determine the corresponding $6 \times 2=12$ coefficients by simultaneously fitting the velocity distribution of {\it all} tertiaries around {\it all} primaries. These coefficients are then used in \Basilisk to model the line-of-sight velocity distribution of the infalling population of interlopers.

Note that this rather elaborate model for the phase-space distribution of interlopers has zero degrees of freedom. The only degrees of freedom for the interloper-modelling is with regard to their number density, which is modelled via the effective bias described by equation~(\ref{beff}). 

\subsection{Modelling the number of secondaries}
\label{sec:modelNs}

As mentioned in \S\ref{sec:Pnull}, the data vector $\bD_{\rm NS}$ that describes the number of secondaries for a random subset of $N_{\rm NS}$ primaries, including those with zero secondaries, contains valuable information regarding the occupation statistics of satellite galaxies, and hence, the CLF. 

Similar in spirit to how we compute the likelihood for the satellite kinematics by marginalizing over halo mass (cf. equation~[\ref{Mmarg}]), the likelihood for $\bD_{\rm NS}$ given a model $\theta$ is given by
\begin{equation}\label{LNsec}
\calL_{\rm NS} = \prod_{i=1}^{N_{\rm NS}}  \int \rmd \Mh \, P(\Mh|\Lci, \zci) \, P(\Nsi | \Mh, \Lci, \zci)\,.
\end{equation}
Here $P(\Ns|M,L,z)$ is given by equation~(\ref{PNsec}), while
\begin{equation}\label{PMbarLz}
P(M|L,z) = \frac{P(L, M, z)}{\int \rmd M  P(L,M,z)}\,,
\end{equation}
with $P(L,M,z)$ given by equation~(\ref{PMLz}). 

\subsection{Modelling the galaxy luminosity function}
\label{sec:modelLF}
 
The final observational constraint that we use to constrain the halo occupation statistics is the comoving number density of SDSS galaxies, $n_{\rm gal}(L_1,L_2)$ in ten 0.15~dex bins in luminosity, $[L_1,L_2]$  covering the range $9.5 \leq \log L/(h^{-2} \Lsun) \leq 11.0$ (see \S\ref{sec:galnumdens}). We include this data in our inference problem by defining the corresponding log-likelihood
\begin{equation}
\ln\calL_{\rm LF}(\bn_{\rm obs}|\btheta) = -\frac{1}{2} \, [\bn(\btheta) - \bn_{\rm obs}]^t \, \boldsymbol{\Psi} \, [\bn(\btheta) - \bn_{\rm obs}]\,.
\end{equation}
Here $\bn_{\rm obs}$ is the data vector and $\bn(\btheta)$ is the corresponding model prediction, computed from the CLF and the halo mass function using
\begin{equation}\label{numdens}
n_{\rm gal}(L_1,L_2) = \int_{L_1}^{L_2} \rmd L \int_{0}^{\infty} \Phi(L|M) \, n(M,z_{\rm SDSS}) \, \rmd M\,,
\end{equation}
where $z_{\rm SDSS} = 0.1$ is a characteristic redshift for the SDSS data used,\footnote{We have verified that our results do not depend significantly on this choice.} and $\boldsymbol{\Psi}$ is the precision matrix, which is the inverse of the covariance matrix, with \citet{Hartlap.etal.07} correction (see \S\ref{sec:galnumdens}).

\subsection{Numerical implementation}
\label{sec:implementation}

The fiducial model used by \Basilisk is characterized by a total of 16 free parameters: 6 parameters describing $\Phi_\rmc(L|M)$ (namely $\log M_1$, $\log L_0$, $\gamma_1$, $\gamma_2$, $\sigma_{13}$ and $\sigma_\rmP$), 4 parameters describing $\Phi_\rms(L|M)$ (namely $\alpha_{\rm s}$, $b_0$, $b_1$, and $b_2$), 1 parameters ($\beta$) to quantify the average velocity anisotropy of satellite galaxies, 2 parameters ($\gamma$ and $\calR$) that describe the radial number density profiles of satellite galaxies (see equation~[\ref{nsatprof}]), and 3 nuisance parameters  ($\eta_0$, $\eta_1$ and $\eta_2$) that specify the abundance of interlopers (see equation~[\ref{beff}]). We assume broad uniform priors on all parameters, except for $\beta$. The value of the anisotropy parameter $\beta$ formally ranges from $-\infty$, for maximal azimuthal anisotropy, to $+1$, for maximal radial anisotropy, which is difficult to probe with our MCMC sampler. Hence, in order to assure roughly equal amounts of parameter space for radially and azimuthally anisotropic models, we sample $\mathcal{B} \equiv -\log(1-\beta)$, rather than $\beta$. In particular, we adopt uniform priors over the range $-1 \leq \mathcal{B} \leq 1$, which corresponds to $-9 \leq \beta \leq 0.9$.
\begin{figure*}
\centering
\includegraphics[width=0.9\textwidth]{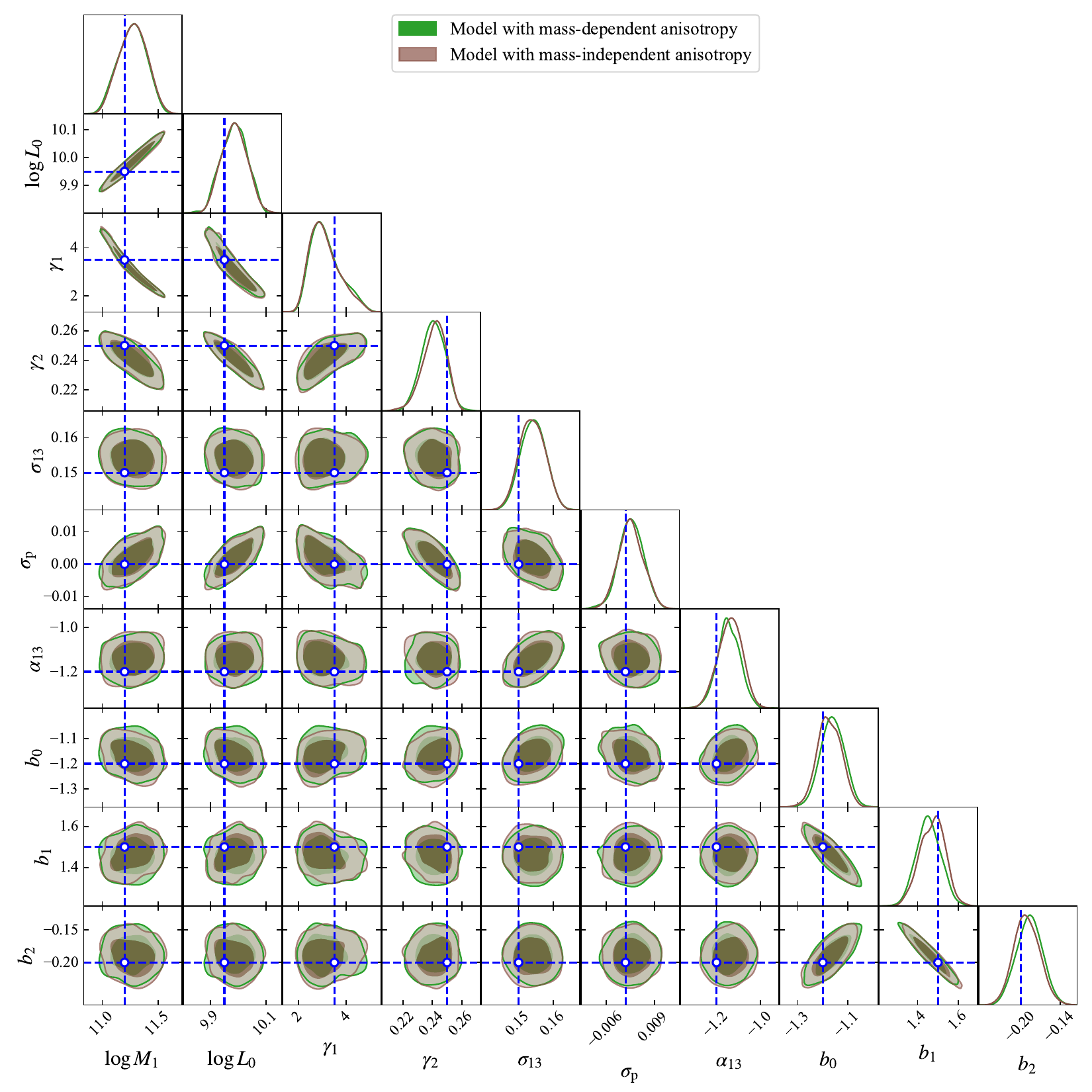}
\caption{Marginalized posteriors obtained by \Basilisk for the Tier-3 mock (assuming the best-fit radial profile for the satellites, with $\gamma = 0$ and $\calR = 2.5$). Results are shown for only those 10 parameters that characterize the CLF. Panels along the diagonal show marginalized 1D posteriors while off-diagonal panels show 2D posteriors. In the latter case, contours demarcate the 68 and 95~\% containment of the posterior, while the blue dashed lines indicate the true input values used to create the mock data set. Brown contours correspond to our fiducial model that assumes a mass-independent velocity anisotropy, $\beta$; and the green contours show the constraints for a model in which $\beta$ is allowed to depend on halo mass (discussed in \S\ref{sec:anisotropy_mock}). The posteriors for all parameters are in good agreement with the input values, and virtually independent of whether the velocity anisotropy is mass dependent or not.  For all parameters we adopted very wide and flat prior with bounds that have no impact on the posteriors.}
\label{fig:T3_corner}
\end{figure*}

Probing the posterior $P(\btheta|\bD)$ over the fiducial 16-dimensional parameter space requires close to a million likelihood evaluations, each of which involves thousands of numerical integrations (see \paperI for details). In order to make this problem feasible we perform the Bayesian inference under the assumption of a {\it fixed} normalized, radial number density distribution of satellite galaxies, i.e., fixed values for $\gamma$ and $\calR$. This has the advantage that $f_{\rm ap}(M,L,z)$ and $P_{\rm int}(\dV, \Rp|L,z)$ are all independent of the model, $\btheta$, while $P_{\rm sat}(\dV, \Rp|M,L,z)$ only depends on a single anisotropy parameter (see \S\ref{sec:Psat}). We compute $P_{\rm sat}(\dV, \Rp|M,L,z)$ for each central-satellite pair for 10 values of $\mathcal{B}$ between $-1$ and $1$ (or $\beta$ between $-9$ and $0.9$), and we interpolate it for intermediate values. Combined with the fact that we perform all integrations over halo mass using Gaussian quadrature with {\it fixed} abscissas (see \papI) implies that we only need to compute all these quantities once for each primary and/or secondary, which we then use to find the posterior $P(\btheta|\bD)$ in the 14-dimensional parameter space at fixed $(\gamma,\calR)$. As a consequence, for a single evaluation of the full likelihood 
\begin{equation}\label{Ltot}
\ln \calL_{\rm tot} = \ln \calL_{\rm SK} + \ln \calL_{\rm NS} + \ln \calL_{\rm LF}\,, 
\end{equation}
for the full SDSS data consisting of 18,373 primaries with at least one secondary and a total of 30,431 secondaries (see \S\ref{sec:SDSS}), it only takes of the order of 200 milliseconds using a single run-of-the-mill CPU. This is sufficiently fast, that it allows one to run different Monte-Carlo Markov Chains for different assumptions regarding  $\bar{n}_{\rm sat}(r|M,z)$, and to find the best-fit radial profile, marginalized over all other model parameters. First, we combine the posteriors from separate MCMC runs on $15\times 15$ grid in $(\gamma, \log \calR)$-space, each time marginalizing over the other 14 parameters, to constrain $\bar{n}_{\rm sat}(r|M,z)$ (see Appendix~\ref{App:nprof}). Having determined the values of $\gamma$ and $\calR$ that maximize $\calL_{\rm tot}$, we then run a MCMC sampler to infer the full posterior $P(\btheta|\bD)$ keeping $\gamma$ and $\calR$ fixed at these best-fit values. The MCMC sampler used to probe our 14 dimensional parameter space is the affine invariant stretch-move algorithm of \citet{Goodman.Weare.10}. Throughout we use 1,000 walkers and the proposal density advocated by \cite{Goodman.Weare.10}. This results in typical acceptance fractions between 0.3 and 0.4, and the MCMC chain is typically converged after about 500 steps (i.e., $5 \times 10^5$ likelihood evaluations). We have experimented at length with other initial guesses, and find the results to be extremely robust, and to always fully converge well under 1 million likelihood evaluations. Finally, throughout we adopt flat priors on all parameters, with very broad prior bounds that do not affect our inference.
\begin{figure*}
\centering
\includegraphics[width=\textwidth]{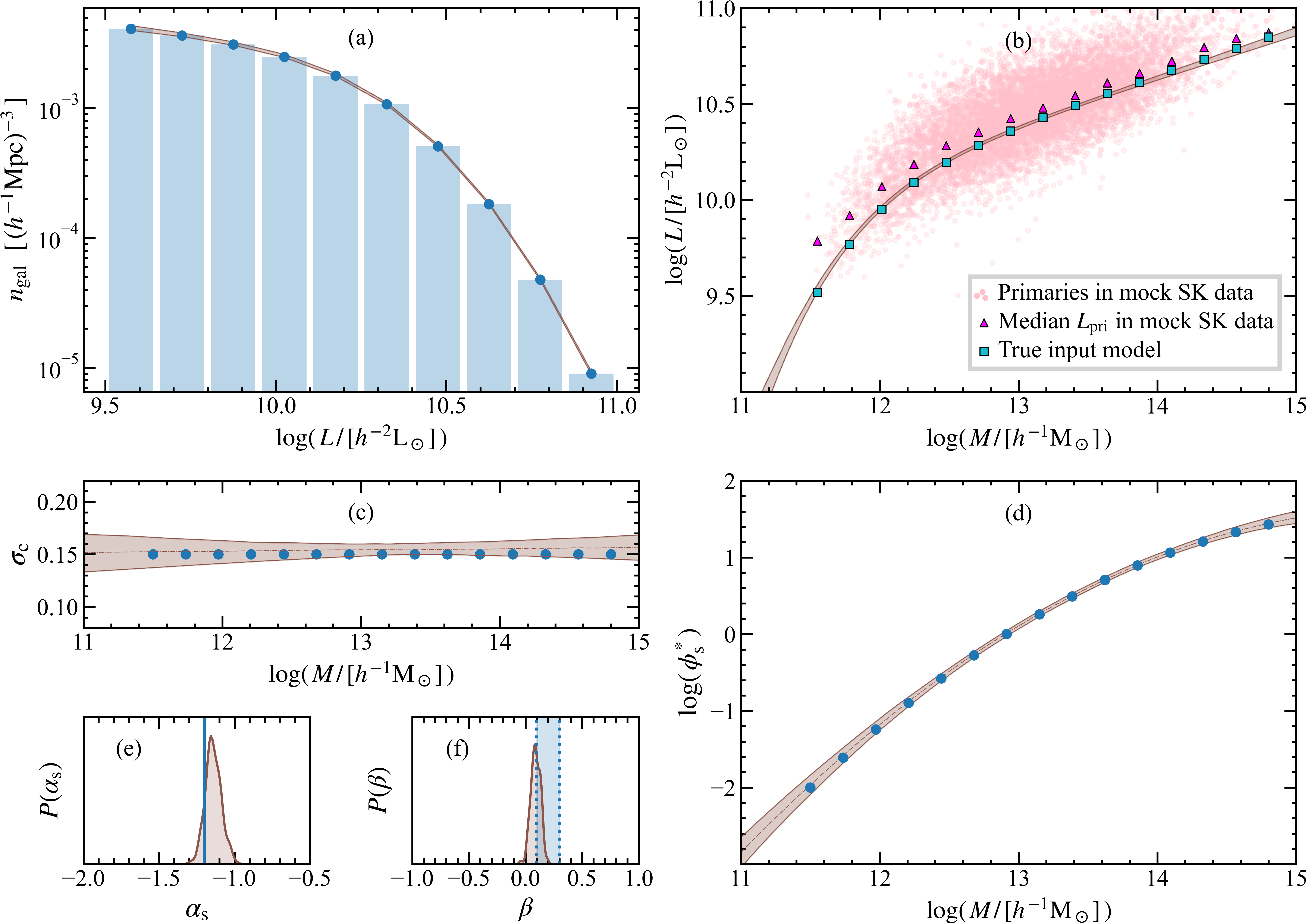}
\caption{Posterior constraints on the galaxy-halo connection in the Tier-3 mock inferred with \Basilisc. In each panel blue dots, lines or shaded regions indicate the true values in the Tier-3 mock, while the brown shaded regions or histograms indicate the 95 percentile posterior ranges inferred with \Basilisk (panels a-d), or the full posterior distributions (panels e and f). 
\textbf{\textit{Panel (a)}}: Number density of galaxies in 0.15~dex bins of galaxy luminosity.
\textbf{\textit{Panel (b)}}: Luminosity of central galaxies as a function of halo mass. For comparison, the pink points show the true $\log M_{\rm vir}$ versus $\log \Lpri$ of all selected primaries, including impurities, while the magenta triangles show their corresponding median $\log \Lpri$ in bins of halo mass. Note that, due to selection effects, this median deviates from the true underlying relation (blue squares).
\textbf{\textit{Panel (c)}}: The scatter, $\sigma_\rmc$ in the luminosity of central galaxies as a function of halo mass.
\textbf{\textit{Panel (d)}}: The normalization, $\phi_{\rms}^{*}(M)$, of the satellite CLF as a function of halo mass.
\textbf{\textit{Panel (e)}}: The faint-end slope, $\alpha_{\rms}$, of the satellite CLF.
\textbf{\textit{Panel (f)}}: The (average) orbital anisotropy, $\beta$, of satellite galaxies.}
\label{fig:T3summary}
\end{figure*}


\section{Validation with Mock data}
\label{sec:validation}

\subsection{Tier-3 mock data}
\label{sec:tier3}

We validate the performance of \Basilisk using the Tier-3 mock data introduced in \papI. This mock sample is constructed using the $z=0$ halo catalogue of the high-resolution SMDPL simulation \citep {Klypin.etal.16}, which uses $3840^3$ particles to trace structure formation in a cubic volume of $(400 \mpch)^3$, adopting cosmological parameters consistent with \cite{Planck.14}.

Each host halo in the catalogue with a mass $M_{\rm vir} \ge 3 \times 10^{10} \Msunh$ is populated with mock galaxies with luminosities $L \geq 10^{8.5} \Lsunh$ according to a particular fiducial CLF model. Each central galaxy is given the position and velocity of its halo core, defined as the region that encloses the innermost 10\% of the halo virial mass. Satellite galaxies are assigned the phase-space coordinates of the subhaloes with the highest peak halo masses. If the number of satellites, drawn from the input CLF, exceeds the number of resolved subhaloes in a specific host halo, we randomly assign the excess satellites the halo-centric positions and velocities of subhaloes hosted by other haloes of similar mass. Note that no assumption of quasi-equilibrium dynamics has been made in the mock making procedure. Therefore, our Tier-3 mock satellites obey the Jeans equations only as much as the live subhaloes do in the SMDPL simulation.

Once all haloes have been populated with mock galaxies, we construct a mock SDSS survey as follows.  First, we place a virtual observer at a random position within the simulation volume. We use this virtual observer to convert the $(x, y, z)$ coordinates of each galaxy into a cosmological redshift, $z_{\rm cosm}$, and sky coordinates (using a random orientation). If necessary, the simulation box is repeated with random sets of right angled rotations until the entire cosmological volume out to $z_{\rm cosm} = 0.20$ is covered. Next, we overlay the SDSS DR7 footprint on the simulated sky, and only keep galaxies with $m_r \leq 17.6$ that lie within the SDSS DR7 survey window. Redshift-space distortions are simulated by adding $(1 + z_{\rm cosm}) v_{\rm los} / c$ to the redshift of each galaxy, with $v_{\rm los}$ the galaxy's peculiar velocity along the line-of-sight. Spectroscopic redshift errors in the SDSS are simulated by adding a random $\Delta z$ from a Gaussian with scatter $\sigma_{\rm err} = 15\kms$ \citep{Guo.etal.15b}. Finally, we simulate the effect of fibre collisions induced spectroscopic incompleteness following the method of \citet{Lange.etal.19a}. Once the mock spectroscopic survey is completed, we select primaries and secondaries using the selection cones described in \S\ref{sec:selection}, and assign spectroscopic weights to all secondaries using the method described in \S\ref{sec:incompleteness}. Similar to what we do for the real data, we remove primaries with an aperture completeness $w_{\rm app}< 0.8$ and exclude secondaries that are located within $55''$ from their primary. Finally, we use the mock data to compute the comoving abundances of galaxies in the ten luminosity bins described in \S\ref{sec:galnumdens} using the same method as used for the real SDSS data (i.e., by taking into account the SDSS DR7 footprint with its mask and window functions as well incompleteness caused by fiber collision).
\begin{figure}
\centering
\includegraphics[width=0.48\textwidth]{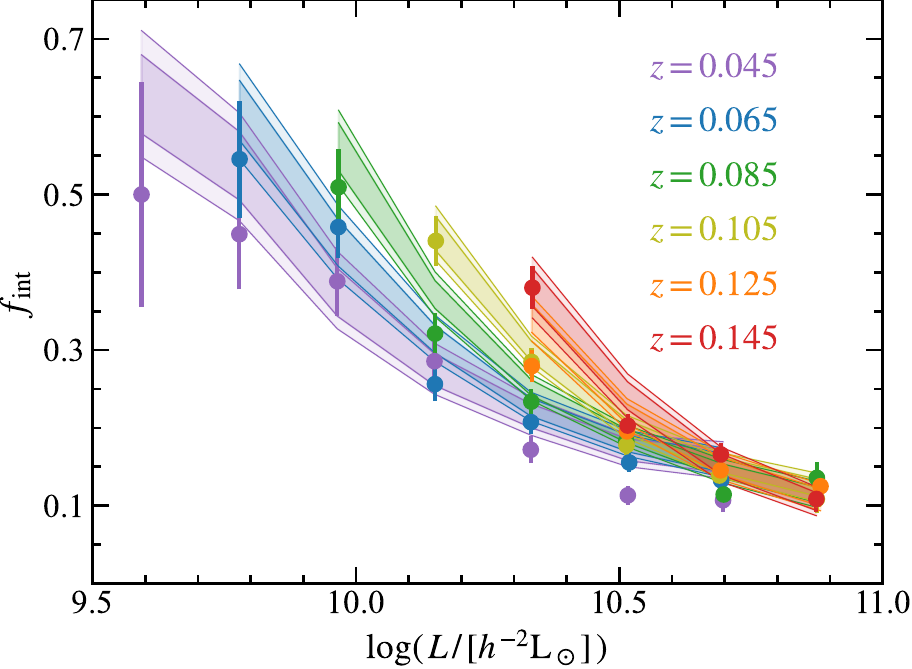}
\caption{The points with error-bars show the true interloper fractions in our Tier-3 mock sample as a function of primary luminosity for 6 different redshift bins, as indicated. The shaded bands of corresponding colour show the 68 and 95 percent confidence intervals of the posterior inferred with \Basilisc. The good agreement indicates that \Basilisk can accurately distinguish, in a statistical sense, between true satellites and interlopers.}
\label{fig:T3fint}
\end{figure}

\begin{figure}
\centering
\includegraphics[width=0.48\textwidth]{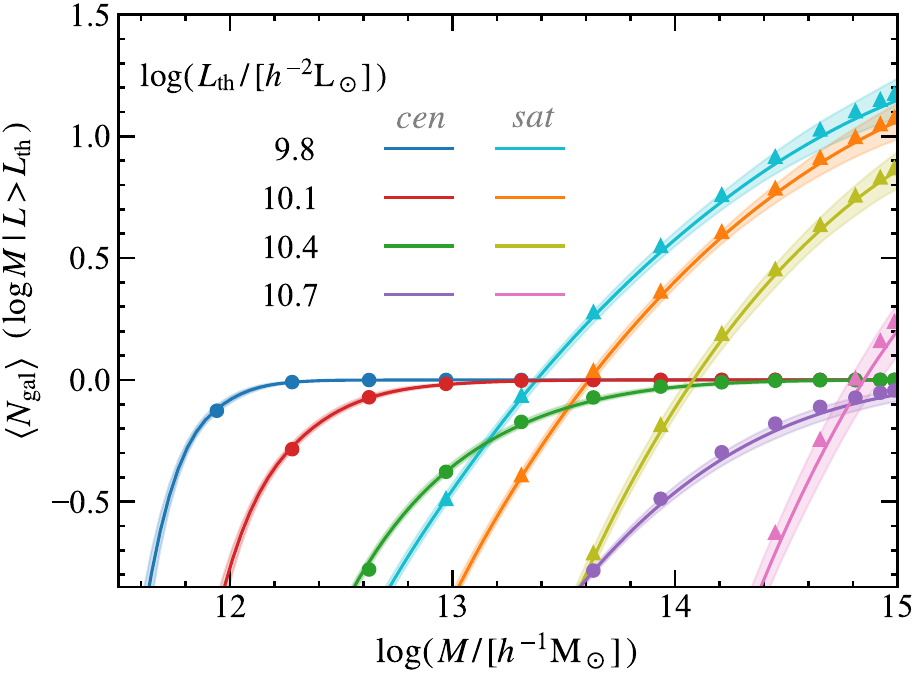}
\caption{The mean halo occupation statistics of Tier-3 mock data, and its recovery. The circles (triangles) show the true, average number of centrals (satellites) per halo, with luminosities above some luminosity threshold values, $L_{\rm th}$, as a function of host halo mass. Different colours corresponds to different $L_{\rm th}$, as indicated.  The shaded bands of corresponding colour show the 95 percent confidence intervals of the posterior inferred with \Basilisc. Note that \Basilisk accurately and precisely recovers the halo occupations statistics across all luminosities.}
\label{fig:T3MOF}
\end{figure}

\subsection{Inference from the Tier-3 mock}
\label{sec:inf3}

The Tier-3 mock data described above is used to test the performance of \Basilisc. As described in \S\ref{sec:implementation}, we first determine the best-fit values of $\gamma$ and $\calR$, which characterize the radial number density profile of satellite galaxies, properly marginalized over all other model parameters. In the Tier-3 mock, satellites are placed in subhaloes which are known to have a radial profile that differs starkly from that of the dark matter. The true radial number density distribution closely follows the generalised NFW shape (equation~\ref{nsatprof}) with $(\gamma,\calR) \approx (0.0, 2.6)$. Hence, the radial profile of satellites, in our Tier-3 mock, is cored and has a scale-radius that is significantly larger than that of their dark matter host haloes. This feature of the DM-only simulation is consistent with many previous studies \citep[e.g.,][]{Ghigna.etal.98, Diemand.Moore.Stadel.04, Springel.etal.08, Jiang.vdBosch.17}. The best-fit values obtained by \Basilisc, when applied on the Tier-3 mock data is $(\gamma,\calR) = (0.0, 2.5)$, in almost perfect agreement with the profile inferred directly from the $N$-body simulation. Thus, in agreement with what we reported in \papI, \Basilisk can accurately recover the radial profile of satellite galaxies. Next, we keep $(\gamma,\calR)$ fixed at the best-fit values, $(0.0, 2.5)$, and run \Basilisk to infer the posterior distribution of the remaining 14 parameters. The brown lines in Fig.~\ref{fig:T3_corner} show the CLF constraints thus obtained for our fiducial model\footnote{No results are shown for the three nuisance parameters $\eta_0$, $\eta_1$ and $\eta_2$ that characterize the abundance of interlopers, or for the velocity anisotropy parameter $\beta$, which is discussed in detail in \S\ref{sec:anisotropy_mock}.}. Note that the posteriors of all CLF parameters are in excellent agreement with the input values, shown as blue dashed lines. For comparison, the posteriors indicated in green correspond to a model that allows for mass-dependence of the orbital anisotropy, and will be discussed in \S\ref{sec:anisotropy_mock}.

Panel (a) of Fig.~\ref{fig:T3summary} demonstrates that \Basilisk accurately fits the galaxy luminosity function. The blue circles show the mean number density of galaxies in logarithmic bins of luminosity of width $0.15$ dex (roughly the width of the blue vertical bars), while the brown band indicates the $95\%$ confidence interval as inferred from \Basilisc's posterior.
\begin{figure*}
\centering
\includegraphics[width=0.95\textwidth]{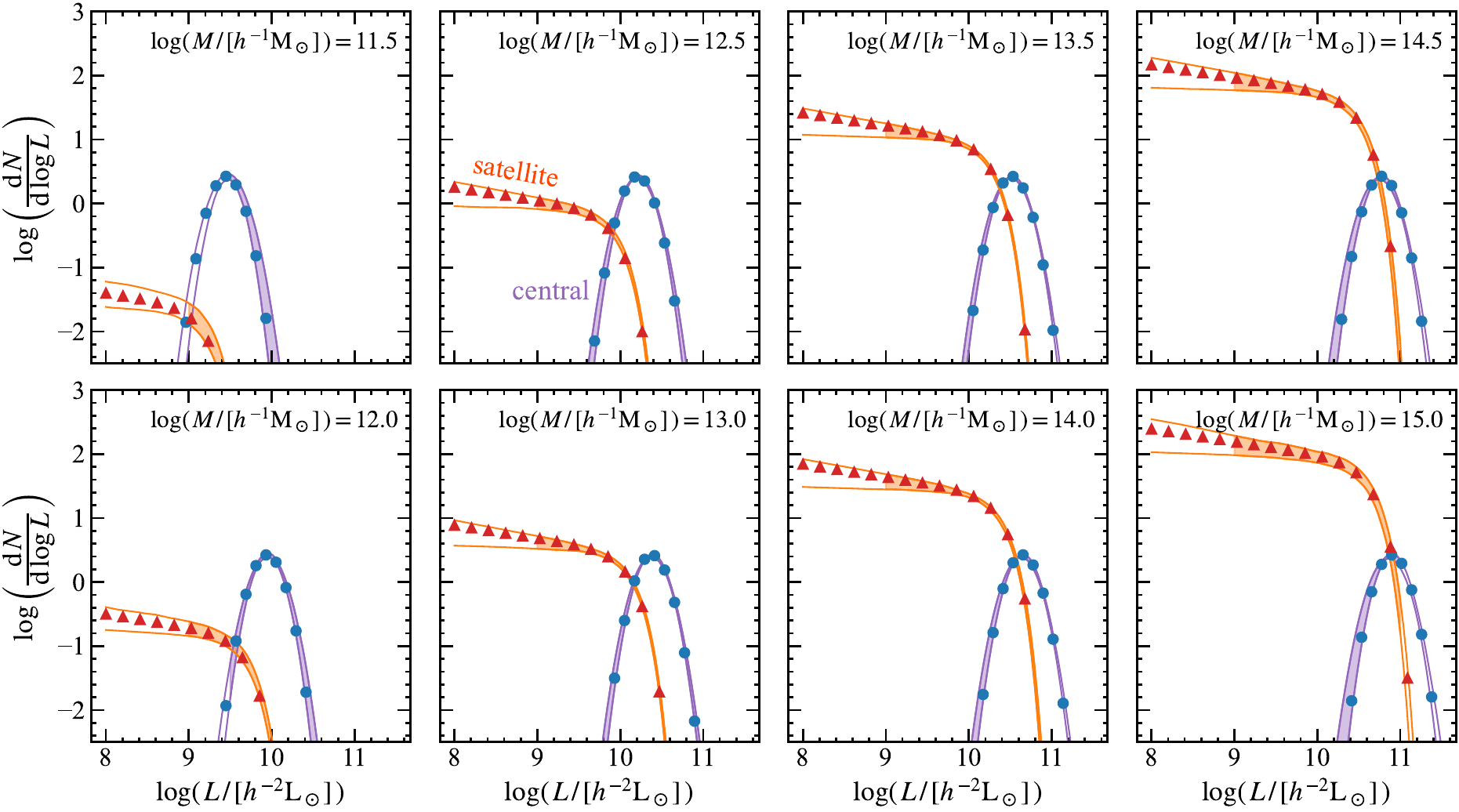}
\caption{Different panels show the true input CLF of central (blue circles) and satellite (red triangles) galaxies for different halo masses (as indicated in the top-right corner of each panel). The purple and orange shaded bands show the corresponding 95 percent confidence intervals inferred from \Basilisc's posterior. Parts of the bands that are not shaded correspond to luminosities outside the ranges covered by the sample of primaries and secondaries.  \Basilisk successfully recovers the input CLF over the entire range in halo mass probed, from $10^{11.5} \Msunh$ to $10^{15} \Msunh$.}
\label{fig:T3clf}
\end{figure*}

Fig.~\ref{fig:T3summary} panel (b) shows the relationship between halo mass and median central galaxy luminosity. Note that the true input model (blue squares) is perfectly recovered by \Basilisk (brown band, indicating the 95~\% confidence interval from the posterior). The pink dots indicate the true halo masses and luminosities of all primaries (both true centrals and impurities) in the Tier-3 satellite kinematics sample. The magenta triangles show their median values of $\log \Lpri$ in bins of $\log M$. Note that, due to selection bias, these do not agree with the true input model (blue squares). A more luminous primary has a larger secondary selection volume associated with it,  as well as a larger luminosity range between $\Lpri$ and $L_{\rm min}(\zb)$. Therefore, at any given redshift, brighter primaries are more likely to have at least one secondary than less luminous primaries in haloes of the same mass, hence the brighter primaries get preferentially selected in the $\bD_{\rm SK}$ data. This causes the median luminosity of primaries in the satellite kinematics sample to be biased high with respect to that of true centrals. This effect is especially pronounced at the low halo mass end, where the expectation value for the number of secondaries is lowest.  Despite this strong and unavoidable selection bias, \Basilisk perfectly recovers the {\it true} input relation between central luminosity and halo mass, and not the biased relation! This indicates that \Basilisc, in its forward modelling, accurately accounts for selection bias and other systematics such as the presence of impurities. 

Panels (c) and (d) of Fig.~\ref{fig:T3summary} show, respectively, the logarithmic scatter in central galaxy luminosity at fixed halo mass, $\sigma_{\rmc}$, and the normalization of the satellite CLF, $\phi_{\rms}^*$. Once again, blue dots show the true input values, while the brown shaded regions mark the 95\% confidence interval of the posterior distribution inferred with \Basilisc. Finally, the brown-shaded histograms in panels (e) and (f) show, respectively, the posterior distributions for the faint-end slope of the satellite CLF, $\alpha_\rms$, and that of the orbital anisotropy parameter, $\beta$. The blue, vertical line in panel (e) indicates the true input value of $\alpha_\rms$, while the blue-shaded region in panel (f) show the range of mean velocity anisotropy of subhaloes in different halo masses in the SMDPL simulation used to construct the Tier-3 mock (discussed further in \S\ref{sec:anisotropy_mock}). As is evident, in each case the posterior constraints are in excellent agreement with the input values, which indicates that \Basilisk can put tight and accurate constraints on the more intricate aspects of the galaxy-halo connection, beyond the mere relation between halo mass and central luminosity. 

Fig.~\ref{fig:T3fint} shows the interloper fraction, $f_{\rm int}$, as a function of primary galaxy luminosity in six redshift bins (indicated by different colours). Solid circles indicate the true interloper fractions in the Tier-3 mock sample with the error bars computed assuming Poisson statistics. The coloured bands show the corresponding posterior predictions as inferred by \Basilisc, which are in good agreement with the true interloper fractions. Hence, \Basilisk correctly distinguishes satellites from interlopers, at least in a statistical sense, and accurately recovers their relative prevalence as a function of luminosity and redshift of the primary.

Finally, Figs.~\ref{fig:T3MOF} and~\ref{fig:T3clf} compare the posterior predictions for the HOD and CLF, respectively, to their true input used to construct the Tier-3 mock data (solid dots). In particular, Fig.~\ref{fig:T3MOF}, shows the average number of central (circles) and satellite (triangles) galaxies per halo, above four different luminosity thresholds ($L_{\rm th}$), as a function of host halo mass. Interestingly, the true satellite mean occupation in our mock data deviates significantly from a simple power-law (which is a common assumption in the literature), and \Basilisk accurately recovers that complexity in its shape across all luminosities and halo masses. Fig.~\ref{fig:T3clf} plots the central (purple) and satellite (orange) CLFs for 8 different halo masses, as indicated in the top-right corner of each panel. Note that in all cases the halo occupation statistics are recovered with exquisite precision and accuracy.

\subsection{Velocity Anisotropy in the Tier-3 Mock}
\label{sec:anisotropy_mock}

Unlike previous studies of satellite kinematics, which used satellite velocity dispersions in bins of primary luminosity, a unique aspect of \Basilisk is that it models the full probability distribution $P(\dV, \Rp|M,L,z)$. By modelling the full 2D phase-space information, rather than just the second moment of velocity, \Basilisk has the potential to break the mass-anisotropy degeneracy that hampers dynamical models which only rely on measurements of the satellite velocity dispersions. Hence, \Basilisk is able to simultaneously constrain the halo masses of the primaries, as well as the velocity anisotropy of its secondaries. 

\begin{figure}
\centering
\includegraphics[width=0.48\textwidth]{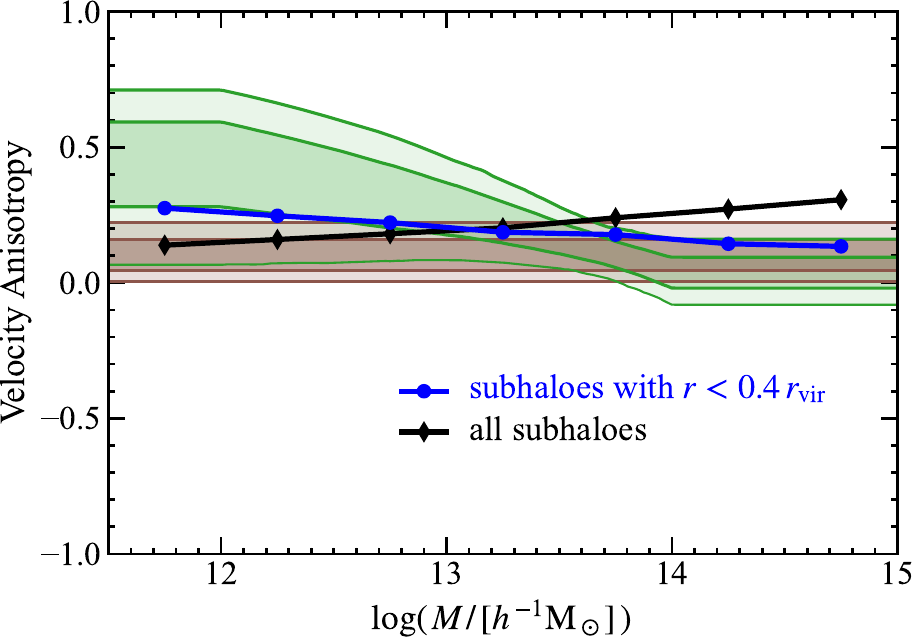}
\caption{Satellite velocity anisotropy in the Tier-3 mock. Brown shaded contours indicate the 68 and 95 per cent confidence intervals on the orbital anisotropy parameter $\beta$ as a function of halo mass inferred from our fiducial model in which $\beta$ is assumed to be global constant (independent of halo mass). The green shaded bands show the confidence intervals obtained when allowing for halo mass dependence as described in the text (see equation~[\ref{eqn:beta_M}]). The black and blue lines show the host halo mass dependence of the true, mean velocity anisotropy of {\it all} subhaloes in the SMDPL simulation, and of those with a halo-centric radius $r < 0.4 \rvir$, respectively.}
\label{fig:betamock}
\end{figure}

The brown shaded contours in Fig.~\ref{fig:betamock} show the 68 and 95 percent confidence intervals on the orbital anisotropy parameter $\beta$ as inferred by \Basilisk from the Tier-3 mock data. Note that the constraints ($\beta = 0.11 \pm 0.05$) are remarkably tight and indicate a mild, radial anisotropy. Recall that the Tier-3 mock was constructed by placing satellite galaxies inside subhaloes in the SMDPL simulation. Hence, these satellite galaxies have the same orbital anisotropy as those subhaloes. The solid, black line indicates the average anisotropy parameter of subhaloes in the SMDPL simulation as a function of host halo mass. It, too, indicates a mild radial anisotropy in reasonable agreement with the constraints obtained with \Basilisc. However, this comparison is not entirely fair. After all, our secondary selection criteria typically only selects secondaries with projected separations $R_{\rm ap}^{\rm sec} \lta 0.4 \rvir$ (see \S\ref{sec:selection}). Therefore, it is more meaningful to compare our posterior constraints with the orbital anisotropy of subhaloes located in the inner regions of their host haloes. The blue, solid line in Fig.~\ref{fig:betamock} shows the orbital anisotropy of subhaloes with a 3D halo-centric radii less than $0.4 \rvir$. These are in better agreement with \Basilisc's posterior constraints, especially given that the satellite kinematics data is dominated by haloes above $10^{12} \Msunh$. Hence, we conclude that \Basilisc, by using the full line-of-sight velocity distributions of the secondaries, is indeed able to break the mass-anisotropy degeneracy and obtain simultaneous, reliable constraints on both halo mass and orbital anisotropy. 

Thus far we have only considered models in which the orbital anisotropy is a   `universal' constant, independent of halo mass or halo-centric radius. However, a detailed analysis of the orbital anisotropy of dark matter subhaloes in the SMDPL simulation (and hence our Tier-3 mock satellites) reveals a rather complicated dependence on both halo mass and halo-centric radius (see Fig.~C1 in \papI). Indeed, the blue and black curves in Fig.~\ref{fig:betamock} indicate some dependence on halo mass, albeit weak. 
We therefore also analysed the Tier-3 mock data using a more flexible model with a mass-dependent orbital anisotropy parameter, given by $\beta(M) = 1 - 10^{-\mathcal{B}(M)}$, with
\begin{equation}\label{eqn:beta_M}
\mathcal{B}(M) =
\begin{cases}
\mathcal{B}_{12}, & \text{if $M < 10^{12} \Msunh$}\\
\mathcal{B}_{14}, & \text{if $M > 10^{14} \Msunh$}\\
\mathcal{B}_{12} + \frac{1}{2} (\mathcal{B}_{14}-\mathcal{B}_{12}) \log M_{12}, & \text{otherwise}
\end{cases}
\end{equation}
Here both $\mathcal{B}_{12}$ and $\mathcal{B}_{14}$ are free parameters for which we adopt uniform priors ranging from $-1$ to $+1$. Hence, by replacing our fiducial model, in which $\beta$ is independent of halo mass, with this mass-dependent model we add one extra free parameter to the mix.

Remarkably, we find that this extra degree of freedom has no discernible impact on the constraints of any of the other parameter. This is evident from Fig.~\ref{fig:T3_corner}, where the green contours show the posterior constraints from the mass-dependent $\beta$ model, which are indistinguishable from those of our fiducial model (brown contours). The green shaded contours in Fig.~\ref{fig:betamock} show the 68 and 95 percent confidence intervals on  $\beta(M)$ of the corresponding model. It reveals a weak hint for the orbital anisotropy to become more radially anisotropic in lower mass haloes, in agreement with the weak trend for the SMDPL subhaloes with $r < 0.4 \rvir$. However, the uncertainties at the low mass end are rather large, and  \Basilisc's inference is consistent with a constant, mass independent $\beta$ at the $2\sigma$ level. 

\begin{figure*}
\centering
\includegraphics[width= 0.8\textwidth]{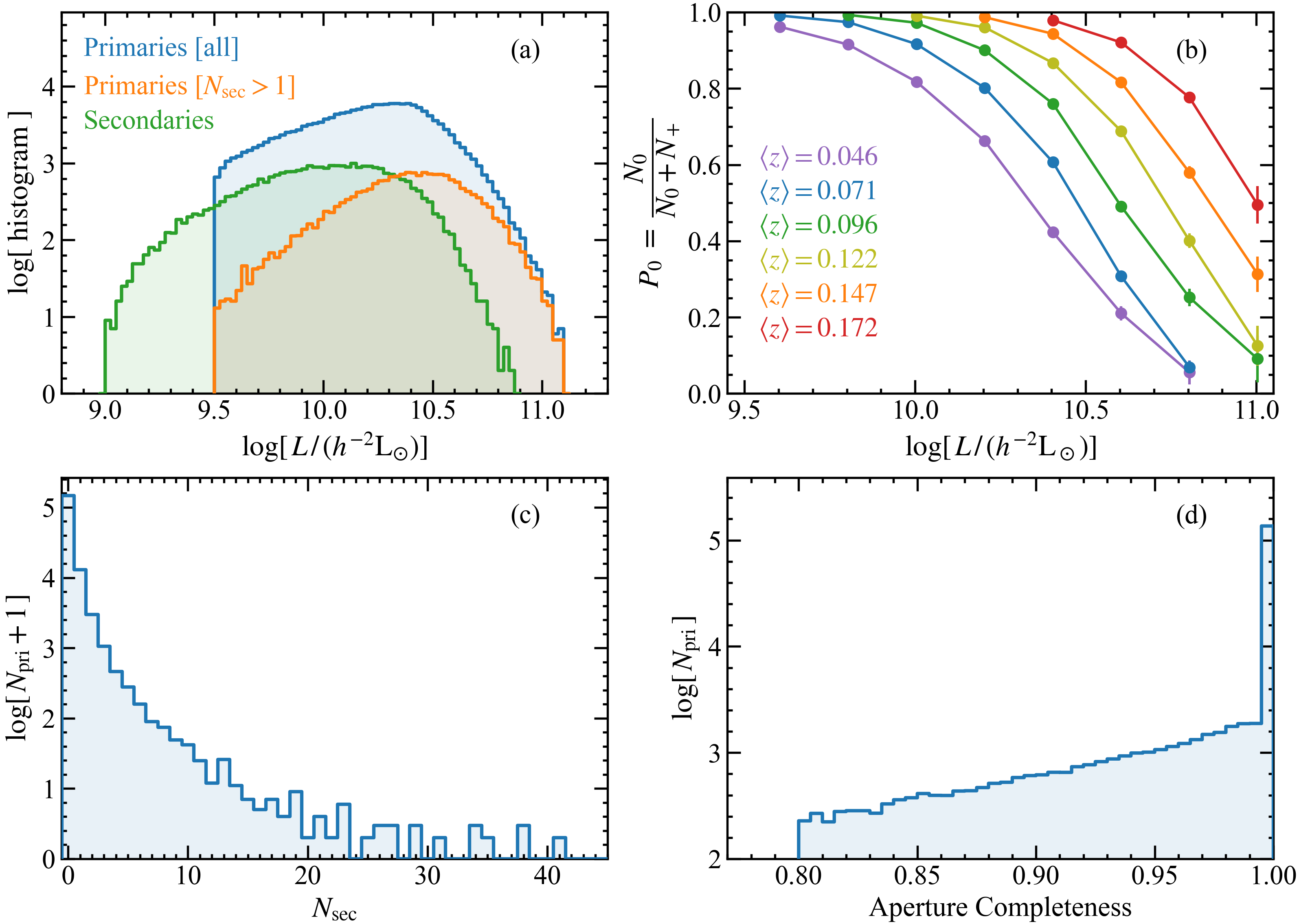}
\caption{Statistics of our sample of primaries and secondaries selected from the SDSS-DR7. \textbf{\textit{Top-left:}} The numbers of all primaries (blue), primaries with at least one secondary (orange), and secondaries (green) as a function of their luminosities. \textbf{\textit{Top-right:}} Probability, $P_0$, that a primary of luminosity $\Lpri$ has zero secondaries. Results are shown for 6 redshift bins, as indicated. \textbf{\textit{Bottom-left:}} Multiplicity function, indicating the number of primaries (as $\log[ N_{\rm pri} + 1]$) with $N_{\rm sec}$ secondaries in our sample. \textbf{\textit{Bottom-right:}} the distribution of aperture completeness, $w_{\rm app}$, for all primaries.}
\label{fig:data_stat}
\end{figure*}

\begin{figure*}
\centering
\includegraphics[width=0.9\textwidth]{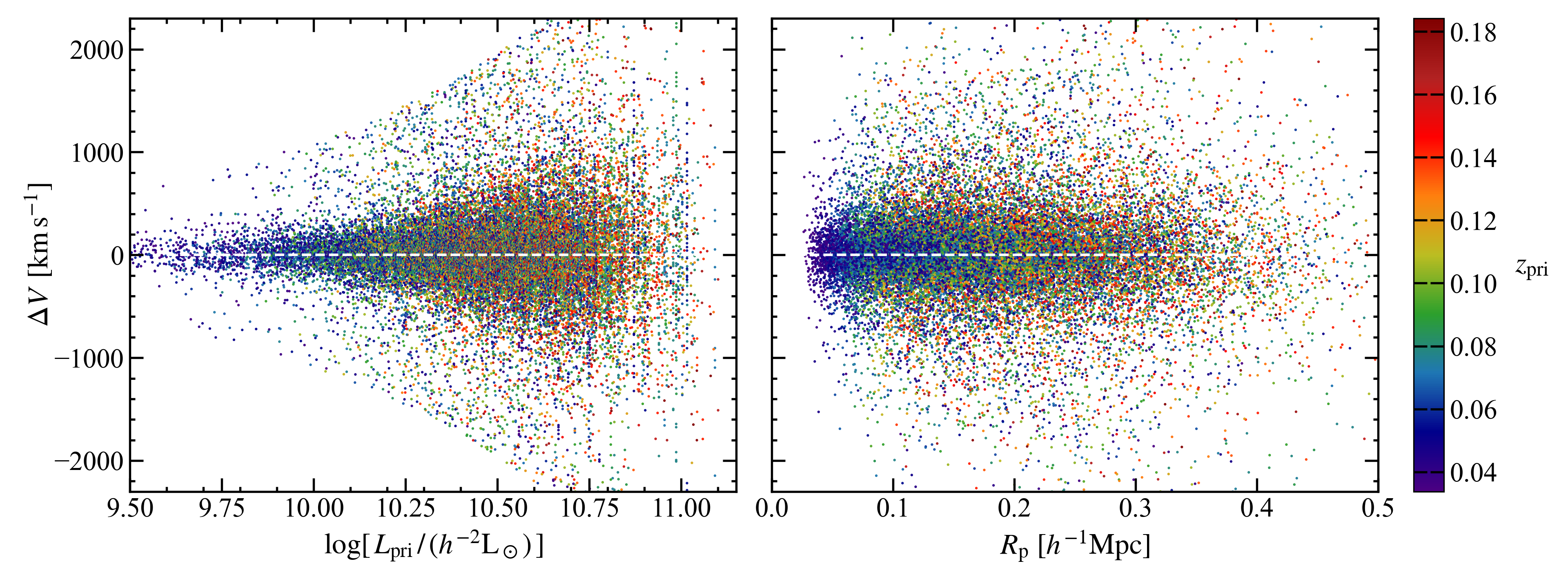}
\caption{The velocity difference, $\Delta V$ for all the 30,431 primary-secondary pairs in our SDSS sample as function of the luminosity of the primary, $L_\rmc$, (left-hand panel) and as a function of the projected separation between primary and secondary, $R_\rmp$ (right-hand panel). Colours indicate the redshift of the primary, as indicated in the colour-bar on the right.}
\label{fig:data}
\end{figure*}

\section{Application to SDSS}
\label{sec:SDSS}

\subsection{The Data}
\label{sec:SDSSdata}

Having demonstrated the \Basilisk can accurately infer the galaxy-halo connection from kinematic data of primaries and secondaries that can be extracted from a galaxy redshift survey, we now apply \Basilisk to data from the SDSS. In particular, we use the New York University Value-Added Galaxy Catalog \citep[VAGC;][]{Blanton.etal.05}, which derives from the Seventh Data Release of the SDSS \citep[SDSS DR7;][]{Abazajian.etal.09}. More specifically, we use the VAGC \texttt{bright0} sample\footnote{\url{http://sdss.physics.nyu.edu/lss/dr72/bright/0/}}, which includes $\sim 570,000$ galaxies with a limiting Petrosian magnitude of $m_r < 17.6$. We use this data to identify primaries and secondaries using the selection criteria outlined in \S\ref{sec:selection}.  As already mentioned there, we limit our analysis to galaxies in the redshift range $0.02\leq z \leq 0.20$. In order to assure that the secondary selection cones fit entirely within this redshift range, the redshifts of primaries are restricted to $0.034 \leq z \leq 0.184$ (see Fig.~\ref{fig:logL_z_selection}). Primaries are also limited to have luminosities in the range $9.504 \leq \log[\Lpri/(h^{-2}\Lsun)] \leq 11.104$. We end up with a total of $\sim 165,000$ primaries\footnote{For the computation of  $\calL_{\rm NS}$, the number of primaries is downsampled by an order of magnitude to $16,491$, as discussed in \S\ref{sec:modelNs}.}, of which $N_+=18,373$ primaries have at least one secondary. The total number of secondaries, and thus the total number of primary-secondary pairs for which kinematic data is available, is $30,431$.

The upper left-hand panel of Fig.~\ref{fig:data_stat} shows the luminosity distributions of all primaries (blue), primaries with at least one secondary (orange), and secondaries (green). Note that primaries with at least one secondary are brighter, on average, than those with zero secondaries. This simply follows from the fact that, on average, brighter centrals reside in more massive haloes, which host more satellites. Also, given a fixed halo mass the brighter primary is assigned a larger secondary selection volume, making it more likely to have a secondary. This latter effect, though, is subdominant.  Note also that there are no satellites with $L < 10^{9} \Lsunh$. As discussed in \S\ref{sec:selection}, this is a consequence of the apparent magnitude limit of the SDSS survey combined with the fact that we only allow for primaries brighter than $10^{9.504} \Lsunh$. The upper right-hand panel shows the probability, $P_0$, that a primary of luminosity $\Lpri$ contains zero secondaries. It is simply defined as the fraction of primaries, in a given luminosity and redshift bin, with zero secondaries, i.e., $P_0 = N_0 / (N_0 + N_+)$, where $N_0$ and $N_+$ are the number of primaries in our sample with zero and at least one detected secondary, respectively. These probabilities have been computed using a $8\times 6$ uniformly-spaced grid in $(\log \Lpri, \zpri)$ covering the ranges $[9.504,11.104]$ and $[0.034,0.184]$, respectively. Different colours correspond to different redshift bins, as indicated. We emphasize that this binned data are not used in our inference; it is merely used here to illustrate how $P_0$ scales with luminosity and redshift. Errorbars are computed assuming Poisson statistics, and are smaller than the data points in most cases. Note that $P_0$ increases with decreasing luminosity and increasing redshift, as expected from the Malmquist bias resulting from the apparent magnitude limit of the spectroscopic SDSS data. The lower left-hand panel shows the multiplicity function, i.e., the number of primaries each with $N_{\rm sec}$ secondaries. Note that most primaries have zero secondaries, and that there are very few primaries with more than 20 secondaries. Finally, the lower right-hand panel shows the distribution of the aperture completeness $w_{\rm app}$ for all primaries. Note that the vast majority of primaries have $w_{\rm app}=1$, i.e., their entire secondary selection cone completely falls within the SDSS survey footprint. As mentioned in \S\ref{sec:incompleteness}, primaries with $w_{\rm app} < 0.8$ have been removed from the sample.

Figure~\ref{fig:data} plots the line-of-sight velocity difference $\dV$ as a function of the luminosity of the primary galaxies (left-hand panel) and as a function of the projected distance between primary-secondary pairs (right-hand panel). Data points are colour-coded according to the redshift of the primary, as indicated. This constitutes the satellite kinematics data in SDSS, that we attempt to forward-model with \Basilisk in order to constrain the galaxy halo connection. The deficit of data points at small $R_{\rm p}$ reflects that we have removed secondaries with a projected separation less than $\theta_{\rm fc} = 55''$ because of fiber collision issues (see \S\ref{sec:incompleteness}). Similarly, the absence of data points for low $\Lpri$ and large $|\Delta V|$ reflects the luminosity dependence of the secondary selection criteria (see \S\ref{sec:selection}). Evidently, that the velocity dispersion of secondaries is a strong function of primary luminosity, consistent with the expectation that more luminous centrals reside in more massive dark matter haloes. The low-density, high velocity wings of $P(\dV)$ at any given $\Lpri$ reflects the contribution of foreground and background interlopers, i.e., galaxies selected as secondaries that do not reside in the dark matter halo of the primary. 
\begin{table*}
\renewcommand{\arraystretch}{1.2}
\caption{Galaxy-halo connection parameters with brief descriptions (see \S\ref{sec:CLF} for details), along with \Basilisc's inference for SDSS DR7 galaxies. Best-fit values and $1 \sigma$ confidence intervals of all parameters quoted in the table, excluding the last two, are for the mass-independent velocity anisotropy model.}
\smallskip
\begin{tabular}{l c c c c}
\hline
  $\,$ & parameter & brief description  & best-fit & $1\sigma$ interval \\ [0.4ex] 
\hline  Central CLF 
& $\log M_1$ & characteristic mass of $\bar{L}_\rmc(M)$ relation  & 11.39 & [11.27, 11.50] \\ (Eqn.~\ref{CLFcen}-\ref{scatter}) 
& $\log L_0$ & normalization of $\bar{L}_\rmc(M)$ relation & 10.04 & [10.00, 10.08] \\
& $\gamma_1$ &  slope $\rmd\log\bar{L}_\rmc/\rmd\log M$ at the low-mass end  & 2.32 & [2.02, 2.78]\\
& $\gamma_2$ & slope $\rmd\log\bar{L}_\rmc/\rmd\log M$ at the massive end  & 0.204 & [0.194, 0.212] \\
& $\sigma_{13}$ & scatter in $\log L_\rmc$ at $M = 10^{13} \Msunh$  & 0.177 & [0.175, 0.180] \\
& $\sigma_\rmp$ & slope $\rmd\sigma_{\rmc}/\rmd \log M$ of halo mass dependence of scatter &  0.001 & [-0.002, 0.005] \\ \\ \cdashline{1-5} \\
Satellite CLF & $\alpha_{\rm s}$ & faint-end slope of satellite CLF & -0.80 & [-0.85, -0.75] \\  (Eqn.~\ref{satCLF}-\ref{satCLFnorm}) 
& $b_0$ & normalization of satellite CLF at $M=10^{12} \Msunh$ & -0.64 & [-0.67, -0.61] \\
& $b_1$ & linear $\log M$-dependence of satellite CLF normalization &  1.06 & [ 1.02, 1.09] \\
& $b_2$ & quadratic $\log M$-dependence of satellite CLF normalization & -0.12 & [-0.13, -0.10] \\ \\ \cdashline{1-5} \\
Nuisance parameters  & $\eta_0$ &  normalization of effective bias of interlopers & 1.65 & [1.50, 1.82] \\ (Eqn.~\ref{beff})
& $\eta_1$ & luminosity dependence of the effective bias of interlopers & 0.32 & [0.27, 0.37] \\
& $\eta_2$ & redshift dependence of the effective bias of interlopers & -0.41 & [-0.52, -0.31] \\ \\ \cdashline{1-5} \\
Constant anisotropy (Eqn.~\ref{eqn:beta})  & $\beta_{\rm avg}$ & typical orbital velocity anisotropy of satellites & 0.29 & [0.25, 0.34] \\ \\ \cdashline{1-5} \\
Mass-dependent  & $\mathcal{B}_{12}$ & controls orbital anisotropy of satellites in low-mass haloes & 0.37 & [0.29, 0.46]  \\
 anisotropy model (Eqn.~\ref{eqn:beta_M}) & $\mathcal{B}_{14}$ & controls orbital anisotropy of satellites in high-mass haloes &  0.09 & [0.05, 0.12] \\
\hline
\end{tabular}
\label{table:SDSS_params}
\end{table*}

\begin{figure*}
\centering
\includegraphics[width=0.85\textwidth]{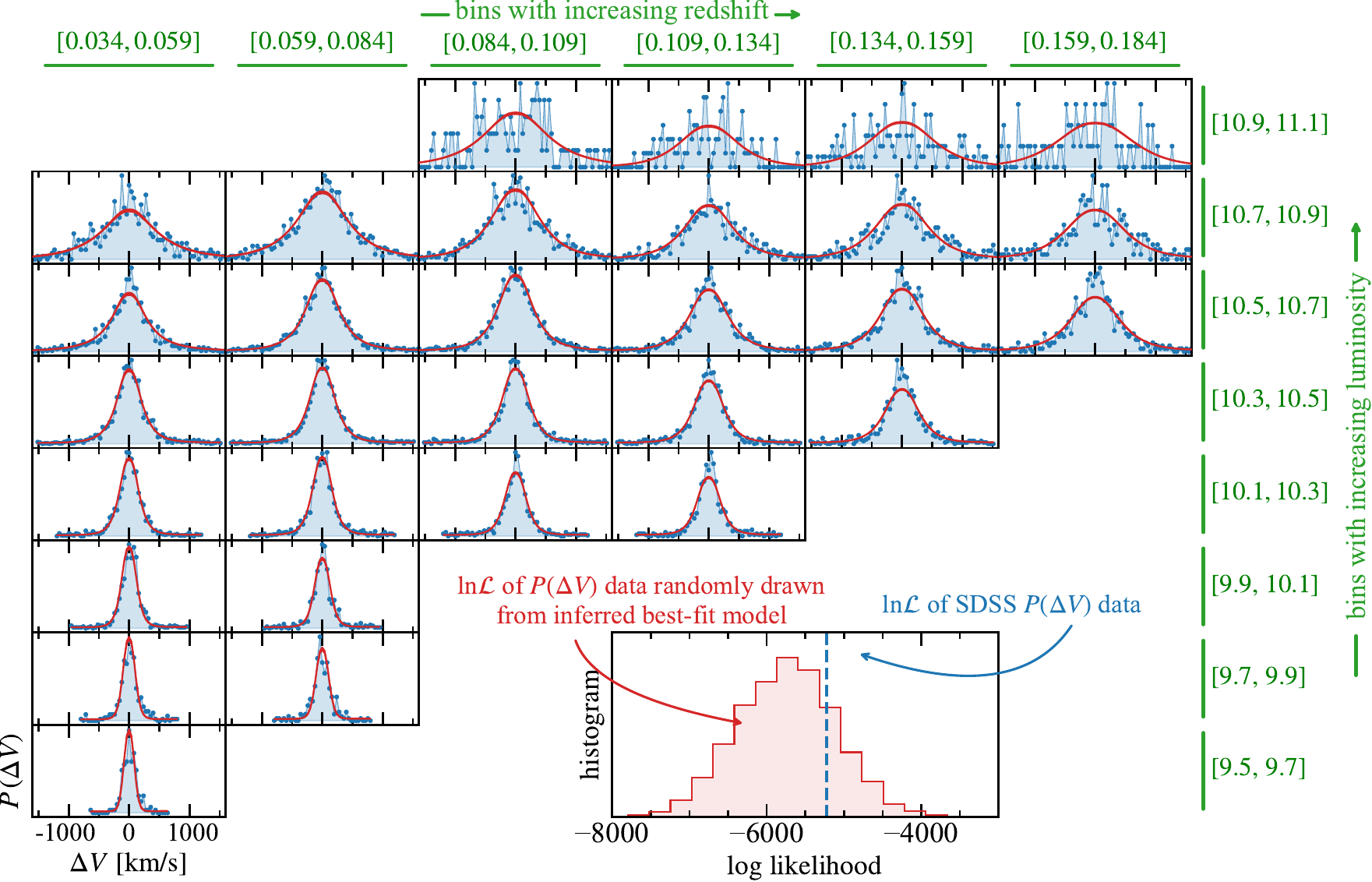}
\caption{The LOSVDs of secondaries around primary galaxies stacked in bins of luminosity (increases vertically) and redshift (increases horizontally), as indicated. Only bins that fall entirely above the SDSS flux limit are shown. The SDSS data is shown as the blue shaded histograms, while the red bands indicate the 95\% confidence intervals obtained using the posterior distributions inferred with \Basilisc. The red histogram in the inset-panel in the lower-right corner shows the likelihood distribution of $10^4 \times N_{\rm pair}$ random realizations of $P(\Delta V)$ data for the inferred best-fit model, while the blue, dashed line shows the corresponding value for the actual SDSS data (see the text for details). Here $N_{\rm pair}$ is total number of primary-secondary pairs in the SDSS dataset. Note that the model inference is an excellent match to the data.}
\label{fig:SK_fit}
\end{figure*}

\begin{figure*}
\centering
\includegraphics[width=0.85\textwidth]{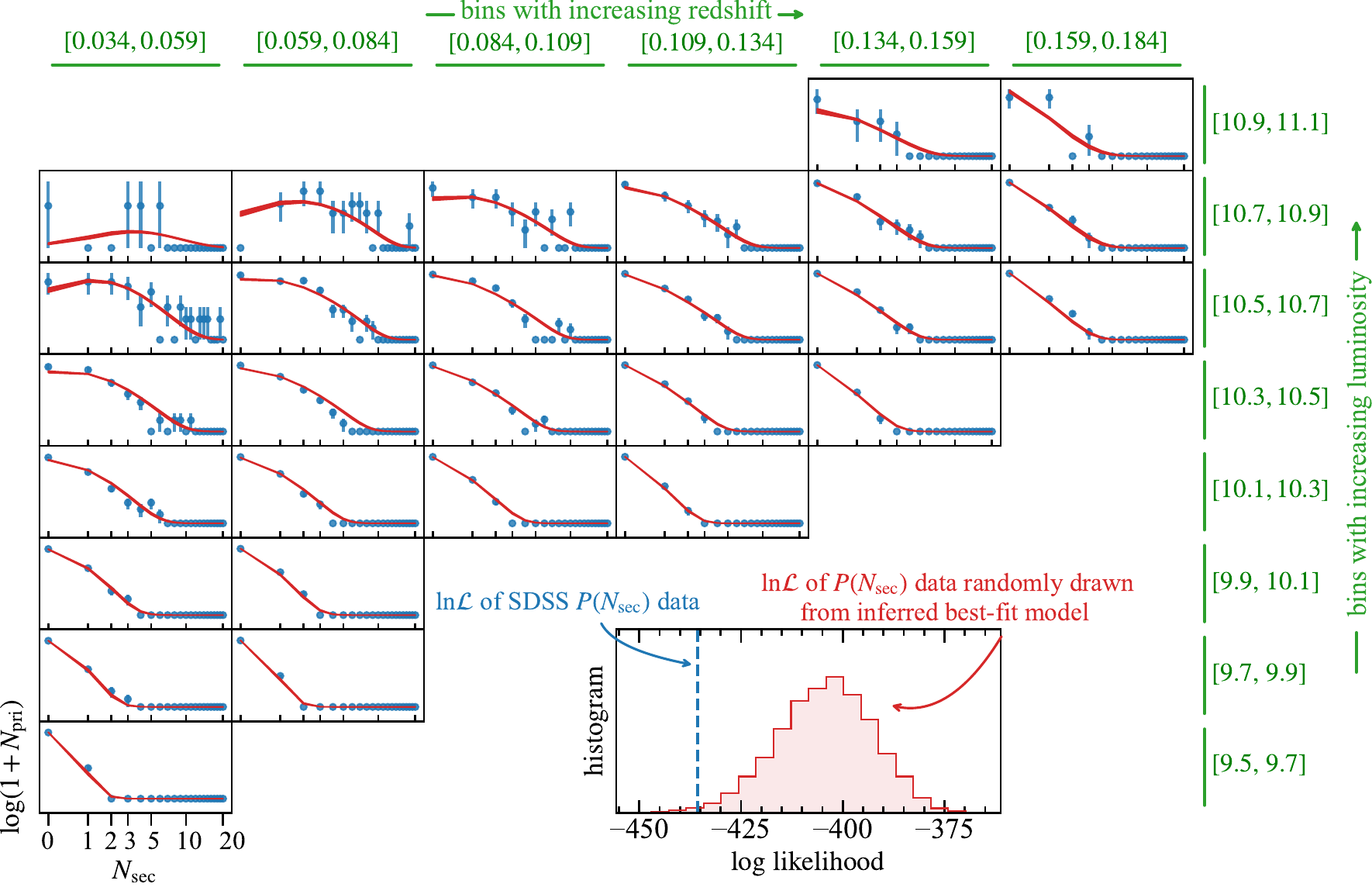}
\caption{Multiplicity functions for the numbers of secondaries in the same bins of luminosity (increasing vertically) and redshift (increasing horizontally) as used in Fig.~\ref{fig:SK_fit}.  Only bins lying entirely above the flux limit of the survey, and with at least 5 primaries in the down-sampled `$\mathcal{L}_{\rm NS}$ data', are shown. The blue points show the SDSS data, with Poisson error-bars, while the red bands indicate the 95\% confidence intervals of the posterior prediction from \Basilisc. The red histogram in the inset-panel in the lower-right corner shows the likelihood distribution of $10^4$ random realizations of these multiplicity functions for the inferred best-fit model, while the blue, dashed line shows the corresponding value for the actual SDSS data. Although the fit to the data looks very reasonable at first sight, this more detailed comparison of likelihoods shows that the best-fit model is not a perfect fit to the data. As discussed in the text, we attribute this to limited freedom in the satellite CLF model used to describe the galaxy-halo connection.}
\label{fig:Ns_fit}
\end{figure*}

\subsection{Results}
\label{sec:SDSSresults}

We start our analysis of the SDSS data presented above by constraining the satellite radial number density profile (equation~[\ref{nsatprof}]). Using a $15\times 15$ grid in $(\gamma, \log \calR)$-space, we obtain $(\gamma,\calR) = (0.94, 1.7)$ (see Appendix~\ref{App:nprof} for details), in good agreement with the results of \cite{Lange.etal.19b}
and \citet{Wojtak.Mamon.13}. Note that this central slope, $\gamma$, is significantly steeper than what we inferred for the Tier-3 mock data ($\gamma=0.0$, see \S\ref{sec:inf3}), in which the satellites were positioned on subhaloes in numerical simulations. This indicates that the radial distribution of real satellite galaxies is more centrally concentrated than that of subhaloes in DM-only numerical simulations. This discrepancy is most likely is due to a combination of artificial disruption in simulations \citep[][]{Penarrubia.etal.10, vdBosch.etal.18a, vdBosch.etal.18b, Errani.Penarrubia.20, Errani.Navarro.21} and failures of the subhalo finders being used \citep[e.g.,][]{Knebe.etal.11, Han.etal.12, Diemer.etal.24}.

Next, keeping $\calR$ and $\gamma$ fixed at $1.7$ and $0.94$, respectively, we run \Basilisk to constrain the posterior distribution of the remaining 14 parameters that characterize the CLFs of central and satellite galaxies, the interlopers, and the average orbital anisotropy of satellite galaxies. Once again, we adopt very broad non-informative priors for all parameters. Table~\ref{table:SDSS_params} lists the best-fit parameters plus their 68\% confidence intervals thus obtained. We emphasize that, as shown in \paperI and Appendix~\ref{App:nprof}, all these results are extremely robust to modest changes in $\calR$ and $\gamma$.

Before showing the key results on the CLF, we first demonstrate that the best-fit model of \Basilisk is an excellent fit to the data. In order to illustrate this, we bin the data in 2D-bins of luminosity and redshift of the primaries. We emphasize that no such binning was used in the analysis; it is merely used here for the purpose of visualization of the data and its corresponding prediction from \Basilisc. The various panels in Fig.~\ref{fig:SK_fit} show the LOSVDs of primary-secondary pairs for bins in $\log\Lpri$ (different rows) and $\zpri$ (different columns). We only show panels for which the luminosity lower bound of the $\{\log\Lpri,\zpri\}$ bin falls above the flux limit at the redshift upper bound of that bin.  Blue dots and shaded histograms show the stacked data, while the red shaded bands show to 95\% confidence intervals obtained using the inferred posterior distributions. In order to quantify the level of agreement between the data and the model, we proceed as follows. Let $N_\rms(\Lpri,\zpri)$ be the actual number of secondaries in the SDSS data for each of the various $\{\log\Lpri,\zpri\}$ bins. Using the best-fit model's predicted $P(\Delta V)$ for each $\{\log\Lpri,\zpri\}$ bin we draw $N_\rms(\Lpri,\zpri)$ values of $\Delta V$, and compute the likelihood of this fake data representing the best-fit model. We repeat this $10^4$ times. The red-shaded histogram in the inset-panel in the lower-right corner of Fig.~\ref{fig:SK_fit} shows the resulting distribution of likelihoods, which we then compare to the analogous likelihood for the actual SDSS data (blue, vertical dashed line). The fact that the latter is perfectly consistent with the distribution of likelihoods (red histogram) indicates that  the LOSVDs obtained from \Basilisk are in excellent agreement with the data, fitting not only the roughly Gaussian LOSVDs centered on $\Delta V=0$, but also the extended wings due to the interlopers.

Fig.~\ref{fig:Ns_fit} uses the same $(\log\Lpri,\zpri)$ binning and panels\footnote{Note that bins with five or fewer primaries in the down-sampled 
$\calL_{\rm NS}$ data have been omitted.} as Fig.~\ref{fig:SK_fit}, but this time we plot the distributions of the numbers of secondaries per primary (red dots with Poisson errorbars).  More specifically, the $x$-axis indicates the number of secondaries, $N_{\rm sec}$, and the $y$-axis indicates $\log(1+N_{\rm pri})$, where $N_{\rm pri}$ is the number of primaries that each have $N_{\rm sec}$ secondaries. In most cases the distributions clearly peak at $N_{\rm sec}=0$, as most primaries in SDSS DR7 do not have a spectroscopically detected secondary. The only exceptions are a few high-$\log L_{\rm pri}$ bins. Once again, in each panel the red shaded bands indicate 95\% confidence intervals obtained using the posterior distributions inferred with \Basilisc.
The inset-panel in the lower-right compares the likelihood of the SDSS data given the best-fit model, to the distribution of expected likelihoods computed by drawing $10^4$ random realisations of the best-fit multiplicity function. This time the likelihood of the SDSS data falls at the edge of the expected range, indicating that the fit to the data is not optimal. Indeed, upon closer inspection one can notice that the best-fit model overpredicts the multiplicity of primaries with $N_{\rm sec} \sim 3-6$, especially for some intermediate $\Lpri$ and $\zpri$ bins. As we demonstrate in a forthcoming paper, this small discrepancy arises from certain limitations in the satellite CLF model, and can be resolved by adopting a slightly more flexible halo-occupation modelling without significantly affecting any of the main relations presented here.

\begin{figure}
\centering
\includegraphics[width=0.48\textwidth]{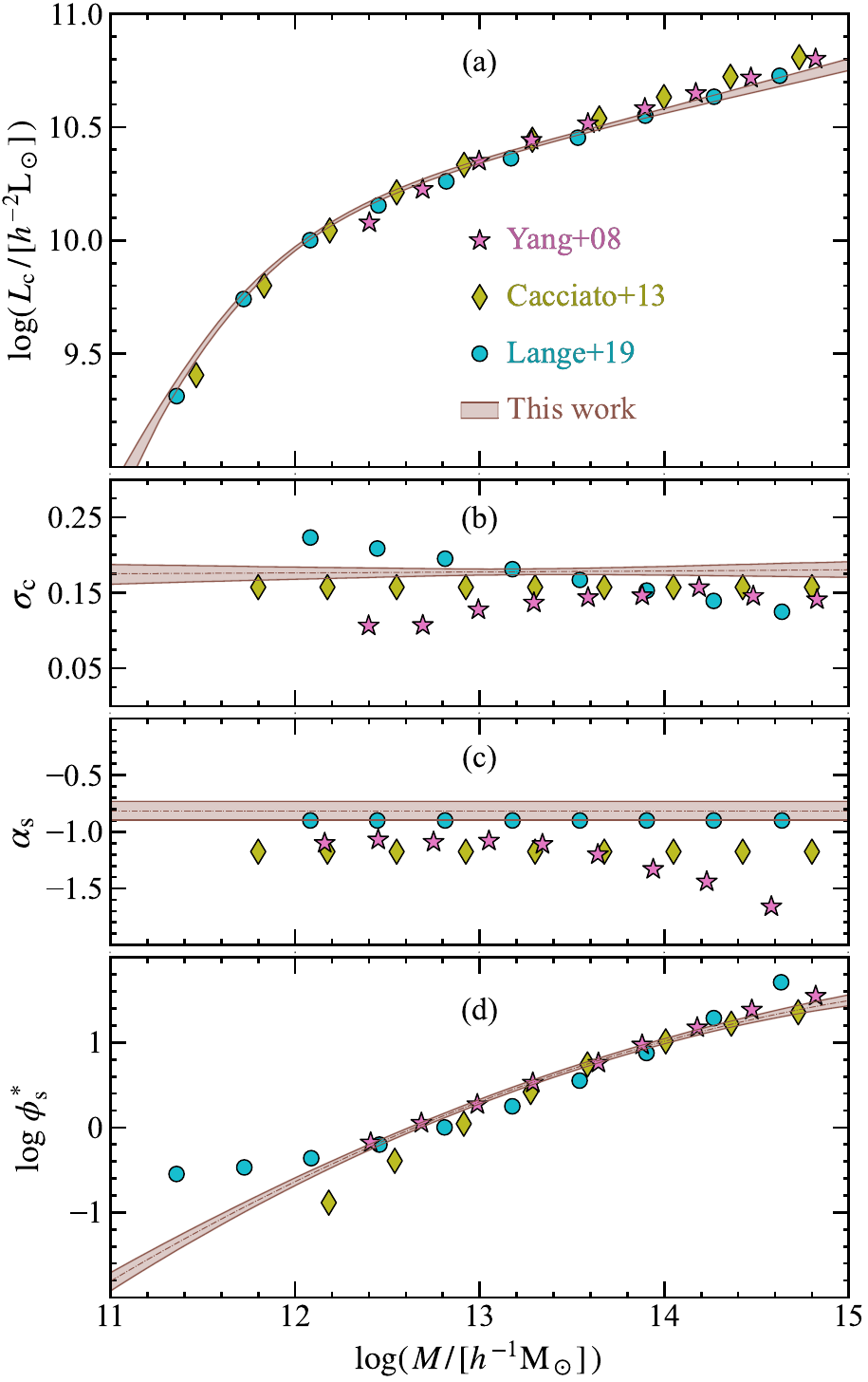}
\caption{Key halo mass dependencies that characterize the galaxy-halo connection in our model, as constrained by \Basilisk using SDSS data. In each panel, the brown shaded regions show the 95 per cent confidence intervals inferred by \Basilisc. The coloured symbols show constraints from previous studies using galaxy group catalogues \citep[][]{Yang.etal.08}, galaxy clustering \citep[][]{Cacciato.etal.13} and satellite kinematics \citep[][]{Lange.etal.19b}.  From top to bottom, the panels plot the median luminosity of centrals ($\bar{L}_\rmc$), the log-normal scatter in central luminosity ($\sigma_\rmc$), the faint-end slope of the satellite CLF ($\alpha_\rms$), and the normalization of the satellite CLF ($\phi_\rms^{*}$), each as a function of halo mass. see the text for a detailed discussion.}
\label{fig:SDSS_summary}
\end{figure}

Fig.~\ref{fig:SDSS_summary} shows several key halo mass dependencies that characterize the galaxy-halo connection inferred by \Basilisk from the SDSS data. In each panel, the brown shaded bands show the inferred 95~\% confidence intervals, while the coloured symbols show best-fit constraints from previous SDSS-based studies\footnote{Where needed we have converted these results to our definition of halo mass.}. In particular, we compare our inference to the results from an analysis of galaxy group catalogues by \citet{Yang.etal.08}, to results based on a simultaneous analysis of galaxy clustering and galaxy-galaxy lensing by \citet{Cacciato.etal.13}, and to results of the most recent analysis of satellite kinematics by \citet{Lange.etal.19b}. Note that the latter did not use the Bayesian hierarchical methodology of \Basilisc, but rather was based on the standard summary statistics of host-weighted and satellite-weighted  velocity dispersions as a function of binned primary luminosity.

Panel (a) plots the median central luminosity, $L_\rmc$, as a function of halo mass. As is evident, the constraints obtained with \Basilisk are in excellent agreement with previous results, though we emphasize that our constraints are significantly tighter.

Panel (b) plots the posterior constraints on $\sigma_\rmc(M)$, characterizing the scatter in $\log L_\rmc$ at fixed halo mass. Most studies in the past have assumed $\sigma_\rmc$ to be independent of halo mass, and inferred values that lie roughly in the range $0.15-0.2$~dex \citep[e.g.,][]{More.etal.09b, Cacciato.etal.13, Shankar.etal.14}. \Basilisc, on the other hand, allows for a mass dependence as characterized by equation~(\ref{scatter}). Yet, despite this extra degree of freedom, our inference is statistically consistent with a constant $\sigma_\rmc = 0.17$~dex over the entire range of halo masses probed.  Note that this is different from \citet{Lange.etal.19b} who, using a similar mass-dependent characterization of the scatter in the $\log L_\rmc$ - $\log M$ relation, inferred that $\sigma_\rmc$ increases with decreasing halo mass, as depicted by the blue circles in Panel (b). As discussed in more detail in \S\ref{sec:scatter}, the reason for this discrepancy can be attributed to the fact that all previous analyses, including \citet{Lange.etal.19b}, invariably assumed the brightest halo galaxy to be the central.

Panel (c) shows the posterior constraints on the the faint-end slope, $\alpha_\rms$, of the satellite CLF. Throughout we have assumed a global, mass-independent $\alpha_\rms$ similar to \citet{Cacciato.etal.13}, \citet{Lange.etal.19b} and most other previous work. Our inference that $\alpha_\rms = -0.87 \pm 0.06$ is in excellent agreement with \citet{Lange.etal.19b}, and is largely consistent with the constraints obtained by \citet{Yang.etal.08} and \citet{Cacciato.etal.13} given their uncertainties. Note, though, that \citet{Yang.etal.08} inferred that $\alpha_\rms$ becomes significantly steeper at the massive end, reaching values as low as $-1.5$ for groups with an inferred halo mass $M \gta 3 \times 10^{14} \Msunh$. In future work we plan to allow for a mass-dependent $\alpha_\rms$, to see if satellite kinematics reveal a similar mass dependence as that inferred from the galaxy group catalogue of \citet{Yang.etal.05}, and to study how this extra degree of freedom impacts the other parameters that characterize the galaxy-halo connection. 

Finally, panel (d) of Fig.~\ref{fig:SDSS_summary} shows the constraints on the normalization, $\phi_\rms^*$, of the CLF of satellite galaxies, as a function of host halo mass. The constraints obtained by \Basilisc, as depicted by the brown bands, are in fair agreement with previous constraints, especially if the uncertainties on the latter are taken into account (note that the coloured symbols only indicate the best-fit values). At the low mass end, the results of \citet{Lange.etal.19b}, which are also based on satellite kinematics, seem to suggest significantly larger values of $\phi_\rms^*$ (i.e., more satellites per halo). This is likely due to the fact that \ \citet{Lange.etal.19b} have assumed that interlopers have a uniform distribution of line-of-sight velocities, \citep[as did many other previous studies, such as][]{McKay.etal.02, Prada.etal.03, vdBosch.etal.04, Conroy.etal.07b, Norberg.etal.08, More.etal.09b, More.etal.11}. As discussed in \S\ref{sec:pdvrp}, this oversimplified assumption implies that some of the splash-back galaxies and infalling interlopers, which have a LOSVD that resembles that of true satellites, will be incorrectly `counted' as satellite galaxies. Because of this, and since we have demonstrated that \Basilisk can accurately recover the interloper fraction as well as the CLF normalization, $\phi_\rms^*(M)$, we reckon that our results for the abundance of satellite galaxies are likely to be more accurate. In all fairness, we emphasize that most previous studies of satellite kinematics were not aiming to accurately recover satellite abundances; rather they mainly focused on constraining halo masses as a function of primary galaxy luminosity. As discussed in detail in \citet{vdBosch.etal.04}, this aspect of the analysis is not significantly impaired by the oversimplified assumption that the LOSVD of interlopers is uniform.

\subsection{Scatter in central galaxy luminosity}
\label{sec:scatter}

Different empirical estimates of the galaxy-halo connection, at present, broadly agree on the relation between central galaxy luminosity (or stellar mass) and halo mass. However, the constraints on the scatter in this relation, especially as a function of halo mass, has yet to attain similar convergence, and therefore is considered a key parameter at the forefront of empirical modelling \citep[see e.g.,][]{Wechsler.Tinker.18} and is highly informative for testing physical models \citep[see e.g.,][]{Porras-Valverde.etal.2023}.

\begin{figure}
\centering
\includegraphics[width=0.48\textwidth]{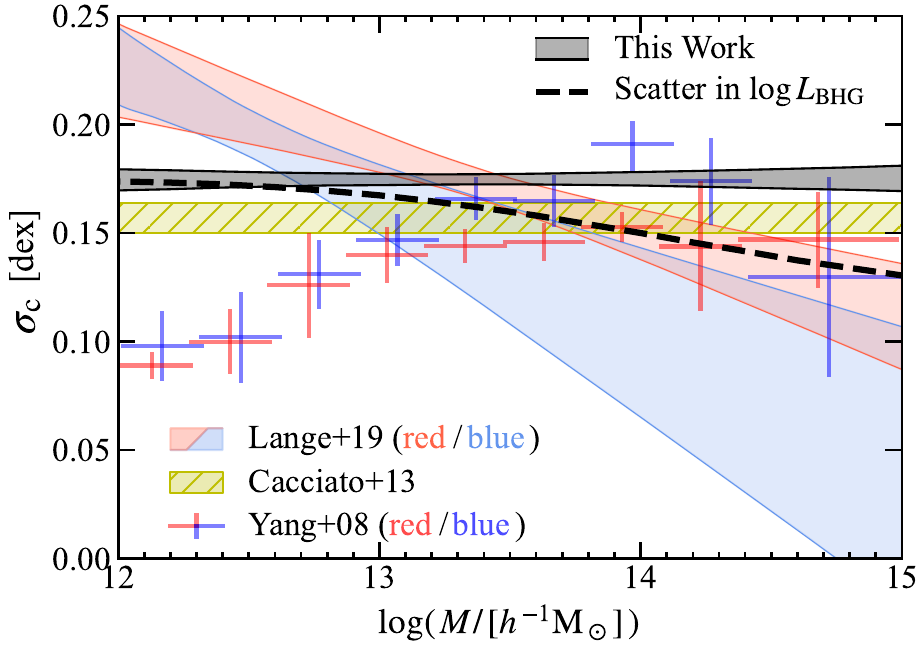}
\caption{The scatter in central galaxy luminosity, as inferred by \Basilisk (grey band), shows no mass dependence. This is in tension with the scatter for red and blue centrals, as inferred by a recent analysis of satellite kinematics in the SDSS by \citet{Lange.etal.19b}, shown by the red and blue shaded regions.  This tension can be resolved by noting that \citet{Lange.etal.19b} assumed that all their primaries are centrals. Consequently, their constraint on $\sigma_\rmc$ is really a constraint on the scatter in the brightest halo galaxies (BHGs). The black dashed line shows the predicted scatter in $\log L_{\rm BHG}$ for our best-fit model, computed using \Basilisc. Note that this is in excellent agreement with the results of  \citet{Lange.etal.19b} for the red centrals, which are the dominant population of centrals in massive haloes. For comparison, the yellow hatched region shows the constraints on $\sigma_\rmc$ (assumed to have no mass dependence) obtained by \citet{Cacciato.etal.13} based on an analysis of galaxy clustering plus galaxy-galaxy lensing, while the red and blue plus signs show the constraints for red and blue centrals, respectively, inferred by \citet{Yang.etal.08} using a galaxy group catalogue. All uncertainty bands and error-bars shown correspond to 68\% confidence intervals. see the text for a more detailed discussion.}
\label{fig:scatter_compare}
\end{figure}

\Basilisc's inference of the logarithmic scatter in central luminosity, $\sigma_\rmc(M_{\rm vir})$, shows no significant halo mass dependence despite having the freedom in the model. Fig.~\ref{fig:scatter_compare} compares our constraints on $\sigma_\rmc$ (grey band) with estimates from previous studies (all error-bars and uncertainties in this plot are 68\% confidence intervals). Crucially, our constraints disagree with \citet{Lange.etal.19b}, who also used satellite kinematics extracted from the SDSS DR-7 data. They split their sample into red and blue centrals, and the scatter in the two sub-populations are shown by the shaded regions of corresponding colour. At the high halo mass end, their red-fraction of centrals approaches unity, and thus the red-shaded region should be a good approximation of the overall scatter. As is evident, it reveals weak but significant mass-dependence with ${\rm d}\sigma_\rmc/{\rm d} \log M \approx -0.04$, which, at first sight appears inconsistent with our results. However, \citet{Lange.etal.19b}, as all other previous studies, have simply assumed that their primary, defined as the brightest galaxy in the selection cone, is always the central galaxy. Hence, their $\sigma_\rmc$ has to be interpreted as the scatter in the brightest halo galaxy, $\log L_{\rm BHG}$,  rather than that in $\log L_\rmc$.

\Basilisc, on the other hand, accounts for type-I impurities, which are BHG-satellites that are misclassified as primaries. In particular, we have demonstrated that by directly forward modelling these BHG-satellites, \Basilisc's inferred $\sigma_\rmc$ is an unbiased estimate of the intrinsic luminosity scatter of true centrals. The probability that the BHG is a satellite, rather than a central, increases strongly with halo mass \citep[][]{Skibba.etal.11,Lange.etal.18}. Therefore, the inferred scatter at the high mass end, from studies that did not account for type-I impurities, may have been biased. We can directly test this with \Basilisc, which makes it straightforward to compute the expected scatter in $\log L_{\rm BHG}$ and compare it to that in $\log L_\rmc$. The black dashed curve in Fig.~\ref{fig:scatter_compare} shows the predicted scatter in BHG luminosity, $\sigma_{\rm BHG}$, as a function of host halo mass for our best-fit CLF model. Note that $\sigma_{\rm BHG}$ drops significantly below $\sigma_\rmc$ at the high-mass end. This is because it is mostly the fainter centrals that are `replaced' by a brighter satellite, which causes the distribution of BHG luminosities to be narrower than that of the true centrals. For $M \gta 10^{13.5} \Msunh$, the mass-dependence of the inferred $\sigma_{\rm BHG}$ is in excellent agreement with \citet{Lange.etal.19b} (recall that the vast majority of all centrals in massive haloes are red, and the comparison should thus be with the red-shaded region). Above $10^{13} \Msunh$, the black dashed line also shows improved agreement with previous results from an analysis of galaxy clustering and galaxy-galaxy lensing by \cite{Cacciato.etal.13}, who assumed that $\sigma_\rmc$ is mass-independent, and that from an analysis of a SDSS galaxy group catalogue by  \citep{Yang.etal.08}.

As is evident from Fig.~\ref{fig:scatter_compare}, the various results disagree strongly at the low-mass end ($M \lta 10^{13} \Msunh$). This, however,  has to be interpreted with caution, as none of the constraints are particularly reliable there. For example, the results of \citet{Yang.etal.08} can not be trusted at the low-mass end, since their halo masses are estimated from the total group luminosity. Since this is dominated by the central luminosity in low mass haloes, their inferred scatter at the low mass end is guaranteed to be an underestimate \citep[see e.g.,][]{Campbell.etal.15}. In the case of \citet{Lange.etal.19b}, most of the constraining power comes from haloes with $M \geq 10^{13} \Msunh$. As they assume a simple linear dependence of $\sigma_\rmc$ on $\log M$, the constraints at the low-mass end mainly reflect extrapolation of the assumed linear relation. Our results are also affected this way, but less so, since we have used a flux-limited sample, rather than a more restricted volume-limited sample as in \citet{Lange.etal.19b}. This allows for a better sampling of fainter centrals, that reside in lower-mass haloes. 

\subsection{Orbital Anisotropy of Satellite Galaxies in SDSS}
\label{sec:anisotropy}

\begin{figure}
\centering
\includegraphics[width=0.48\textwidth]{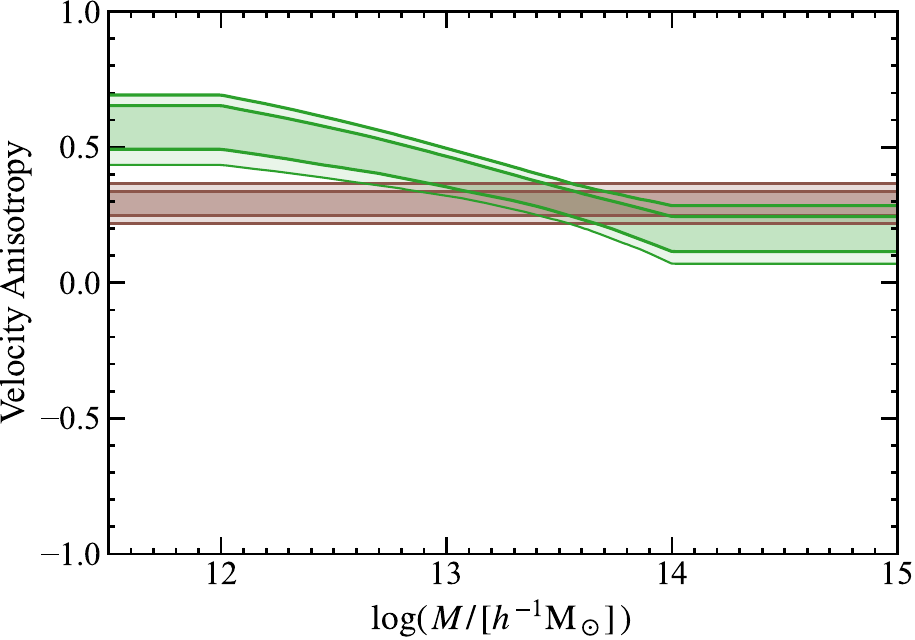}
\caption{Same as Fig.~\ref{fig:betamock} but for the SDSS data. Brown and green contours indicate the posterior constraints on the velocity anisotropy of satellite galaxies as inferred from our fiducial (constant-$\beta$) model, and the mass-dependent $\beta(M)$ model, respectively.}
\label{fig:advanced_anisotropy}
\end{figure}

The brown contours in Fig.~\ref{fig:advanced_anisotropy} show the 68 and 95 percentile constraints on the orbital anisotropy parameter, $\beta$, for satellite galaxies in the SDSS data as inferred from our fiducial model. We infer a significant radial anisotropy with $\beta = 0.29^{+0.05}_{-0.04}$. These global constraints on the average orbital anisotropy of satellite galaxies, across a large range of halo masses, are perfectly consistent with, but significantly tighter than, the results of \citet{Wojtak.Mamon.13} who also analysed the kinematics of satellite galaxies in SDSS data to infer  $\beta=0.26 \pm 0.09$. We emphasize that, unlike \Basilisc, the analysis of Wojtak \& Mamon did not account for mass mixing, and was based on a much smaller sample of primary-secondary pairs than used here.

Interestingly, our constraints on the orbital anisotropy are also consistent with the typical orbital anisotropy of subhaloes in numerical simulations of structure formation in a $\Lambda$CDM cosmology. In fact, the green contours show the constraints we obtain using a model in which the orbital anisotropy is allowed to depend on halo mass as given by equation~[\ref{eqn:beta_M}]. We find a weak indication that the orbits of satellite galaxies become more radially anisotropic towards lower halo mass. Most importantly, these results for the SDSS data (Fig.~\ref{fig:advanced_anisotropy}) are consistent with those for the Tier-3 mock data (Fig.~\ref{fig:betamock}), in which the satellite orbits reflect those of subhaloes in the $\Lambda$CDM-based SMDPL. The fact that the orbital anisotropy of satellite galaxies in the SDSS appears to be consistent with that of subhaloes in $N$-body simulations can be heralded as yet another success for the $\Lambda$CDM model.

Although the weak mass-dependence of the orbital anisotropy inferred here is intriguing, especially in light of the agreement with the Tier-3 results, we emphasize that these results have to be interpreted with caution. The reason is that we have excluded data on projected separations $< 55''$ because of fibre collision issues. As a consequence, the range in radii probed, in terms of the halo virial radius, in low mass haloes is different than that probed in more massive haloes. Hence, any potential radial dependence of the orbital anisotropy of satellite galaxies can, in principle, masquerade as a mass dependence in our analysis. We intend to address this `degeneracy' in a forthcoming paper (Mitra et al., in prep) in which we consider models in which the orbital anisotropy is allowed to depend on halo-centric radius, as well as halo mass. In particular, we will consider Osipkov-Merritt model \citep[][]{Osipkov.79, Merritt.85}, as well as more realistic simulation-inspired models such as those used by \citet{Mamon.Lokas.05}.


\section{Summary and Conclusion}
\label{sec:summary}

In \paperI we presented \Basilisc, a novel, Bayesian hierarchical method for analysing the kinematics of satellite galaxies. Based on the spherically symmetric Jeans equations it models the kinematics of large ensembles of satellite galaxies associated with central galaxies that span a wide range in halo mass and luminosity. The halo masses of the individual centrals act as latent variables in a hierarchical Bayesian framework that uses the data to  constrain the detailed galaxy–halo connection as characterized by the CLF.  Unlike traditional methods for analysing satellite kinematics, \Basilisk does not make use of any summary statistic, such as velocity dispersions of satellite galaxies in central galaxy luminosity bins. Rather, it leaves the data in its raw form, which has the advantage that all data are used optimally while avoiding systematics that arise from binning. In addition, whereas traditional methods typically require volume-limited samples, \Basilisk can be applied to flux limited samples, thereby greatly enhancing the quantity and dynamic range of the data. And finally, \Basilisk is the only available method that simultaneously solves for halo mass and orbital anisotropy of the satellite galaxies, while properly accounting for `mass-mixing'.

In this paper we have presented a number of important improvements to \Basilisc, required for an unbiased recovery of all parameters when using large samples of data comparable to what can be achieved with existing SDSS catalogues. In particular, 
\begin{itemize}

\item We introduced an improved selection of primaries and secondaries that assures that the secondaries associated with each individual primary are volume-limited, even-though the overall sample is still flux-limited. This facilitates a more accurate modelling of the abundance and velocity distribution of the secondaries.

\item We forward model the contribution of impurities among the primaries, where impurities are predominantly those satellites that are brighter than their corresponding centrals. 

\item We slightly modified the selection criteria of primaries to minimize the effect of other kinds of impurities that are extremely difficult to forward-model.

\item We extended the satellite kinematics model to higher-order, by using the fourth-order Jeans equation to compute the kurtosis of the LOSVD. Incorporating this, in the modelling of the full 2D phase-space distribution of satellites, allows \Basilisk to break the mass-anisotropy degeneracy, and to put tight constraints on the global, average velocity anisotropy of satellite galaxies.

\item We drastically improved the modelling of interlopers by (i) accounting for the fact that the selection volume of secondaries is conical rather than cylindrical, (ii) accounting for splash-back galaxies, and (iii) using a data-driven method to model the line-of-sight velocity distribution of the interlopers.

\item Instead of discarding primaries with zero secondaries, \Basilisk utilizes their information to further constrain the galaxy-halo connection. Congruous with the satellite kinematics methodology, we introduced a similar Bayesian hierarchical framework to model the abundance of secondary galaxies around each primary, which improves significantly on the stacking-based approach used in \papI. This allows \Basilisk to put unprecedented constraints on the satellite CLF.

\end{itemize}

Using realistic mock data of similar quality and volume as the SDSS DR7, we have demonstrated that, with this improved methodology, \Basilisk can break the mass-anisotropy degeneracy, and simultaneously constrain the host masses and average orbital velocity anisotropy of satellite galaxies. In particular, \Basilisk achieves an unbiased recovery of all 10 CLF parameters that characterize the galaxy-halo connection covering almost four orders of magnitude in halo mass (from $\sim 10^{11}$ to $10^{15} \Msunh$), and with unprecedented accuracy. In addition, it simultaneously recovers the orbital anisotropy parameter, $\beta$, the luminosity and redshift dependence of the interloper fraction, and the radial number density profile of satellite galaxies. It is worth emphasizing that the recovery is unbiased despite the fact that the selection of primaries and secondaries is (unavoidably) plagued by biases, incompleteness, and impurities.

We applied \Basilisk to a sample of $18,373$ primaries and $30,431$ secondaries extracted from the SDSS DR-7 data, yielding some of the tightest constraints on the galaxy-halo connection to date (Table~\ref{table:SDSS_params}). The model accurately reproduces both the abundance and line-of-sight velocity distributions of secondaries (Figs.~\ref{fig:SK_fit} and Fig.~\ref{fig:Ns_fit}), and is in good agreement with previous constraints on the galaxy-halo connection derived from galaxy group catalogues, galaxy clustering, galaxy-galaxy lensing and previous analyses of satellite kinematics. 

Assuming that the orbital anisotropy of satellite galaxies is independent of halo mass and halo-centric radius, our analysis of SDSS data reveals a significant radial anisotropy of $\beta = 0.29^{+0.05}_{-0.04}$, in excellent agreement with, but significantly tighter than, previous results \citep[][]{Wojtak.Mamon.13}. We also find a weak indication that $\beta$ is slightly larger in lower mass haloes, in good agreement with the orbital anisotropy of subhaloes in dark-matter only simulations of structure formation in a $\Lambda$CDM cosmology \citep[e.g.,][]{Diemand.etal.04, Cuesta.etal.08, Sawala.etal.17, vdBosch.etal.19}. Since satellite are believed to reside in subhaloes, this may be considered another success of the standard model for structure formation.

We find that the radial number density profile of satellite galaxies, $n_{\rm sat}(r|M)$, is tightly constrained and well characterized by a generalized-NFW profile (equation~[\ref{nsatprof}]) with a central cusp-slope $\gamma=0.94$ (compared to $\gamma=1$ for a pure NFW profile), and a characteristic scale radius that is roughly two times larger than what is expected for the dark matter. Within the uncertainties, this is consistent with several previous studies \citep[e.g.,][]{Yang.etal.05a, Chen.08, More.etal.09b, Guo.etal.12a, Lange.etal.19b}. Consistent with paper~I, we find our results to be extremely robust to modest changes in $n_{\rm sat}(r|M)$; the only parameter that displays some dependence is the anisotropy parameter, $\beta$ (see Appendix~\ref{App:nprof}). This is to be expected given that both $\beta$ and $n_{\rm sat}(r|M)$ appear in the Jeans equation used to model the kinematics of the satellite galaxies.

Interestingly, we find no evidence for a significant halo mass dependence of the scatter in central luminosity, and at any given halo mass we find the luminosity scatter to be around $\sigma_\rmc = 0.17 \, {\rm dex}$. This is inconsistent with the latest analysis of satellite kinematics by \citet{Lange.etal.19b}, also based on SDSS DR-7 data, who inferred that the scatter decreases with increasing halo mass, albeit only weakly with $\rmd\sigma_\rmc/\rmd\log M \sim -0.04$. As discussed in \S\ref{sec:scatter}, this discrepancy arises primarily from the fact that \citet{Lange.etal.19b} and all previous studies simply assumed the brightest galaxy in the halo to be the central. We, however, take into account the existence of brightest halo galaxy satellites, and forward-model the probability of misidentifying them as primaries. By doing so, we demonstrate that \Basilisc's inference of $\sigma_\rmc$ is an unbiased recovery of the intrinsic luminosity scatter of true centrals. From our best-fit model we can predict what the scatter in brightest halo galaxy luminosity should be, as a function of halo mass, and that is in good agreement with the scatter inferred by \citet{Lange.etal.19b} and other previous analyses.

For completeness, we point out that several studies that used stellar mass, rather than $r$-band luminosity, to characterize the galaxy-halo connection also inferred that the scatter in stellar mass of central galaxies decreases with increasing halo mass \citep[][]{Moster.etal.10, Zu.Mandelbaum.15, Behroozi.etal.19}. However, most of these inferences were only significant below the $\sim 2 \sigma$ level. Hence, observationally  it remains unclear whether or not the scatter in the galaxy-halo connection has a significant mass dependence. Taking our results at face value, it seems that scatter is fairly mass-independent, at least for $\log M \gta 10^{12} \Msunh$, and that previous indications for a significant mass dependence at the massive end are likely a result of confounding true centrals with brightest halo galaxies. 

On the theory side, the situation is even more higgledy-piggledy, with a clear lack of consensus \citep[e.g., see Fig.~2 in][]{Porras-Valverde.etal.2023}. In general, semi-analytical models \citep[e.g.,][]{Somerville.etal.12, Lu.etal.14b, Henriques.etal.15, Croton.etal.16} predict a weak mass dependence with a small, negative value for $\rmd\sigma_\rmc/\rmd\log M$, but the magnitude of the overall scatter is typically much larger than what is inferred observationally \citep[][]{Wechsler.Tinker.18}. Hydrodynamical simulations of galaxy formation, typically predict a significantly lower scatter, at least for haloes with $M \gta 10^{12} \Msunh$, in much better agreement with observations. Typically, though, they predict that the scatter rapidly increases for $M \lta 10^{12} \Msunh$ \citep[e.g.,][]{Matthee.etal.17, Pillepich.etal.18}. Finally, empirical models such as the \texttt{UniverseMachine} \citep[][]{Behroozi.etal.19} and \texttt{EMERGE} \citep[][]{Moster.etal.18} seem to predict $\sigma_\rmc(M)$ relations that fall roughly in between the predictions from semi-analytical models and hydrodynamical simulations.

To conclude, we have demonstrated that satellite kinematics extracted from galaxy redshift surveys contain a wealth of information regarding the statistical relation between galaxies and their associated dark matter haloes. The Bayesian hierarchical framework \Basilisc, developed here and in \papI, is able to analyse such data in an unbiased way, yielding accurate constraints on the galaxy-halo connection over a wide range of halo mass, and with unprecedented precision. Hence, satellite kinematics is complementary to other techniques that are used to constrain the galaxy-halo connection, in particular galaxy clustering and galaxy-galaxy lensing. Importantly, by only probing the smallest, most non-linear scales (i.e., the 1-halo term) it is insensitive to halo assembly bias, which hampers an unambiguous interpretation of the 2-halo term in clustering and galaxy-galaxy lensing. Hence, it is to be expected that by combining all these methods, degeneracies can be broken which opens up new avenues to test our cosmological paradigm and our models for galaxy formation. To this end, we plan to use, and where necessary further develop, \Basilisk in future work. In particular, among others, we intend to explore additional degrees of freedom in the characterization of the galaxy-halo connection (for example, mass dependence in the faint-end slope of the satellite CLF and in the {\it ratio} of the characteristic luminosities of the centrals and satellites in haloes of any given mass), the impact of baryonic effects on the halo potential (which may introduce systematic errors in the inference from satellite kinematics), and the impact of scatter in the halo concentration-mass relation (and the expected correlation with the abundance of subhaloes/satellite galaxies). In addition, we are excited about the prospects of using \Basilisk to probe the galaxy-halo connection as a function of secondary galaxy properties, such as galaxy colour and/or size, and to put constraints on cosmological parameters by combining satellite kinematics with other observables. 


\section*{Acknowledgments}

We are grateful to the anonymous referee for an insightful referee report that has resulted in significant improvements of the manuscript. FvdB has been supported by the National Aeronautics and Space Administration through Grant No. 19-ATP19-0059 issued as part of the Astrophysics Theory Program and by the National Science Foundation (NSF) through grant AST-2307280, and received additional support from the Klaus Tschira foundation. This work was performed in part at the Aspen Center for Physics, which is supported by National Science Foundation grant PHY-1607611, and at the KITP Santa Barbara, which is supported in part by the National Science Foundation under Grant No. NSF PHY-174895.

This work utilized, primarily for plotting purposes, the following python packages: \texttt{Matplotlib} \citep{Matplotlib_Hunter2007}, \texttt{SciPy} \citep{SciPy_Virtanen2020}, \texttt{NumPy} \citep{numpy_vanderWalt2011}, and \texttt{PyGTC} \citep{PyGTC_Bocquet2016}.


\section*{Data Availability}

The SDSS DR7 New York University Value Added Galaxy Catalog \texttt{bright0} sample, used in this work, is publicly available at \url{http://sdss.physics.nyu.edu/lss/dr72/bright/0/}. All derived data, such as kinematic data of primaries and secondaries, and the data corresponding to each plot, will be made available on reasonable request to the corresponding author.


\bibliographystyle{mnras}
\bibliography{references_vdb}


\appendix

\section{Optimizing Selection Criteria}
\label{App:selcrit}
\begin{figure*}
\centering
\includegraphics[width=\textwidth]{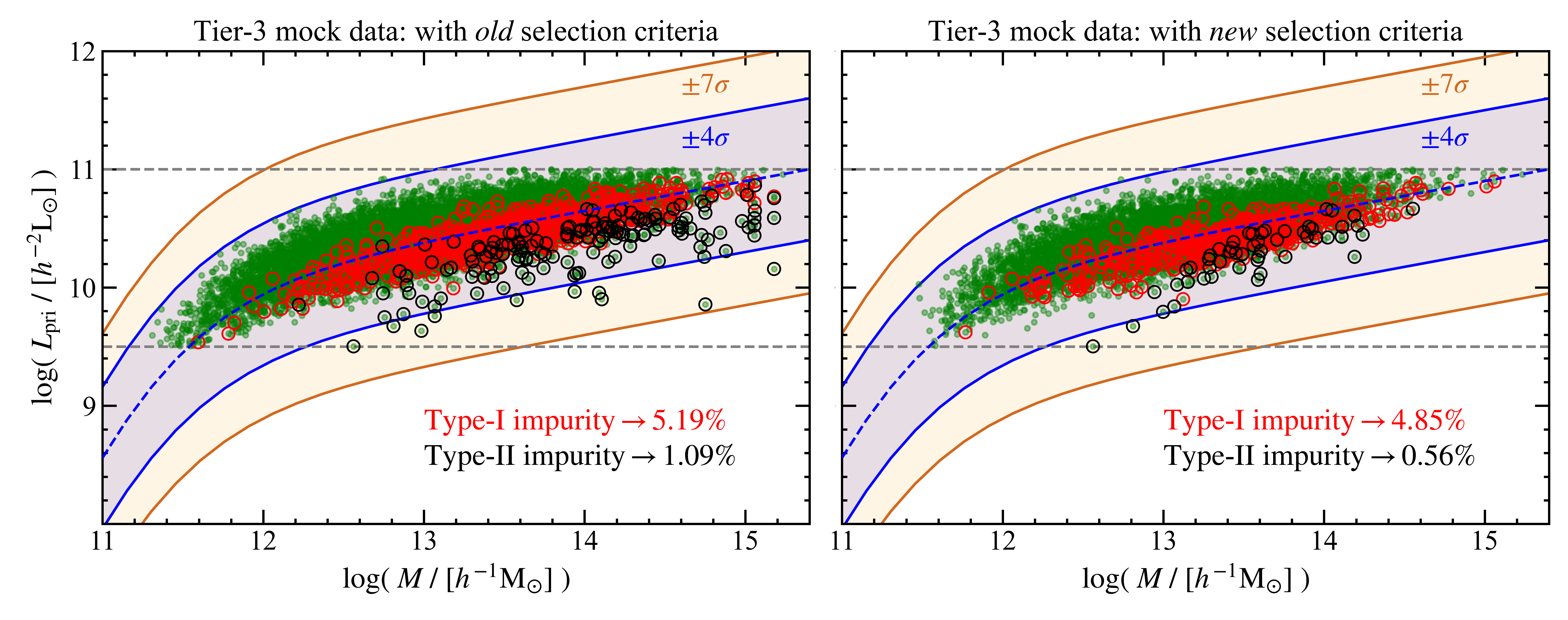}
\caption{Luminosity and halo mass of primaries in our Tier-3 mock satellite kinematics sample, identified using old (left) and new (right) selection criteria. We introduce the new, more restrictive, selection criteria to reduce the number of type-II impurities which are marked with black circles. Note that the new selection criteria especially eliminate the most problematic impurities which are offset from the input mass-luminosity relation by more than $4 \sigma_\rmc$ (blue shaded region) and up to as much as $7 \sigma_\rmc$ (yellow shaded region). It also distinctively reduces impurities at the high-mass end, which is crucial for obtaining unbiased constraints.}
\label{fig:selection_compare}
\end{figure*}

As described in \S\ref{sec:selection}, the selection of primaries and secondaries makes use of conical selection volumes. In this appendix we describe modifications to the parameters characterizing these selection volumes that result in an improved purity of the sample. Following \cite{vdBosch.etal.04}, \cite{More.etal.09b} and \cite{Lange.etal.19b} the selection cones are characterized by  $\Rh = a_\rmh \,\sigma_{200} \mpch$, $\Rs = a_\rms \, \sigma_{200} \mpch$,  $\dVh = b_\rmh \,\sigma_{200} \kms$, and $\dVs = b_\rms \,\sigma_{200} \kms$ (see Fig.~\ref{fig:conical_cylinders}). Here $\sigma_{200}$ is an estimate for the satellite velocity dispersion in units of $200 \kms$, which scales with the luminosity of the primary as $\log \sigma_{200} = c_0 + c_1 \log L_{10} + c_2 (\log L_{10})^2$, where $L_{10} = \Lpri / (10^{10} \Lsunh)$. In \paperI we adopted exactly the same parameters as \cite{Lange.etal.19b}: $a_\rmh=0.5$, $a_\rms = 0.15$, $b_\rmh = 1000$, $b_\rms = 4000/\sigma_{200}$ and $(c_0, c_1, c_2) = (-0.04, 0.38, 0.29)$. As discussed in \papI, this results in impurity fractions of $\sim 5$ percent. These impurities cause a slightly bias estimate of the scatter in the galaxy-halo connection. In the case of the small mock data samples used in \papI, the effect was not significant. However, when using an 8-times larger, full-size SDSS sample the systematic bias in scatter becomes $>3 \sigma$ significant.

In \S\ref{sec:PMLz} we classified impurities as either Type-I (BHG satellites) or Type-II (neither a central nor a BHG satellite). The green dots in the left-hand panel of Fig.~\ref{fig:selection_compare} show the luminosities as function of host halo mass of the primaries selected from the Tier-3 mock data using the selection criteria used in \papI. Blue and orange contours mark the $4$ and $7\sigma_\rmc$ ranges around the median relation between halo mass and central luminosity used to construct the mock data. Red and black circled dots mark impurities of Type-I and~II, respectively. As is evident, Type-I impurities have luminosities that are comparable to those of true centrals at the same host halo mass. That is because Type-I impurities are BHG satellites which are brighter than their corresponding central galaxy (hence they must have a luminosity in the typical range of $\Phi_\rmc(L|M)$ for a true central to be fainter). Being the brightest one in the corresponding halo, a Type-I impurity is impossible to avoid in the selection procedure. However, as we have demonstrated in \S\ref{sec:PMLz}, we actually forward model the contribution of Type-I impurities.

Type-II impurities, though, are a much bigger concern. As is evident from Fig.~\ref{fig:selection_compare}, these can have luminosities that are much lower than that of a typical central at the corresponding halo mass (by as much as $7\sigma$). Since lower luminosities are indicative of a lower halo mass, a too-large contribution of Type-II impurities can give rise to significant, systematic errors in the inference. Since the kinematic information from secondaries associate with Type-II impurities still reflect a high velocity dispersion consistent with the actual halo mass, the main effect of Type-II impurities is to cause a systematic overestimate in the scatter of central luminosities at a fixed halo mass (i.e., an overestimate of $\sigma_\rmc$). Since we are not aware of a reliable method to forward model the impact of Type-II impurities, it is prudent that we minimize their incidence by tuning our selection criteria accordingly.

After extensive testing with different mock data sets similar to the Tier-3 mock discussed in the main text, we finally settled on the following set of parameters: $a_\rmh = 0.6$, $a_\rms =0.15$, $b_\rmh = b_\rms = 1000$ and $\{c_0, c_1, c_2\} = \{0.04, 0.48, 0.05\}$. With these new selection criteria we are able to reduce the fraction of Type-II impurities from $\sim 1.1 \%$ to $\sim 0.5 \%$. The impact of this reduction can be seen by comparing the two panels of Fig.~\ref{fig:selection_compare}. Note that in addition to dropping the fraction of Type-II impurities to sub-percent levels, the new selection criteria preferentially removes the most dramatic outliers and also drastically reduces the contribution of Type-II impurities at the high mass end, where the old selection criteria caused the fraction of Type-II impurities to be very high.

With these new and improved selection criteria we find that Type-II impurities no longer cause a significant overestimate of $\sigma_\rmc$. Although the new selection criteria reduces the number of primaries in the satellite kinematics sample by almost 40 percent, we find that this does not significantly compromise the precision with which \Basilisk can infer the galaxy-halo connection. The reason is that the main reduction of primaries occurs at the low-luminosity end, where most of the secondaries are interlopers that do little to constrain the halo occupation model. 

\section{The completeness of centrals}
\label{App:compl}

\begin{figure}
\centering
\includegraphics[width=0.48\textwidth]{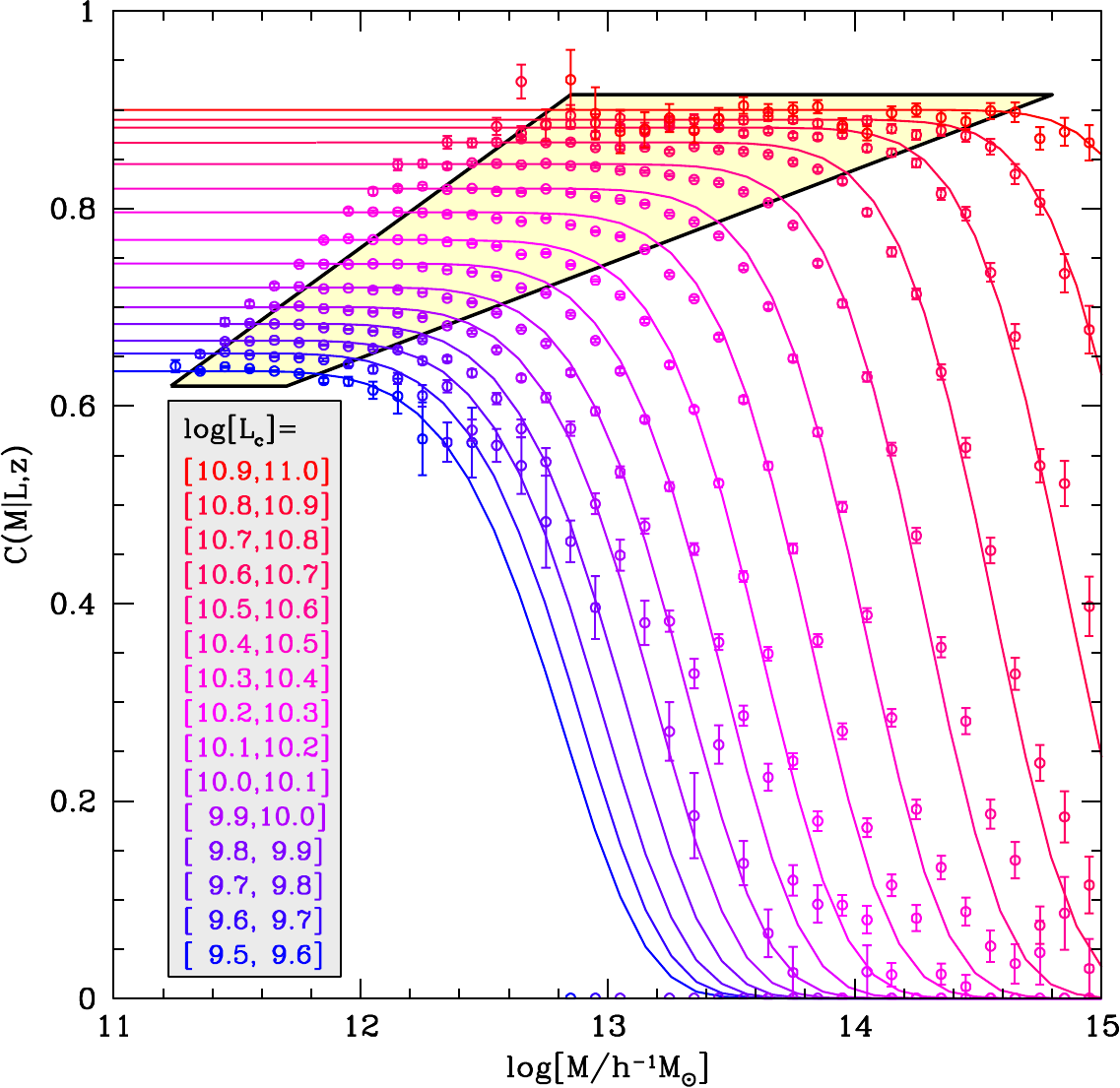}
\caption{Halo completeness, $\calC(M|L,z)$, defined as the fraction of haloes of mass $M$, hosting centrals of luminosity $L$ at redshift $z$, whose centrals are selected as primaries by our selection criteria. Points with Poisson error-bars indicate halo completeness as a function of halo mass in our Tier-3 mock sample. Results are shown for different bins of central galaxy luminosity (different colours, as indicated). For each luminosity bin, the completeness is roughly independent of halo mass at the low-mass end, but with a steep, almost exponential, decline at the high mass end. The yellow shaded region roughly indicates, for each luminosity bin, the 5 to 95 percentile range of halo masses. Note that the exponential decline of $\calC(M|L,z)$ at large masses only affects the top $\sim 5$ percent of centrals of a given luminosity. For the vast majority of centrals the mass dependence of $\calC(M|L,z)$ is therefore negligible (which was the assumption we made in \papI). The coloured, solid lines show the theoretically predicted $\calC(M|L,z)$ computed under the assumption that the steep decline at the high-mass end is entirely due to centrals being fainter than their corresponding brightest satellites. As is evident, this is an excellent fit to the data, indicating that we can actually take the full mass-dependence of $\calC(M|L,z)$ into account by forward-modelling the halo occupation of brightest halo galaxies, instead of centrals. see the text for details.}
\label{fig:compl}
\end{figure}

The selection of centrals as primaries (\S\ref{sec:sample_selection}) is not complete; i.e., not every central is selected as a primary. This incompleteness owes to two different reasons: (i) incompleteness in the SDSS redshift survey, for example due to fibre-collisions, or (ii) the central is located inside the selection cone of a brighter galaxy.  Let the completeness $\calC(M,L,z)$ be the fraction of centrals --- of luminosity $L$, at redshift $z$, residing in haloes of mass $M$, in the survey volume of the SDSS --- that are selected as primaries. We can write that $\calC(M,L,z) = \calC(M|L,z) \calC_0(L,z)$. As discussed in the main text (see \S\ref{sec:modelSK}), the modelling in \Basilisk is independent of $\calC_0$, which drops out. In other words, we only need to account for any potential halo mass dependence of the completeness given by $\calC(M|L,z)$.

In order to gauge this mass dependence, we construct 100  mock SDSS redshift surveys similar to the Tier-3 mock survey discussed in the main text to which we apply our primary selection criteria. For each central galaxy in the mock SDSS volumes we assess whether it is selected as a primary. The combined results from all 100 mocks are shown as symbols with Poisson errorbars in Fig.~\ref{fig:compl}. Different colours correspond to different luminosities of the centrals, as indicated. Here we have combined data on all centrals over the entire redshift range, but we emphasize that the redshift dependence is weak. A few trends are evident. First of all, the completeness is lower for fainter centrals. This simply reflects that fainter centrals are more likely to have a brighter galaxy in a neighbouring halo that happens to fall within the primary selection criterion. Modelling this would require accurate knowledge of the clustering of haloes (2-halo term) and is sensitive to assembly bias issues. Fortunately, we do not need to model this luminosity dependence. All we care about is the halo mass dependence as characterized by $\calC(M|L,z)$. 

As is evident from the data, the completeness for centrals of a given luminosity is roughly independent of halo mass at the low-mass end, but then drops drastically at the high-mass end. This transition occurs at higher mass for brighter centrals. The yellow-shaded region indicates, for each luminosity bin, the 5 to 95 percentile range of halo masses. Note that the exponential decline of $\calC(M|L,z)$ at large masses only affects the top $\sim 5$ percent of centrals of a given luminosity. This is why, in \papI, we decided to ignore this mass dependence all together. However, as we demonstrate below, it is actually fairly straightforward to model $\calC(M|L,z)$. As it turns out, this mass dependence owes almost entirely to the fact that a central of a given luminosity in a more massive halo is more likely to have a satellite that is brighter than itself. Recall that $\Phi_\rmc(L|M)$ is modelled as a log-normal distribution. Hence, the central galaxies in haloes of a given mass have a tail of excessively faint centrals. And since we assume that the luminosities of satellite galaxies are independent of that of their central, those faint centrals are more likely to have a brighter satellite, and thus to fail selection as a primary. As shown in Appendix~A of \cite{Lange.etal.18}, the probability that the central galaxy is the brightest galaxy in its halo is given by
\begin{equation}\label{PBNC}
P_{\rm BC}(M|L,z) = \exp\left[-\Lambda_{\rm BTC}\right]\,.
\end{equation}
Here $\Lambda_{\rm BTC}$ is the expectation value for the number of satellites that are brighter than the central, which is given by equation~(\ref{Lambda_for_L_M}).

The solid lines in Fig.~\ref{fig:compl} show the predictions for $P_{\rm BC}(M|L,z)$, computed using equation~(\ref{PBNC}) for the same CLF model as used to construct the mocks. The absolute normalization for each luminosity bin, which represents $\calC_0(L,z)$, is tuned to match the mock data at the low-$\log M$ end. As is evident, equation~(\ref{PBNC}) accurately describes the halo-mass dependence of the completeness of primaries. Hence, the mass-dependence of the completeness of centrals can be modelled as $\calC(M|L,z) = P_{\rm BC}(M|L,z)$. 

Note that $1-\calC(M|L,z)$ is the probability that a central is {\it not} the brightest galaxy in its halo, and thus the probability that the halo gives rise to a Type-I impurity. As discussed in the main text, we forward model the contribution of these Type-I impurities, which effectively means that we already account for the mass-dependence of the completeness of centrals depicted in Fig.~\ref{fig:compl}. Indeed, $P_{\rm BC}$ given by equation~(\ref{PBNC}) is identical to $P(L_{\rm bs}<L|M,z)$ (equation~[\ref{P_Lbs_lt_L}]) used in \S\ref{sec:PMLz} to forward model the Type-I impurities. 

\section{The radial profile of satellites}
\label{App:nprof}

\begin{figure*}
\centering
\includegraphics[width=0.85\textwidth]{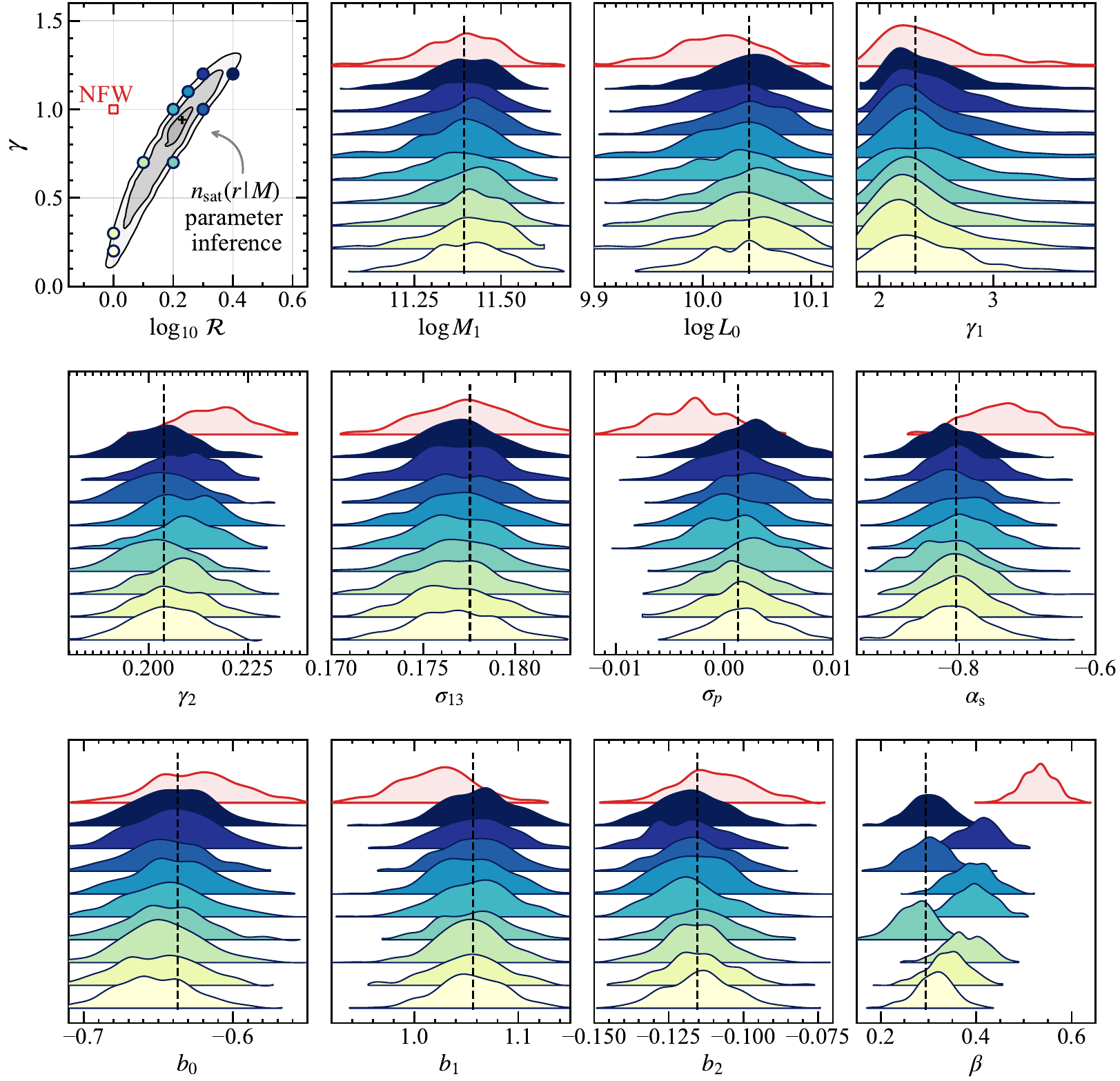}
\caption{\Basilisc's constraints on the radial density profile of satellite galaxies in the SDSS data, and the impact of varying the profile on the posteriors of other parameters. The top-left panel shows the 1, 2 and 3$\sigma$ confidence intervals (grey contours) on $\{ \gamma, \, \calR \}$ as inferred by \Basilisk from the SDSS data. For comparison, the red square labelled 'NFW' corresponds to a model in which the satellite galaxies are an unbiased tracer of the dark matter distribution (i.e., $\{ \gamma, \, \calR \} = \{1,1\}$), and is clearly inconsistent with the data at more than $5\sigma$. Rather, the data prefers a model in which the satellites follow a radial number density profile that is significantly less concentrated than the underlying dark matter. The remaining panels show the inferred posteriors of all CLF parameters and the velocity anisotropy ($\beta$), as inferred by \Basilisc, assuming different values of $\{ \gamma, \, \calR \}$ (marked by corresponding coloured circles in the top-left panel). The vertical black dashed lines show the best-fit parameter values (quoted in table~\ref{table:SDSS_params}) corresponding to the best-fit $\{ \gamma, \, \calR \}$. The red histograms show the posterior if \Basilisk is forced to assume, in its modelling, that the satellites are an unbiased tracer of the dark matter mass distribution. Note that the constraints on the CLF parameters are very robust to moderate changes in $\gamma$ and/or $\calR$. see the text for details.}
\label{fig:calR_gamma}
\end{figure*}

Throughout we assume that $n_{\rm sat}(r|M)$ is characterized by a generalized-NFW form (equation~\ref{nsatprof}) which has two free parameters: the inner logarithmic density slope, $\gamma$, and the concentration ratio $\mathcal{R} = c_{\rm vir} / c_{\rm sat}$ which characterizes the scale radius of the number density profile. As discussed in \S\ref{sec:implementation}, \Basilisk pre-computes and stores essential arrays which are then used in each step of the MCMC chain varying the CLF, anisotropy, and nuisance parameters. This pre-computation drastically speeds up \Basilisc, but requires that $n_{\rm sat}(r|M)$, and thus \{$\gamma,\, \calR$\}, are held fixed. Therefore, instead of keeping the radial profile free, we run separate MCMC chains for each assumed radial profile on a $15 \times 15$ grid of \{$\gamma,\, \log \calR$\}. We then combine the posteriors and likelihoods from each of these runs to compute the marginalized likelihood $\calL(\gamma, \calR | \bD)$.

The grey contours in the top-left panel of Fig.~\ref{fig:calR_gamma} show the 68, 95 and 99 percent confidence intervals for $\gamma$ and $\log \calR$ thus obtained using the SDSS data described in \S\ref{sec:SDSSdata}. The best-fit values, indicated by the black cross, correspond to $\{\gamma,\, \calR\} = \{0.94,\, 1.7\}$, which are the values we adopt for our detailed SDSS-analysis described in \S\ref{sec:SDSSresults}. However, the confidence intervals for $\gamma$ and $\log \calR$ reveal a significant degeneracy along a narrow ridge-line in $\gamma-\calR$ parameter space (see also \papI). To demonstrate the impact this degeneracy has on our inference, the coloured histograms in the other panels of Fig.~\ref{fig:calR_gamma} show the posteriors on our CLF parameters and the anisotropy parameter, $\beta$, for 9 different combinations of $\gamma$ and $\calR$ (indicated by the circles of corresponding colour in the top-left panel) that roughly trace out the boundary of the $3\sigma$ confidence interval. The vertical black dashed line in each of these panels show the best-fit parameters inferred by \Basilisk with the best-fit $\{\gamma, \, \calR\}$ combination, same as the values quoted in table~\ref{table:SDSS_params}.  As is evident, the inferred CLF parameters are extremely robust to changes in $\gamma$ and $\calR$ along the direction of this degeneracy. The only parameter that shows a weak dependence is the orbital anisotropy parameter $\beta$ (bottom-right panel), which is to be expected from the fact that both $n_{\rm sat}(r|M)$ and $\beta$ appear in the expression for the line-of-sight velocity dispersion given by equation~(\ref{sigmalos}). 

Note also that the constraints on $\gamma$ and $\calR$ are inconsistent with satellite galaxies following the same radial profile as the dark matter (i.e., $\gamma = \calR = 1$, indicated by the red square in the top-left panel) at $>5\sigma$ significance. If we had assumed that satellite galaxies are an unbiased tracer of their host halo mass distribution, which is not uncommon in the literature when modelling the galaxy-halo connection, we would have obtained the posteriors indicated by the red histograms. Interestingly, most CLF parameters would still be consistent with the values inferred using our fiducial, best-fit model with $\{\gamma,\, \calR\} = \{0.94,\, 1.7\}$. The main exceptions, though, is the orbital anisotropy parameter, which would be biased high (i.e., we would markedly overestimate the radial velocity anisotropy). A few other parameters like $\gamma_2$, $\alpha_\rms$, and  $\sigma_\rmp$, are also somewhat biased in red histograms. Thus, we would incorrectly infer a steeper $\bar{L}_\rmc (M)$ relation, a shallower faint-end slope of satellite CLF, and a slight decreasing trend in the central luminosity scatter with host halo mass, if we wrongly assumed the satellites to follow the dark matter radial distribution. In conclusion, \Basilisk yields tight constraints on the radial number density profile of satellite galaxies, and whatever degeneracy remains between the central density slope and concentration has no significant impact on any inferred parameter. 


\bsp
\label{lastpage}
\end{document}